\documentclass[german,english]{article}
\usepackage[T1]{fontenc}
\usepackage[latin9]{inputenc}
\usepackage{geometry}
\usepackage{color}
\usepackage{babel}
\usepackage{float}
\usepackage{amsmath}
\usepackage{amssymb}
\usepackage{cancel}
\usepackage{graphicx}
\usepackage[unicode=true,pdfusetitle,
 bookmarks=true,bookmarksnumbered=false,bookmarksopen=false,
 breaklinks=false,pdfborder={0 0 0},pdfborderstyle={},backref=false,colorlinks=true]
 {hyperref}

\makeatletter

\providecommand{\tabularnewline}{\\}

\@ifundefined{date}{}{\date{}}
\usepackage{lmodern}
\usepackage[T1]{fontenc}
\usepackage{caption}
\numberwithin{equation}{section}
\usepackage{upgreek}

\@ifundefined{showcaptionsetup}{}{%
 \PassOptionsToPackage{caption=false}{subfig}}
\usepackage{subfig}
\makeatother

\begin{document}
\title{A model for plasma-neutral fluid interaction and its application to
\\
a study of CT formation in a magnetised Marshall gun}
\author{Carl Dunlea$^{1*}$, Chijin Xiao$^{1}$, and Akira Hirose$^{1}$}

\maketitle
$^{1}$University of Saskatchewan, 116 Science Pl, Saskatoon, SK S7N
5E2, Canada 

$^{*}$e-mail: cpd716@mail.usask.ca 
\begin{abstract}
A model for plasma/neutral fluid interaction was developed and included
in the DELiTE code framework implementation of non-linear MHD equations.
The source rates of ion, electron and neutral fluid momentum and energy
due to ionization and recombination are derived using a simple method
that enables determination of the volumetric rate of thermal energy
transfer from electrons to photons and neutral particles in the radiative
recombination reaction. This quantity can not be evaluated with the
standard formal procedure of taking moments of the relevant collision
operator, and has been neglected in other studies. The plasma/neutral
fluid interaction model was applied to study CT formation in the SMRT
and  SPECTOR magnetized Marshall guns, enabling clarification of the
mechanisms behind the significant increases in CT electron density
that are routinely observed well after formation on the SPECTOR experiment.
Neutral gas, which remains concentrated below the gas valves after
CT formation, diffuses up the gun barrel to the CT containment region
where it is ionized, leading to the observed electron density increases.
This understanding helps account for the exceptionally significant
increase in temperature, and markedly reduced density, observed during
the electrode edge biasing experiment conducted on SPECTOR. It is
thought that edge fueling impediment, a consequence of a biasing-induced
transport barrier, is largely responsible for the observed temperature
increase and density decrease.
\end{abstract}

\section{Introduction}

The DELiTE (Differential Equations on Linear Triangular Elements)
framework \cite{SIMpaper,thesis} was developed for spatial discretisation
of partial differential equations on an unstructured triangular grid
in axisymmetric geometry. The framework is based on discrete differential
operators in matrix form, which are derived using linear finite elements
and mimic some of the properties of their continuous counterparts.
A single-fluid two-temperature MHD code with anisotropic thermal diffusion
was implemented in this framework. The inherent properties of the
operators are used in the code to ensure global conservation of energy,
particle count, toroidal flux, and, in some cases, angular momentum.
As described in \cite{SIMpaper,thesis,exppaper}, the code was applied
to study a novel experiment in which a compact torus (CT), produced
with the SMRT (Super Magnetized Ring Test) magnetized Marshall gun,
is magnetically levitated radially inwards off an insulating wall
and then magnetically compressed through the action of currents in
the levitation/compression coils located outside the wall. The code
also has the capability to start magnetic compression from a Grad-Shafranov
equilibrium. There are simulated diagnostics for magnetic probes,
interferometers, ion-Doppler measurements, and CT $q$ profile. Special
care was taken to simulate the poloidal vacuum field in the insulating
region between the inner radius of the insulating wall and the levitation/compression
coils, and to couple this solution to the MHD solutions in the plasma
domain, while maintaining toroidal flux conservation, enabling a quantitative
model of plasma/wall interaction in various coil configurations.

The initial motivation for including plasma / neutral fluid interaction
in the MHD model was to reduce simulated ion temperature to levels
corresponding to experimental measurements. To maintain conservation
of net system energy, simulations are run with $\nu_{num}\text{ set equal to }\nu_{phys}$,
where $\nu_{num}${[}m$^{2}$/s{]} is the (isotropic) viscosity diffusion
coefficient that is used in the viscous stress tensor in the momentum
equation, and $\nu_{phys}${[}m$^{2}$/s{]} is used in the viscous
stress tensor in the ion energy equation. In practice, for formation
simulations with a practically large timestep, and mesh of reasonable
resolution (element size around 2 mm), it is found that we need to
use a reasonably high value of $\nu_{num}$ to maintain sufficiently
smooth velocity fields. The same value for $\nu_{phys}$ resulted
in an overestimation (compared with experimental measurements) of
simulated ion temperature, due to excessive viscous heating of the
ion fluid during CT formation. Charge exchange collisions are an important
mechanism by which ion temperature is reduced - in a charge exchange
reaction, a hot ion takes an electron from a cold neutral particle,
resulting in a hot neutral particle and a cold ion. However, it was
found that, due to equilibration between the ion and electron temperatures,
ionization rates are high in regions where ions are hot, so that neutral
fluid density, and hence the level of charge-exchange related ion
cooling, is approximately negligible in those key regions. Secondly,
it was expected that ion cooling would be a consequence of the presence
of neutral gas in the gun around the gas valves during the CT formation
process. In the SMRT experiment, as described in \cite{SIMpaper,thesis,exppaper},
the CT formation process is initiated when an electric field is applied
across a cloud of cold gas between the inner and outer machine electrodes.
The gas cloud is concentrated near the gas-puff valves. The gas valves
are typically opened approximately $400\upmu$s before the formation
voltage is applied across the electrodes, so that a neutral gas cloud
has ample time to diffuse away from the valves. The ionization action
of the applied formation field results in seed electrons that lead
to further impact ionization. As the plasma is advected up the gun
by the $\mathbf{J}_{r}\times\mathbf{B}_{\phi}$ force, it ionizes
or displaces the gas cloud. The neutral cloud slows down the advected
plasma through collisions, resulting in a reduction of ion viscous
heating as $\mathbf{v}$ and $\nabla\mathbf{v}$ are reduced. It turned
out that the cause of the temperature reduction observed when neutral
fluid was included in the CT formation model was due to a reduction
in viscous heating associated with bulk inertial effects. Increased
net initial particle inventory results in lower ion temperatures,
regardless of whether part of the initial inventory includes neutral
particles.

The model for plasma / neutral fluid interaction proved to be useful
because it helped clarify the mechanisms behind the significant increases
in electron density that are routinely observed well after CT formation
in the SPECTOR (Spherical Compact Toroid) plasma injector \cite{spectPoster}.
Neutral gas, which remains concentrated below the gas valves after
CT formation, diffuses up the gun barrel to the CT containment region
where it is ionized, leading to the observed electron density increases.
This understanding helps account for the exceptionally significant
increase in temperature, and markedly reduced density, observed during
the electrode edge biasing experiment conducted on SPECTOR \cite{Spector_biasing}.
Impediment of fueling with cool particles, a consequence of a biasing-induced
transport barrier, is largely responsible for the observed biasing-induced
temperature increase and density decrease. Note that density increase
after CT formation was not observed on the SMRT injector \cite{SIMpaper,thesis,exppaper},
as CTs produced in that configuration lasted for less than the time
it takes for neutral gas to diffuse upwards from the valves to the
CT containment region.

Plasma fluid equations are derived in Braginskii\textquoteright s
1965 paper \cite{Braginski} by taking moments of ion and electron
Boltzmann equations, and applying the Chapman- Enskog closures. In
\cite{Braginski}, a multicomponent model which deals with the plasma
and neutral fluids as a combined fluid, without reactive collisions,
is also developed. Models including charge exchange reactions between
hydrogen and ions, neglecting ionisation and recombination reactions,
have been applied to study interaction between the interstellar medium
and the solar wind \cite{Pauls}. A general model for capturing ionization,
recombination and charge exchange reactions in wide range of plasma
physics application is presented in \cite{Meier,MeierPhd}.

Electron impact ionization, and radiative recombination are inelastic
reactive collisions, whereas the resonant charge exchange process
is an elastic reactive collision because the initial and final states
are degenerate. The procedure, described in \cite{Meier,MeierPhd},
of taking moments of the charge exchange collision operator, cannot
be avoided when evaluating the terms in the plasma fluid and neutral
fluid momentum and energy corresponding to charge exchange reactions.
However, to derive the terms in the fluid equations corresponding
to ionization and recombination, instead of taking moments of the
corresponding collision operators, as done in \cite{Meier,MeierPhd},
a more straightforward and intuitive method is used in this work.
We directly evaluate the effect of the particle sources and sinks
arising from to ionization and recombination on the species (ion,
electron, neutral) momentums and energies. This approach also allows
for determination of the terms that must be included in the MHD equations
when external particle sources are present. The strategy has the additional
benefit that it allows for the evaluation of $Q_{e}^{rec}$, the thermal
energy transferred from electrons to photons and neutral particles
due to recombination per volume per second. By contrast, the method
used in \cite{Meier,MeierPhd}, allows for the evaluation of $Q_{i}^{rec}$(rate
of thermal energy transferred from ions to neutral particles due to
recombination), but leads to an entangled integral expression for
$Q_{e}^{rec}$ that can't be easily evaluated, so that $Q_{i}^{rec}$
is retained while the $Q_{e}^{rec}$ term is neglected from the species
energy equations. 

This paper is arranged as follows. An overview of the model for plasma-neutral
fluid interaction is presented in section \ref{sec:Model-overview}.
The scattering terms are collected in section \ref{sec:Scattering-collsion-termsS}
(details of the derivation is presented in appendix \ref{sec:Appendix:AScattering-collision}).
Development of the reactive terms pertaining to ionization, recombination
and charge exchange collisions, that are included in the plasma and
neutral fluid equations, is presented in section \ref{sec:Reactive-collision-terms}.
The resultant set of conservation equations for ions, electrons and
neutral particles is developed in appendix \ref{sec:Appendix: 3-fluid-MHD-equations}.
The equations solved by the model implemented to the DELiTE framework,
including the single plasma fluid conservation equations with terms
pertaining to interaction with the neutral fluid, is presented in
section \ref{sec:2-fluid-MHD-equations}. A demonstration of total
system energy, particle count, and angular momentum conservation with
inclusion of a neutral fluid in the model is presented and discussed
in section \ref{sec:Energy-conservation-with}. Results pertaining
to neutral fluid interaction, from simulations of CT formation, levitation
and magnetic compression \cite{SIMpaper,thesis,exppaper} in the SMRT
plasma injector, and from simulations of CT formation in the SPECTOR
plasma injector \cite{spectPoster}, are presented and discussed in
section \ref{sec:Simulation-results-with}. Section \ref{sec:conclusions}
concludes the paper with an overview of the principal findings.

\section{Model overview\label{sec:Model-overview}}

Ion, electron and neutral particles are added and taken away from
the system due to reactive collisions (ionization, recombination,
and charge exchange). In addition, particles may be added to the system
through external sources, for example ion, electron, or neutral beam
injection, or, as in the case considered here, neutral gas injected
through the machine gas valves. The combination of reactive collisions
and external sources and sinks modify the local mass, momentum and
energy distributions, and appear in the species conservation equations.
Scattering collisions are manifested as intra-species frictional forces,
and change local momentum and energy distributions. 

Based on the work presented in \cite{Meier,MeierPhd}, the model includes
resonant charge exchange, electron impact ionization, and radiative
recombination reactions:
\begin{align}
i^{+}+n & \rightarrow n+i^{+}\nonumber \\
e^{-}+n & \rightarrow i^{+}+2e^{-}-\phi_{ion}\label{eq:520-1}\\
e^{-}+i^{+} & \rightarrow n+h\nu_{p}\nonumber 
\end{align}
Here, $i^{+},\,e^{-}$ and $n$ respectively represent singly charged
ions and electrons, and neutral particles, $h\,[\mbox{m}^{2}\,\mbox{kg/s}]$
is Planck's constant, and $\nu_{p}$ is the wave frequency associated
with the photon emitted in the recombination reaction. The charge
exchange process is \emph{resonant} because the initial and final
states have the same quantum mechanical energy - the exchanged electron's
initial and final energy states are the same so that the combined
kinetic energy and momentum of the ion and neutral is unchanged \cite{Goldston,HazeltineWaelbroeck}.
In the following derivations, singly charged ions and a single neutral
species will be considered. The plasma is assumed to be optically
thin, so that radiation energy $h\nu_{p}$ associated with radiative
recombination is lost from the system. Following from \cite{Meier,MeierPhd},
excited states are not tracked in order to simplify the model. Instead,
an effective ionization potential, $\phi_{ion}$, includes the excitation
energy that is expended on average for each ionization event, as well
as the electron binding energy. 

\section{Scattering collsion terms\label{sec:Scattering-collsion-termsS}}

Mass, momentum and energy are conserved and particles are not created
or destroyed in elastic scattering collisions. The Coulomb interaction
of charged particles is a long range one characterised by multiple
simultaneous interactions. In contrast, the short range fields (within
the electronic shells) of neutral particles result in binary collisions.
Because of the long range nature of the Coulomb force, small-angle
deflections associated with Coulomb collisions are much more frequent
than the large angle deflections associated with binary collisions.
The cumulative effect of many small-angle collisions is much larger
than that of relatively fewer large-angle collisions \cite{Goldston}.
It is possible to deal with multiple collisions by approximating them
as a number of simultaneous binary collisions \cite{bittencourt}.
To find the contribution of scattering collisions to the rates of
change of momentum of the ions, electrons and neutral particles, the
first moments of the operators for scattering collisions can be taken.
As shown in appendix \ref{sec:Appendix:AScattering-collision} (equation
\ref{eq:520.1}), this leads to: 

\begin{align}
\rho_{i}\left(\frac{\partial\mathbf{v}_{i}}{\partial t}\right)_{scatt.}=\mathbf{R}_{i} & =\mathbf{R}_{ie}+\mathbf{R}_{in}\nonumber \\
\rho_{e}\left(\frac{\partial\mathbf{v}_{e}}{\partial t}\right)_{scatt.}=\mathbf{R}_{e} & =\mathbf{R}_{ei}+\mathbf{R}_{en}=-\mathbf{R}_{ie}+\mathbf{R}_{en}\label{eq:521-1}\\
\rho_{n}\left(\frac{\partial\mathbf{v}_{n}}{\partial t}\right)_{scatt.}=\mathbf{R}_{n} & =\mathbf{R}_{ni}+\mathbf{R}_{ne}=-\mathbf{R}_{in}-\mathbf{R}_{en}\nonumber 
\end{align}
where $\rho_{\alpha}$ and $\mathbf{v}_{\alpha}$are the mass density
and fluid velocity of species $\alpha,$ $\mathbf{R}_{\alpha\sigma}$
is the frictional force exerted by species $\sigma$ on species $\alpha$,
and $\alpha=i,\,e,$ $n$ denote ions, electrons, and neutral particles.
Note that $\mathbf{R}_{\alpha\sigma}=-\mathbf{R}_{\sigma\alpha}$,
implying that the frictional force exerted by species $\alpha$ on
species $\sigma$ is balanced by the frictional force exerted by species
$\sigma$ on species $\alpha$. The terms $\mathbf{R}_{in}\mbox{ and }\mathbf{R}_{en}$
can be neglected in many cases - in general, neutral-charged particle
scattering collisions are unimportant when the plasma is ionized by
even a few percent \cite{Meier,Goldston}. 

To find the contribution of scattering collisions to the rates of
change of energy of the ions, electrons and neutral particles, the
second moments of the operators for scattering collisions can be taken,
leading, as outlined in appendix \ref{sec:Appendix:AScattering-collision}
(equation \ref{eq:521.11}) to

\begin{align}
\frac{1}{\gamma-1}\left(\frac{\partial p_{i}}{\partial t}\right)_{scatt.} & =Q_{ie}+Q_{in}\nonumber \\
\frac{1}{\gamma-1}\left(\frac{\partial p_{e}}{\partial t}\right)_{scatt.} & =Q_{ei}+Q_{en}\label{eq:522-1}\\
\frac{1}{\gamma-1}\left(\frac{\partial p_{n}}{\partial t}\right)_{scatt.} & =Q_{ni}+Q_{ne}\nonumber 
\end{align}
Here, $Q_{\alpha\sigma}$ represents the rate at which species $\sigma$
collisionally transfers energy to species $\alpha$ in the frame moving
with velocity $\mathbf{v}_{\alpha}$ \cite{HazeltineWaelbroeck,farside_plasma}.
Again, due to the relative unimportance of neutral-charged particle
scattering collisions \cite{Meier,Goldston}, the terms $Q_{in}$,
$Q_{en}$, $Q_{ni}$, and $Q_{ne}$ can usually be neglected.

\section{Terms corresponding to reactive collisions and external particle
sources\label{sec:Reactive-collision-terms}}

In this section, the terms in the species mass, momentum and energy
conservation equations that pertain to external particle sources,
and to collisions associated with ionization, recombination, and charge
exchange reactions, will be assessed. 

\subsection{Ionization, recombination, and external particle sources\label{subsec:Ionization-and-recombination}}

\subsubsection{Mass Conservation}

As shown in \cite{Meier,MeierPhd}, the sources and sinks in the species
continuity equations that arise from ionization and recombination
reactions can be evaluated by taking the zeroth moments of the collision
operators pertaining to ionization and recombination collisions. Alternatively,
a more intuitive method that doesn't use the collision operators is
presented in \cite{Goldston}. An expression for the mean free path
for impact ionization collisions is evaluated by considering its definition
- there is one particle in the volume swept out, over one mean free
path, by the cross-sectional area for impact ionization collisions:
 $n_{n}\sigma_{ion}(V_{rel})\,\lambda_{mfp}^{ion}=1$. Here, $V_{rel}=|\mathbf{V}_{e}-\mathbf{V}_{n}|,$
the relative particle speed for the ionization reaction, where $\mathbf{V}_{e}$
is the electron particle velocity and $\mathbf{V}_{n}$ is the neutral
particle velocity. Since $m_{n}\gg m_{e}$, $V_{rel}\approx V_{e}$.
The frequency for electron impact ionization collisions is defined
as an average over all velocities in the Maxwellian distribution:
\[
\nu_{ion}=<\frac{V_{rel}}{\lambda_{mfp}^{ion}}>\approx n_{n}<\sigma_{ion}(V_{e})\,V_{e}>
\]
From this, $\Gamma_{i}^{ion}$, the rate of increase of ions per unit
volume due to ionization reactions is 
\[
\Gamma_{i}^{ion}=\Gamma_{e}^{ion}=-\Gamma_{n}^{ion}=n_{n}n_{e}<\sigma_{ion}\,V_{e}>
\]
Here, $\Gamma_{e}^{ion}$ and $\Gamma_{n}^{ion}$ are the rates of
increase of electrons and neutral particles per unit volume due to
ionization reactions. The velocity-space-integrated quantity $<\sigma_{ion}\,V_{e}>[\mbox{m}^{3}/\mbox{s}]$
is the\emph{ ionization rate parameter; }its value can be found as
a function of temperature from the fitting formula given by Voronov
\cite{Voronov}, as 
\begin{equation}
<\sigma_{ion}\,V_{e}>(\mathbf{r},t)=\frac{A\left(1+P\sqrt{U}\right)U^{K}exp\left(-U\right)}{U+X}\label{eq:530-1}
\end{equation}
where $U(\mathbf{r},t)=\phi_{ion}/T_{e}(\mathbf{r},t)$, with $\phi_{ion}$
being the effective ionization potential in the same units as $T_{e}$.
Following from \cite{Meier,MeierPhd}, an effective ionization potential
including the excitation energy that is expended on average for each
ionization event, as well as the electron binding energy, is used
instead of the regular ionization energy, because, for simplicity,
excited states are not tracked. An estimate of the validity of the
formula for the ionization rate parameter is given by Voronov for
each element. For example, for hydrogen, accuracy is to within 5\%
for electron temperatures from 1 $\mbox{eV}$ to 20 $\mbox{keV}$.
The DELiTE MHD code has the option of either hydrogen, deuterium or
helium as the neutral gas and plasma source. From \cite{Voronov},
the coefficients required for equation \ref{eq:530-1} for these options
are shown in table \ref{tab:coef_ionization rate parameter}. Also
included is the atomic diameter ($d_{atom}$) for each atom, which
is used to calculate the viscous and thermal diffusion coefficients
for the neutral fluid. The values for effective ionization potentials
are taken from \cite{Yusupaliev}. 
\begin{table}[H]
\centering{}%
\begin{tabular}{|c|c|c|c|}
\hline 
$\mbox{ion\,\,type}:$ & H & $\mbox{D}$ & $\mbox{He}$\tabularnewline
\hline 
\hline 
$\phi_{ion}[\mbox{eV}]$ & $13.6$ & $33$ & $28$\tabularnewline
\hline 
$A$ & $2.91\times10^{-14}[\mbox{m}^{3}/\mbox{s}]$ & $2.91\times10^{-14}[\mbox{m}^{3}/\mbox{s}]$ & $1.75\times10^{-14}[\mbox{m}^{3}/\mbox{s}]$\tabularnewline
\hline 
$P$ & $0$ & $0$ & $0$\tabularnewline
\hline 
$K$ & $0.39$ & $0.39$ & $0.35$\tabularnewline
\hline 
$X$ & $0.232$ & $0.232$ & $0.18$\tabularnewline
\hline 
$d_{atom}[\mbox{m}]$ & $1.06\times10^{-10}$ & $2.4\times10^{-10}$ & $2.8\times10^{-10}$\tabularnewline
\hline 
\end{tabular}\caption{\label{tab:coef_ionization rate parameter}$\,\,\,\,$Coefficients
for calculating ionization rate parameters}
\end{table}
The rate of increase of ions per unit volume due to recombination
reactions can be evaluated by considering the mean free path and collision
frequency associated with recombination \cite{Goldston} as 
\begin{equation}
\Gamma_{i}^{rec}=\Gamma_{e}^{rec}=-\Gamma_{n}^{rec}=\int C_{e}^{rec}d\mathbf{V}=-n_{i}n_{e}<\sigma_{rec}\,V_{e}>\label{eq:530.001}
\end{equation}
 The velocity integrated quantity $<\sigma_{rec}\,V_{e}>[\mbox{m}^{3}/\mbox{s}]$
is the\emph{ recombination rate parameter - }its value for recombination
to charge state $Z_{eff}-1$ can be estimated as a function of electron
temperature as \cite{McWhirter,MeierPhd,Goldston} 
\begin{equation}
<\sigma_{rec}\,V_{e}>(\mathbf{r},t)=2.6\times10^{-19}\frac{\,Z_{eff}^{2}}{\sqrt{T_{e}(\mathbf{r},t)[eV]}}\label{eq:531.0}
\end{equation}
Hence, the mass continuity equations for the three-fluid system are:

\begin{align}
\frac{\partial n_{i}}{\partial t} & =-\nabla\cdot(n_{i}\mathbf{v}_{i})+\Gamma_{i}^{ion}-\Gamma_{n}^{rec}\nonumber \\
\frac{\partial n_{e}}{\partial t} & =-\nabla\cdot(n_{e}\mathbf{v}_{e})+\Gamma_{i}^{ion}-\Gamma_{n}^{rec}\label{eq:531.2}\\
\frac{\partial n_{n}}{\partial t} & =-\nabla\cdot(n_{n}\mathbf{v}_{n})+\Gamma_{n}^{rec}-\Gamma_{i}^{ion}+\Gamma_{n}^{ext}\nonumber 
\end{align}
Note that all source terms here, as well as each of $n_{\alpha}$
and $\mathbf{v}_{\alpha}$, are functions of $\mathbf{r}$ and $t$.
In the magnetic compression experiment \cite{thesis,exppaper}, the
plasma injector gas puff valves take time to shut, and remain open
for several hundred microseconds after the formation banks are fired,
so neutral particles are being added to the system near the valves.
An additional neutral \emph{external} source term, $\Gamma_{n}^{ext}(\mathbf{r},t)\,[\mbox{m}^{-3}\mbox{ s}^{-1}]$,
has been included on the right side of the expression for $\partial n_{n}/\partial t$,
in order to be able to this simulate neutral particle injection.

\subsubsection{Momentum Conservation}

As shown in \cite{Meier,MeierPhd}, the sources and sinks in the species
momentum equations that arise from ionization and recombination reactions
can be evaluated by taking the first moments of the relevant collision
operators. Here, the formal process of taking first moments (and second
moments for the contributions to the rates of change of energy) is
skipped. The set of equations produced by this alternative simple
method are equivalent to those presented in \cite{Meier,MeierPhd},
with the exception that an expression is found for the $Q_{e}^{rec}$
term, representing the volumetric rate of transfer of thermal energy
transferred from electrons to photons and neutral particles due to
recombination, which was not evaluated in \cite{Meier,MeierPhd}.
Here we introduce some new notation that will be referred to in the
following derivations. Referring to equation \ref{eq:531.2}, the
general form of the expression for the species rates of change of
number density that correspond to the reactive collisions of ionization
and recombination, and also to any external particle sources, may
be expressed as 
\begin{equation}
\left(\frac{\partial n_{\alpha}}{\partial t}\right)_{ire}=\underset{k}{\Sigma}S_{\alpha k}\label{eq:531.40}
\end{equation}
where $\left(\frac{\partial X}{\partial t}\right)_{ire}$ denotes
the time-rate of change of any quantity $X$ that arises due to \emph{ionization},
\emph{recombination,} and to any \emph{external} particle sources.
$S_{\alpha k}\,[\mbox{m}^{-3}\mbox{s}^{-1}]$ represents the $k^{th}$
source (in units of particles per metres cubed per second) for particles
of type $\alpha,$ as determined from equation \ref{eq:531.2}. Here,
$S_{i1}=S_{e1}=-S_{n2}=\Gamma_{i}^{ion},\,S_{i2}=S_{e2}=-S_{n1}=-\Gamma_{n}^{rec}$,
and $S_{n3}=\Gamma_{n}^{ext}$. Note that particle \textquotedbl sources\textquotedbl{}
with a negative sign such as $S_{i2}=-\Gamma_{n}^{rec}$ in the expression
for $\partial n_{i}/\partial t$  in equation \ref{eq:531.2}, are
actually particle sinks. \\
\\
Species momentum conservation, in the absence of reactive collisions,
is described by \cite{farside_plasma}: 
\begin{equation}
\frac{\partial(\rho_{\alpha}\mathbf{v}_{\alpha})}{\partial t}=-\nabla\cdot\underline{\mathbf{p}}_{\alpha}-\nabla\cdot(\rho_{\alpha}\mathbf{v}_{\alpha}\mathbf{v}_{\alpha})+q_{\alpha}n_{\alpha}\left(\mathbf{E}+\mathbf{v}_{\alpha}\times\mathbf{B}\right)+\mathbf{R}_{\alpha}\label{eq:531.401}
\end{equation}
To include the terms that correspond to the reactive collisions of
ionization and recombination, and to any external particle sources,
this can be written as
\[
\frac{\partial(m_{\alpha}n_{\alpha}\mathbf{v}_{\alpha})}{\partial t}=\,\,\,\,...\,\,\,+\left(\frac{\partial(m_{\alpha}n_{\alpha}\mathbf{v}_{\alpha})}{\partial t}\right)_{ire}
\]
where \textquotedbl$...$\textquotedbl{} represents the right side
of equation \ref{eq:531.401}. Particles sourced by $S_{\alpha k}$
add, or (for sources with negative sign), remove, species $\alpha$
momentum $\underset{j}{\Sigma}m_{jk}\mathbf{v}_{0jk}$. Here, $m_{jk}$
and $\mathbf{v}_{0jk}$ are the mass and \textquotedbl initial\textquotedbl{}
($i.e.,$ at the time when the particles are sourced) fluid velocity
of the particles of type $j$ which are introduced or taken away by
source $S_{\alpha k}$.  The summation over sourced particles of type
$j$ is relevant only for $S_{\alpha k}=S_{n1}=\Gamma_{n}^{rec}$;
recombination is a source for total neutral particle momentum, and
each neutral particle added to the neutral population through recombination
initially has momentum $m_{i}\mathbf{v}_{i}+m_{e}\mathbf{v}_{e}$.
The general form of the expression for the species rates of change
of momentum that correspond to the reactive collisions of ionization
and recombination, and also to any external particle sources, is 
\begin{equation}
\left(\frac{\partial(m_{\alpha}n_{\alpha}\mathbf{v}_{\alpha})}{\partial t}\right)_{ire}=\underset{k}{\Sigma}\left(S_{\alpha k}\underset{j}{\Sigma}\left(m_{jk}\mathbf{v}_{0jk}\right)\right)\label{eq:531.41}
\end{equation}
This expression must be retained for the neutral recombination source
term $S_{n1}$. However, for all other source terms, $\underset{k}{\Sigma}\left(S_{\alpha k}\underset{j}{\Sigma}\left(m_{jk}\mathbf{v}_{0jk}\right)\right)\rightarrow m_{\alpha}\underset{k}{\Sigma}\left(S_{\alpha k}\mathbf{v}_{0k}\right)$,
where $\mathbf{v}_{0k}$ is the initial fluid velocity of the particles
of type $\alpha$ which are introduced or taken away due to source
$S_{\alpha k}$, and the general expression (equation \ref{eq:531.41})
can be simplified to
\[
\left(\frac{\partial(m_{\alpha}n_{\alpha}\mathbf{v}_{\alpha})}{\partial t}\right)_{ire}=m_{\alpha}\underset{k}{\Sigma}\left(S_{\alpha k}\mathbf{v}_{0k}\right)
\]
The corresponding additional terms on the right side of the momentum
equations are: 
\begin{align}
\left(\frac{\partial(m_{i}n_{i}\mathbf{v}_{i})}{\partial t}\right)_{ire} & =\Gamma_{i}^{ion}m_{i}\mathbf{v}_{n}-\Gamma_{n}^{rec}m_{i}\mathbf{v}_{i}\nonumber \\
\left(\frac{\partial(m_{e}n_{e}\mathbf{v}_{e})}{\partial t}\right)_{ire} & =\Gamma_{i}^{ion}m_{e}\mathbf{v}_{n}-\Gamma_{n}^{rec}m_{e}\mathbf{v}_{e}\label{eq:531.5}\\
\left(\frac{\partial(m_{n}n_{n}\mathbf{v}_{n})}{\partial t}\right)_{ire} & =\Gamma_{n}^{rec}(m_{i}\mathbf{v}_{i}+m_{e}\mathbf{v}_{e})-\Gamma_{i}^{ion}m_{n}\mathbf{v}_{n}+\Gamma_{n}^{ext}m_{n}\mathbf{v}_{n0}\nonumber 
\end{align}
For example, in the expression above for $\left(\frac{\partial(m_{i}n_{i}\mathbf{v}_{i})}{\partial t}\right)_{ire}$,
ions that are sourced from neutral particles through ionization add
to the total ion momentum, and newly ionized particles are introduced
with velocity $\mathbf{v}_{n}$. Meanwhile, ions with velocity $\mathbf{v}_{i}$,
that are lost to recombination, take away from the total ion momentum.
Each neutral particle introduced by recombination initially has momentum
$m_{i}\mathbf{v}_{i}+m_{e}\mathbf{v}_{e}$. Neutral particles introduced
by external sources such as gas puffing also add neutral particle
momentum - each externally sourced neutral has initial momentum $m_{n}\mathbf{v}_{n0}$,
where $\mathbf{v}_{n0}$ is its initial velocity.\\
\\
Equations \ref{eq:531.41} and \ref{eq:531.40} imply that 

\begin{align}
m_{\alpha}n_{\alpha}\left(\frac{\partial\mathbf{v}_{\alpha}}{\partial t}\right)_{ire} & =\underset{k}{\Sigma}\left(S_{\alpha k}\underset{j}{\Sigma}\left(m_{jk}\mathbf{v}_{0jk}\right)\right)-m_{\alpha}\mathbf{v}_{\alpha}\left(\underset{k}{\Sigma}S_{\alpha k}\right)\label{eq:531.51}
\end{align}
As discussed above in relation to equation \ref{eq:531.41}, for the
ions and electrons (all sources), and for the neutral source terms
corresponding to ionization and external sources, the general expression
(equation \ref{eq:531.51}) can be simplified to 
\begin{equation}
m_{\alpha}n_{\alpha}\left(\frac{\partial\mathbf{v}_{\alpha}}{\partial t}\right)_{ire}=m_{\alpha}\underset{k}{\Sigma}\left(S_{\alpha k}\left(\mathbf{v}_{0k}-\mathbf{v}_{\alpha}\right)\right)\label{eq:531.52}
\end{equation}
Equation \ref{eq:531.52} and (for neutral recombination only) equation
\ref{eq:531.51} lead to:\\
\begin{align}
\rho_{i}\left(\frac{\partial\mathbf{v}_{i}}{\partial t}\right)_{ire} & =\Gamma_{i}^{ion}m_{i}(\mathbf{v}_{n}-\mathbf{v}_{i})\nonumber \\
\rho_{e}\left(\frac{\partial\mathbf{v}_{e}}{\partial t}\right)_{ire} & =\Gamma_{i}^{ion}m_{e}(\mathbf{v}_{n}-\mathbf{v}_{e})\label{eq:531.6}\\
\rho_{n}\left(\frac{\partial\mathbf{v}_{n}}{\partial t}\right)_{ire} & =\Gamma_{n}^{rec}(m_{i}\mathbf{v}_{i}+m_{e}\mathbf{v}_{e}-m_{n}\mathbf{v}_{n})+\Gamma_{n}^{ext}m_{n}(\mathbf{v}_{n0}-\mathbf{v}_{n})\nonumber 
\end{align}

\subsubsection{Energy Conservation\label{subsec:Econ_ionrecomb}}

Corresponding to the assumption that the Chapman-Enskog closures are
a good approximation to the plasmas being considered, Maxwellian distributions
are assumed for each of species $\alpha$,  so that $p_{\alpha}=n_{\alpha}T_{\alpha}$.
Hence, the part of the species energy equation that corresponds to
particle sources due to the reactive collisions of ionization and
recombination, and to external particle sources, can be written as
\begin{equation}
\left(\frac{\partial}{\partial t}\left(\frac{1}{2}m_{\alpha}n_{\alpha}v_{\alpha}^{2}+\frac{p_{\alpha}}{\gamma-1}\right)\right)_{ire}=\frac{1}{2}\underset{k}{\Sigma}\left(S_{\alpha k}\underset{j}{\Sigma}\left(m_{jk}v_{0jk}^{2}\right)\right)+\frac{1}{\gamma-1}\underset{k}{\Sigma}\left(\xi_{\alpha k}S_{\alpha k}\underset{j}{\Sigma}T_{0jk}\right)\label{eq:531.39}
\end{equation}
Here, $\xi_{\alpha k}$ is a particle\emph{ mass ratio} that must
be considered for the ionization source that introduces ions ($S_{i1}=\Gamma_{i}^{ion}$)
and electrons ($S_{e1}=\Gamma_{i}^{ion}$), and for the recombination
source that introduces neutral particles $(S_{n1}=\Gamma_{n}^{rec})$.
To clarify this for the ionization source, when a neutral particle
with thermal energy $\frac{1}{\gamma-1}T_{n}$ is ionized, the resultant
ion and electron have thermal energies $\frac{m_{i}}{m_{n}}\frac{1}{\gamma-1}T_{n}$
and $\frac{m_{e}}{m_{n}}\frac{1}{\gamma-1}T_{n}$ respectively, so
that their combined thermal energy is equal to that of the original
neutral. To clarify the relevance of $\xi_{\alpha k}$ for the recombination
source, simple analysis of the kinematics of the radiative recombination
reaction indicates that the bulk of the electron thermal energy, and
a fraction of the ion thermal energy, is transferred to the emitted
photon. Noting that $m_{i}\sim m_{n}\gg m_{e}$, then, in the rest
frame of the neutral particle (post reaction), the ion (prior to the
reaction) has negligible kinetic energy ($ie.,$ thermal energy, since
we are considering single particles with random velocities), the electron
has approximately the same energy that it has in the laboratory frame,
and the neutral particle has no kinetic energy. Consequently, as an
approximation, the bulk of the electron thermal energy (of the order
$\sim(m_{i}/m_{n})\,T_{e}/(\gamma-1)$) and a negligible portion of
the ion thermal energy ($\sim(m_{e}/m_{n})\,T_{i}/(\gamma-1)$) is
transferred to the photon emitted in the radiative recombination reaction,
while a negligible portion of the electron thermal energy (of the
order $\sim(m_{e}/m_{n})\,T_{e}/(\gamma-1)$) and the bulk of the
ion thermal energy ($\sim(m_{i}/m_{n})\,T_{i}/(\gamma-1)$) is transferred
to the neutral particle. The combined thermal energy of the neutral
particle and photon is equal to the combined thermal energy of the
ion and electron. The partitioning of the electron and ion energies
to the neutral particle and photon can be shown explicitly by considering
the momentum balance associated with the radiative recombination reaction.
First, it is important to recognize that the photon momentum can be
neglected: for example, consider a photon with energy $U_{ph}$ =
100 eV. If the photon's momentum $(2U_{ph}/c)$ is transferred to
an electron, this would impart energy $U_{e}=\frac{1}{2}(U_{ph}/c)^{2}/m_{e}\sim0.01$
eV to the electron, much smaller than electron energies under consideration.
Neglecting photon momentum, the momentum conservation equation is,
to a very good approximation, $m_{i}\mathbf{V}_{i}+m_{e}\mathbf{V}_{e}=m_{n}\mathbf{V}_{n}$.
Squaring both sides, this implies that $m_{i}{}^{2}V_{i}{}^{2}+2m_{i}m_{e}\mathbf{V}_{i}\cdot\mathbf{V}_{e}+m_{e}{}^{2}V_{e}{}^{2}=m_{n}^{2}V_{n}^{2}$.
The average over all angles of the dot product $\mathbf{V}_{i}\cdot\mathbf{V}_{e}$
is zero, so the kinetic energies are related by $m_{i}U_{i}+m_{e}U_{e}=m_{n}U_{n}$,
implying that $U_{n}=\frac{m_{i}}{m_{n}}U_{i}+\frac{m_{e}}{m_{n}}U_{e}$.
The remaining electron and ion energy goes to the photon: $U_{ph}=\frac{m_{e}}{m_{n}}U_{i}+\frac{m_{i}}{m_{n}}U_{e}$
\cite{M Reynolds}.  Note that for $\Gamma_{n}^{ext}$, the external
neutral particle source, $\xi_{\alpha k}=1$. \\

Once again, the summation over $j$ is relevant only for $S_{\alpha k}=S_{n1}=\Gamma_{n}^{rec}$;
recombination is a source for neutral particle energy, and it is assumed
that each neutral particle added to the neutral particle population
through recombination initially has energy $\frac{1}{2}\left(m_{i}v_{i}^{2}+m_{e}v_{e}^{2}\right)+\frac{1}{\gamma-1}\left(\frac{m_{i}}{m_{n}}\,T_{i}+\frac{m_{e}}{m_{n}}\,T_{e}\right)$.
Note that for all other sources (apart from recombination) the summation
over $j$ in equation \ref{eq:531.39} can be neglected; $\underset{j}{\Sigma}T_{0jk}\rightarrow T_{0k}$,
where $T_{0k}$ is the initial temperature of the sourced particle,
and $\underset{k}{\Sigma}\left(S_{\alpha k}\underset{j}{\Sigma}\left(m_{jk}v_{0jk}^{2}\right)\right)\rightarrow m_{\alpha}\underset{k}{\Sigma}\left(S_{\alpha k}v_{0k}^{2}\right).$
We want to obtain an expression for $\left(\frac{\partial p_{\alpha}}{\partial t}\right)_{ire}$,
which will be included on the right side of the species energy equations.
As shown in appendix \ref{sec:Appendix:B},  the resultant forms for
the individual species are:

\begin{align}
\left(\frac{\partial p_{i}}{\partial t}\right)_{ire} & =\Gamma_{i}^{ion}\frac{m_{i}}{m_{n}}\,T_{n}-\Gamma_{n}^{rec}\,T_{i}+(\gamma-1)\frac{1}{2}m_{i}\left(\Gamma_{i}^{ion}\left(\mathbf{v}_{i}-\mathbf{v}_{n}\right)^{2}\right)\nonumber \\
\left(\frac{\partial p_{e}}{\partial t}\right)_{ire} & =\Gamma_{i}^{ion}\frac{m_{e}}{m_{n}}\,T_{n}-\Gamma_{n}^{rec}\,T_{e}+(\gamma-1)\left(\Gamma_{i}^{ion}\left(\frac{1}{2}m_{e}\left(\mathbf{v}_{e}-\mathbf{v}_{n}\right)^{2}-\phi_{ion}\right)\right)\nonumber \\
\left(\frac{\partial p_{n}}{\partial t}\right)_{ire} & =\Gamma_{n}^{rec}\left(\frac{m_{i}}{m_{n}}\,T_{i}+\frac{m_{e}}{m_{n}}\,T_{e}\right)-\Gamma_{i}^{ion}\,T_{n}+\Gamma_{n}^{ext}\,T_{n0}+(\gamma-1)\biggl[\Gamma_{n}^{rec}\biggl(\frac{1}{2}m_{n}v_{n}^{2}+\frac{1}{2}m_{i}v_{i}^{2}\nonumber \\
 & \,\,\,\,\,\,\,+\frac{1}{2}m_{e}v_{e}^{2}-m_{i}\mathbf{v}_{n}\cdot\mathbf{v}_{i}-m_{e}\mathbf{v}_{n}\cdot\mathbf{v}_{e}\biggr)+\Gamma_{n}^{ext}\,\frac{1}{2}m_{n}\left(\mathbf{v}_{n}-\mathbf{v}_{n0}\right)^{2}\biggr]\label{eq:533}
\end{align}
Here, $T_{n0}$ is the initial temperature of the externally sourced
neutral particles. The effective ionization energy has been included
as a sink of electron energy - recalling that $U_{Th}=\frac{p}{\gamma-1}$,
for each electron with energy $\left(\frac{1}{\gamma-1}\frac{m_{e}}{m_{n}}\,T_{n}+\frac{1}{2}m_{e}\left(\mathbf{v}_{e}-\mathbf{v}_{n}\right)^{2}\right)$
Joules that is sourced by ionization, another electron has expended
$\phi_{ion}$ Joules to initiate the ionization process. Note that
the thermal energy of a particle with temperature $T$ {[}J{]} is
$T/(\gamma-1)$. If particles are being added to a system at a volumetric
rate of $\Gamma$ particles per volume per second, then the energy
being added to the system, per volume per second, is $\Gamma\,T/(\gamma-1).$
 Here, the following definitions are made, representing the thermal
energy per unit volume per second transferred between species due
to ionization and recombination processes: 
\begin{flalign}
Q_{n}^{ion} & =\Gamma_{i}^{ion}\frac{1}{\gamma-1}T_{n} &  & \mbox{(neutral particles\ensuremath{\rightarrow} ions and electrons, due to ionization)}\nonumber \\
Q_{i}^{rec} & =\Gamma_{n}^{rec}\frac{1}{\gamma-1}T_{i} &  & \mbox{(ions \ensuremath{\rightarrow} neutral particles (and photons), due to recombination)}\nonumber \\
Q_{e}^{rec} & =\Gamma_{n}^{rec}\frac{1}{\gamma-1}T_{e} &  & \mbox{(electrons \ensuremath{\rightarrow} photons (and neutral particles), due to recombination)}\label{eq:533.0}
\end{flalign}
Hence, equation \ref{eq:533} can be re-expressed as:
\begin{align}
\left(\frac{\partial p_{i}}{\partial t}\right)_{ire} & =(\gamma-1)\left(\frac{m_{i}}{m_{n}}Q_{n}^{ion}-Q_{i}^{rec}+\frac{1}{2}m_{i}\left(\Gamma_{i}^{ion}\left(\mathbf{v}_{i}-\mathbf{v}_{n}\right)^{2}\right)\right)\nonumber \\
\left(\frac{\partial p_{e}}{\partial t}\right)_{ire} & =(\gamma-1)\left(\frac{m_{e}}{m_{n}}Q_{n}^{ion}-Q_{e}^{rec}+\Gamma_{i}^{ion}\left(\frac{1}{2}m_{e}\left(\mathbf{v}_{e}-\mathbf{v}_{n}\right)^{2}-\phi_{ion}\right)\right)\nonumber \\
\left(\frac{\partial p_{n}}{\partial t}\right)_{ire} & =(\gamma-1)\biggl[\frac{m_{i}}{m_{n}}\,Q_{i}^{rec}+\frac{m_{e}}{m_{n}}\,Q_{e}^{rec}-Q_{n}^{ion}\nonumber \\
 & \,\,\,\,\,\,\,+\Gamma_{n}^{rec}\biggl(\frac{1}{2}m_{n}v_{n}^{2}+\frac{1}{2}m_{i}v_{i}^{2}+\frac{1}{2}m_{e}v_{e}^{2}-m_{i}\mathbf{v}_{n}\cdot\mathbf{v}_{i}-m_{e}\mathbf{v}_{n}\cdot\mathbf{v}_{e}\biggr)\nonumber \\
 & \,\,\,\,\,\,\,+\Gamma_{n}^{ext}\frac{1}{2}m_{n}\left(\mathbf{v}_{n}-\mathbf{v}_{n0}\right)^{2}\biggr]+\Gamma_{n}^{ext}T_{n0}\label{eq:533.1}
\end{align}

\subsection{Charge exchange\label{subsec:Charge-exchange}}

In order to find the terms in the MHD equations that correspond to
the charge exchange reactions, the process of taking moments of the
charge exchange collision operators can't be avoided due to, and is
complicated by, the degeneracy associated with the charge exchange
reaction. In this work, the details of the process won't be reproduced
- only the required results that were originally achieved in \cite{Pauls},
and then very well detailed in \cite{Meier,MeierPhd}, will be presented. 

Taking the zeroth moment of the charge exchange collision operator
leads to the source rate of neutral particles, equal to the source
rate of ions, for the charge exchange reaction: 
\begin{equation}
\Gamma^{cx}=n_{i}n_{n}\sigma_{cx}(V_{cx})V_{cx}\label{eq:535}
\end{equation}
Note $\sigma_{cx}${[}m$^{2}${]}, the cross section for charge exchange
reactions, is evaluated at $V_{cx,}$where $V_{cx}$ is a representative
speed for charge exchange collisions \cite{Pauls,Meier}:
\begin{equation}
V_{cx}=\sqrt{\frac{4}{\pi}V_{thi}^{2}+\frac{4}{\pi}V_{thn}^{2}+v_{in}^{2}}\label{eq:536}
\end{equation}
where $V_{thi}$ and $V_{thn}$ are the thermal speeds of the ions
and neutral particles, and $v_{in}=|\mathbf{v}_{in}|=|\mathbf{v}_{i}-\mathbf{v}_{n}|$.
A formula for $\sigma_{cx}(V_{cx})\,[\mbox{m}^{2}]$ can be found
based on charge exchange data from Barnett \cite{Barnett,MeierPhd}.
For hydrogen and deuterium the fitting formulae are 
\begin{align}
\sigma_{cx-H}(V_{cx}) & =1.12\times10^{-18}-7.15\times10^{-20}\,ln(V_{cx})\nonumber \\
\sigma_{cx-D}(V_{cx}) & =1.09\times10^{-18}-7.15\times10^{-20}\,ln(V_{cx})\label{eq:536.1}
\end{align}
Taking the first moment of the moment of the charge exchange collision
operator leads to 
\begin{align}
\rho_{i}\left(\frac{\partial\mathbf{v}_{i}}{\partial t}\right)_{cx} & \approx-m_{i}\mathbf{v}_{in}\Gamma^{cx}-\mathbf{R}_{ni}^{cx}+\mathbf{R}_{in}^{cx}\nonumber \\
\rho_{n}\left(\frac{\partial\mathbf{v}_{n}}{\partial t}\right)_{cx} & \approx m_{i}\mathbf{v}_{in}\Gamma^{cx}+\mathbf{R}_{ni}^{cx}-\mathbf{R}_{in}^{cx}\label{eq:537}
\end{align}
where the notation $\left(\frac{\partial X}{\partial t}\right){}_{cx}$
is introduced here to represent the rate of change of any quantity
$X$ that arises due to charge exchange collisions, the term $m_{i}\mathbf{v}_{in}\Gamma^{cx}$
represents the transfer of momentum due to bulk fluid effects \cite{Meier,MeierPhd},
and $\mathbf{R}_{ni}^{cx}\mbox{ and }\mathbf{R}_{in}^{cx}$ represent
frictional drag forces that arise due to charge exchange:

\begin{align}
\mathbf{R}_{in}^{cx}\approx-m_{i}\sigma_{cx}(V_{cx})n_{i}n_{n}\mathbf{v}_{in}V_{thn}^{2}\left(4\left(\frac{4}{\pi}V_{thi}^{2}+v_{in}^{2}\right)+\frac{9\pi}{4}V_{thn}^{2}\right)^{-\frac{1}{2}}\nonumber \\
\mathbf{R}_{ni}^{cx}\approx m_{i}\sigma_{cx}(V_{cx})n_{i}n_{n}\mathbf{v}_{in}V_{thi}^{2}\left(4\left(\frac{4}{\pi}V_{thn}^{2}+v_{in}^{2}\right)+\frac{9\pi}{4}V_{thi}^{2}\right)^{-\frac{1}{2}}\label{eq:538}
\end{align}
Such frictional terms do not arise for the ionization and recombination
processes, in which the electron thermal speed is assumed to be far
higher than the relative particle motion \cite{Meier,MeierPhd}. Taking
the second moment of the moment of the charge exchange collision operator
leads to 
\begin{align}
\left(\frac{\partial p_{i}}{\partial t}\right)_{cx} & \approx(\gamma-1)\left(\Gamma^{cx}\frac{1}{2}m_{i}\,v_{in}^{2}-\mathbf{v}_{in}\cdot\mathbf{R}_{in}^{cx}+Q_{in}^{cx}-Q_{ni}^{cx}\right)\nonumber \\
\left(\frac{\partial p_{n}}{\partial t}\right)_{cx} & \approx(\gamma-1)\left(\Gamma^{cx}\frac{1}{2}m_{i}\,v_{in}^{2}+\mathbf{v}_{in}\cdot\mathbf{R}_{ni}^{cx}-Q_{in}^{cx}+Q_{ni}^{cx}\right)\label{eq:540.1}
\end{align}
where $Q_{in}^{cx}$ and $Q_{ni}^{cx}$ represent the transfer of
thermal energy \cite{Meier,MeierPhd} associated with charge exchange
reactions:

\begin{align}
Q_{in}^{cx}\approx m_{i}\sigma_{cx}(V_{cx})n_{i}n_{n}\frac{3}{4}V_{thn}^{2}\sqrt{\frac{4}{\pi}V_{thi}^{2}+\frac{64}{9\pi}V_{thn}^{2}+v_{in}^{2}}\nonumber \\
Q_{ni}^{cx}\approx m_{i}\sigma_{cx}(V_{cx})n_{i}n_{n}\frac{3}{4}V_{thi}^{2}\sqrt{\frac{4}{\pi}V_{thn}^{2}+\frac{64}{9\pi}V_{thi}^{2}+v_{in}^{2}}\label{eq:540}
\end{align}
Note that the term $-\mathbf{v}_{in}\cdot\mathbf{R}_{in}^{cx}$ in
the ion energy equation represents the rate of frictional work done
by neutral fluid on the ion fluid as a result of charge exchange reactions,
and the similar term $\mathbf{v}_{in}\cdot\mathbf{R}_{ni}^{cx}$ in
the neutral fluid energy equation represents the rate of frictional
work done by $\mathbf{R}_{ni}^{cx}$, which acts on the neutral fluid
with relative velocity $\mathbf{v}_{in}$.

\section{2-fluid MHD equations and other principal equations implemented in
the model\label{sec:2-fluid-MHD-equations}}

As shown in appendix \ref{sec:Appendix: 3-fluid-MHD-equations},  expressions
for the contibutions to the time rates of change of species number
density, fluid velocity and pressure due to scattering collisions,
reactive collsions, and external particle sources, which were developed
in appendix \ref{sec:Appendix:AScattering-collision} and section
\ref{sec:Reactive-collision-terms}, can be incorporated into the
three-fluid conservation equations for ion, electrons and neutral
particles. The single plasma fluid MHD equations use single fluid
center of mass velocity $\mathbf{v}$, and current density $\mathbf{J}$,
to describe average motion, where $\mathbf{v}=\frac{1}{\rho}\underset{\alpha=i,e}{\Sigma}\rho_{\alpha}\mathbf{v}_{\alpha}$
(with $\rho=\underset{\alpha=i,e}{\Sigma}\rho_{\alpha}$), and $\mathbf{J}=\underset{\alpha=i,e}{\Sigma}q_{\alpha}n_{\alpha}\mathbf{v}_{\alpha}$.
The assumptions $n=n_{i}=n_{e}$ and $m_{e}\rightarrow0$ are made,
so that $\mathbf{v}=\mathbf{v}_{i}$ and $m_{i}=m_{n}$. For the assumption
of singly charged ions, $q_{i}=-q_{e}=e$. The single fluid plasma
pressure is $p=p_{i}+p_{e}$, the heat flux density is $\mathbf{q}=\mathbf{q}_{i}+\mathbf{q}_{e}$,
and the single fluid viscous stress tensor is $\overline{\boldsymbol{\pi}}=\overline{\boldsymbol{\pi}}_{i}+\overline{\boldsymbol{\pi}}_{e}$.
The single plasma fluid mass, momentum, and energy conservation equations
are constructed by summing the corresponding ion and electron equations.
The procedure for the summation of terms is well documented in \cite{Meier,Braginski,bittencourt,farside_plasma}.
The charged-neutral particle frictional forces $\mathbf{R}_{in}=-\mathbf{R}_{ni}\mbox{ and }\mathbf{R}_{en}=-\mathbf{R}_{ne}$,
and heat exchange terms $Q_{in},\,Q_{en},\,Q_{ni}$, and $Q_{ne}$,
can be neglected, as mentioned in section \ref{sec:Scattering-collsion-termsS}. 

The resulting set of conservation equations, in continuous form, for
the two-fluid (plasma and neutral fluids) system, including ionization
\& recombination, charge exchange terms, as well as neutral source
terms, is:

\selectlanguage{german}%
\texttt{
\begin{align*}
\dot{n} & =-\nabla\cdot(n\mathbf{v})++\Gamma_{i}^{ion}-\Gamma_{n}^{rec}+\ensuremath{\nabla\cdot}\left(\zeta\nabla n\right)\\
\dot{\mathbf{v}} & =-\mathbf{v}\cdot\nabla\mathbf{v}+\frac{1}{\rho}\left(-\nabla p-\nabla\cdot\overline{\boldsymbol{\pi}}+\mathbf{J\times}\mathbf{B}-\Gamma_{i}^{ion}m_{i}\mathbf{v}_{in}-\Gamma^{cx}m_{i}\mathbf{v}_{in}\mathbf{-R}_{ni}^{cx}+\mathbf{R}_{in}^{cx}+\mathbf{f}_{\zeta}\right)\\
\dot{p} & =-\mathbf{v}\cdot\nabla p-\gamma p\,\nabla\cdot\mathbf{v}+(\gamma-1)\biggl(-\overline{\boldsymbol{\pi}}:\nabla\mathbf{v}-\nabla\cdot\mathbf{q}+\eta'J^{2}+\Gamma_{i}^{ion}\frac{1}{2}m_{i}v_{in}^{2}+Q_{n}^{ion}-\Gamma_{i}^{ion}\phi_{ion}\\
 & \,\,\,\,\,\,\,-Q_{i}^{rec}-Q_{e}^{rec}-\mathbf{R}_{in}^{cx}\cdot\mathbf{v}_{in}+Q_{in}^{cx}-Q_{ni}^{cx}+\Gamma^{cx}\frac{1}{2}m_{i}v_{in}^{2}\biggr)\\
\dot{n}_{n} & =-\nabla\cdot(n_{n}\mathbf{v}_{n})-\Gamma_{i}^{ion}+\Gamma_{n}^{rec}++\Gamma_{n}^{ext}+\ensuremath{\nabla\cdot}\left(\zeta_{n}\nabla n_{n}\right)\\
\dot{\mathbf{v}}_{n} & =-\mathbf{v}_{n}\cdot\nabla\mathbf{v}_{n}+\frac{1}{\rho_{n}}\biggl(-\nabla p_{n}-\nabla\cdot\overline{\boldsymbol{\pi}}_{n}+\Gamma_{n}^{rec}m_{i}\mathbf{v}_{in}-\mathbf{R}_{in}^{cx}+(\mathbf{R}_{ni}^{cx}+\Gamma^{cx}m_{i}\mathbf{v}_{in}{\color{blue})}\\
 & \,\,\,\,\,\,\,+\Gamma_{n}^{ext}m_{n}(\mathbf{v}_{n0}-\mathbf{v}_{n})+\mathbf{f}_{\zeta_{n}}\biggr)\\
\dot{p}_{n} & =-\mathbf{v}_{n}\cdot\nabla p_{n}-\gamma p_{n}\,\nabla\cdot\mathbf{v}_{n}+(\gamma-1)\biggl(-\overline{\boldsymbol{\pi}}_{n}:\nabla\mathbf{v}_{n}-\nabla\cdot\mathbf{q}_{n}-Q_{n}^{ion}+\Gamma_{n}^{rec}\frac{1}{2}m_{i}v_{in}^{2}+Q_{i}^{rec}{\color{red}-}\\
 & \,\,\,\,\,\,\,+\left(\mathbf{R}_{ni}^{cx}\cdot\mathbf{v}_{in}+\Gamma^{cx}\,\frac{1}{2}m_{i}v_{in}^{2}+Q_{ni}^{cx}\right)-Q_{in}^{cx}+\Gamma_{n}^{ext}\frac{1}{2}m_{n}(\mathbf{v}_{n}-\mathbf{v}_{n0})^{2}\biggr)+\Gamma_{n}^{ext}T_{n0}
\end{align*}
}

\selectlanguage{english}%
Note that the artificial density diffusion terms $\ensuremath{\nabla\cdot}\left(\zeta\nabla n\right)$
and $\ensuremath{\nabla\cdot}\left(\zeta_{n}\nabla n_{n}\right)$
have been included in the plasma fluid and neutral fluid continuity
equations. $\zeta$ and $\zeta_{n}$ {[}m$^{2}$/s{]} are the constant
density diffusion coefficients for the plasma and neutral fluids respectively.
Density diffusion smooths the density fields, by effectively removing
particles from high density regions and re-allocating them to low
density regions. A certain minimum level of density diffusion (around
50 to 100 m$^{2}$/s) is required for numerical stability in conjunction
with an acceptably large timestep for simulations involving rapid
acceleration of the fluids, such as in the CT formation and CT magnetic
compression scenarios presented in section \ref{sec:Simulation-results-with}.
The correction terms $\mathbf{f}_{\zeta}\ensuremath{/\rho}$ and $\mathbf{f}_{\zeta_{n}}\ensuremath{/\rho_{n}}$
are included in the plasma and neutral fluid momentum conservation
equations in order to maintain energy conservation and, in some simulation
scenarios, angular momentum conservation, as described in detail in
\cite{SIMpaper,thesis}. For the plasma fluid, anisotropic heat conduction
is implemented in the model; the continuous form for the plasma heat
flux density vector is $\mathbf{q=}n\left(\chi_{\parallel}\nabla_{\parallel}T+\chi_{\perp}\nabla_{\perp}T\right)$.
For the neutral fluid, $\mathbf{q}_{n}=n_{n}\,\chi_{n}\nabla_{\parallel}T_{n}$.
For simplicity, we have implemented isotropic viscosity and resistivity.
Further details of the model, with the discrete forms of the equations
implemented, and analytical demsonstrations of the conservation properties
of the numerical scheme, are presented in detail in \cite{SIMpaper,thesis}.

The code has the option to evolve the single plasma-fluid energy equation
or to evolve separate energy equations for the ions and electrons.
For the latter option, when plasma-neutral interaction is included,
the ion and electron energy equations are obtained by partitioning
the single fluid energy equation:\foreignlanguage{german}{\texttt{
\begin{align*}
\dot{p}_{i} & =-\mathbf{v}\cdot\nabla p_{i}-\gamma p_{i}\,\nabla\cdot\mathbf{v}+(\gamma-1)\biggl(-\overline{\boldsymbol{\pi}}:\nabla\mathbf{v}-\nabla\cdot\mathbf{q}_{i}+Q_{ie}+\Gamma_{i}^{ion}\frac{1}{2}m_{i}v_{in}^{2}+Q_{n}^{ion}-Q_{i}^{rec}\\
 & -\mathbf{R}_{in}^{cx}\cdot\mathbf{v}_{in}+Q_{in}^{cx}-Q_{ni}^{cx}+\Gamma^{cx}\frac{1}{2}m_{i}v_{in}^{2}\biggr)\\
\dot{p}_{e} & =-\mathbf{v}\cdot\nabla p_{e}-\gamma p_{e}\,\nabla\cdot\mathbf{v}+(\gamma-1)\left(+\eta'J^{2}-\nabla\cdot\mathbf{q}_{e}-Q_{ie}-\Gamma_{i}^{ion}\phi_{ion}-Q_{e}^{rec}\right)
\end{align*}
}}Here, $Q_{ei}=-Q_{ie}+\eta'J^{2}=-\frac{3m_{e}}{m_{i}}\frac{n(T_{e}-T_{i})}{\tau_{e}}+\eta'J^{2}$
determines the thermal energy per unit volume per second transferred
from ions to electrons due to ion-electron collisions \cite{Braginski,bittencourt,farside_plasma}.
Note that $p=p_{i}+p_{e}$, and $\mathbf{q}=\mathbf{q}_{i}+\mathbf{q}_{e}$,
so that the sum of the component energy equations yields the single
plasma-fluid energy equation. The magnitudes of species viscous tensor
components are proportional to $\mu_{\alpha}=\rho_{\alpha}\nu_{\alpha}$,
where $\mu_{\alpha},\,\rho_{\alpha},\,\mbox{and }\nu_{\alpha}$ are
the dynamic viscosity, mass density, and kinematic viscosity respectively
for species $\alpha=i,\,e$. Assuming comparable orders of density
and temperature for ions and electrons, then since $\nu_{\alpha}$
scales with $T_{\alpha}/m_{\alpha}$ \cite{Braginski,farside_plasma},
and $m_{e}\ll m_{i}$, it is reasonable to drop the viscous heating
term from the electron energy equation. In the two-temperature, single
plasma-fluid model, anisotropic thermal dissusion is evaluated as
$\mathbf{q}_{i}=n\left(\chi_{\parallel i}\nabla_{\parallel}T_{i}+\chi_{\perp i}\nabla_{\perp}T_{i}\right)$
and $\mathbf{q}_{e}=n\left(\chi_{\parallel e}\nabla_{\parallel}T_{e}+\chi_{\perp e}\nabla_{\perp}T_{e}\right)$.
The perpendicular and parallel thermal diffusion coefficients, $\chi_{\parallel\alpha}$
and $\chi_{\perp\alpha}$, are usually set to constant values which
lead to simulated CT lifetime and simulated ion temperature approximately
matching the experimentally observed counterparts - typical values
used are given in section \ref{sec:Neutral_SMRT}. 

With axisymmetry, the magnetic field can be represented in divergence-free
form in terms of the poloidal flux (per radian) function $\psi(r,z)$
and toroidal function $f(r,z)=rB_{\phi}$ as 
\begin{equation}
\mathbf{B=\nabla\psi\times\nabla\phi+}f\nabla\phi\label{eq:480}
\end{equation}
The reduced Ohms' law, $\mathbf{E}+\mathbf{v}\times\mathbf{B}=\eta'\mathbf{J}$
(here, $\eta'$ {[}$\Omega$-m{]} is the plasma resistivity), together
with Maxwell's equations and the expression for the axisymmetric magnetic
field, are used to derive expressions for $\dot{\psi}$ and $\dot{f}$:
\begin{eqnarray}
\dot{\psi} & = & -\mathbf{v}\cdot\nabla\psi+\eta\Delta^{*}\psi\label{eq:480.3}\\
\dot{f} & = & r^{2}\,\nabla\cdot\left(-\left(\frac{f}{r^{2}}\mathbf{v}\right)+\omega\mathbf{B}+\frac{\eta}{r^{2}}\nabla f\right)\label{eq:480.4}
\end{eqnarray}
Here, the eliptical operator $\Delta^{*}$ is defined as $\Delta^{*}\psi=r\frac{\partial}{\partial r}\left(\frac{1}{r}\frac{\partial\psi}{\partial r}\right)+\frac{\partial^{2}\psi}{\partial z^{2}}$,
$\omega=v_{\phi}/r$ is the angular speed, and $\eta\,[\mbox{m}^{2}\mbox{/s}]=\eta'/\mu_{0}$
is the resistive diffusivity. The divergence form of $\dot{f}$, along
with naturally imposed boundary conditions for $f$, enables conservation
of net system toroidal flux, as described in detail in \cite{SIMpaper,thesis}. 

\section{Conservation properties with inclusion of neutral fluid\label{sec:Energy-conservation-with}}

As detailed in \cite{SIMpaper,thesis}, the inherent properties of
the differential operators matrices that form the core of the DELiTE
code framework are used to enable, with appropriate boundary conditions,
global conservation of energy, particle count, toroidal flux, and
angular momentum. Because the equations for the neutral fluid are
analogous, in terms of conservation properties, to the equations for
the plasma fluid, it is to be expected that the conservation properties
will be maintained when the evolution of a neutral fluid is simulated
along with that of a plasma fluid. 
\begin{figure}[H]
\begin{centering}
\subfloat[]{\includegraphics[width=7cm,height=7cm]{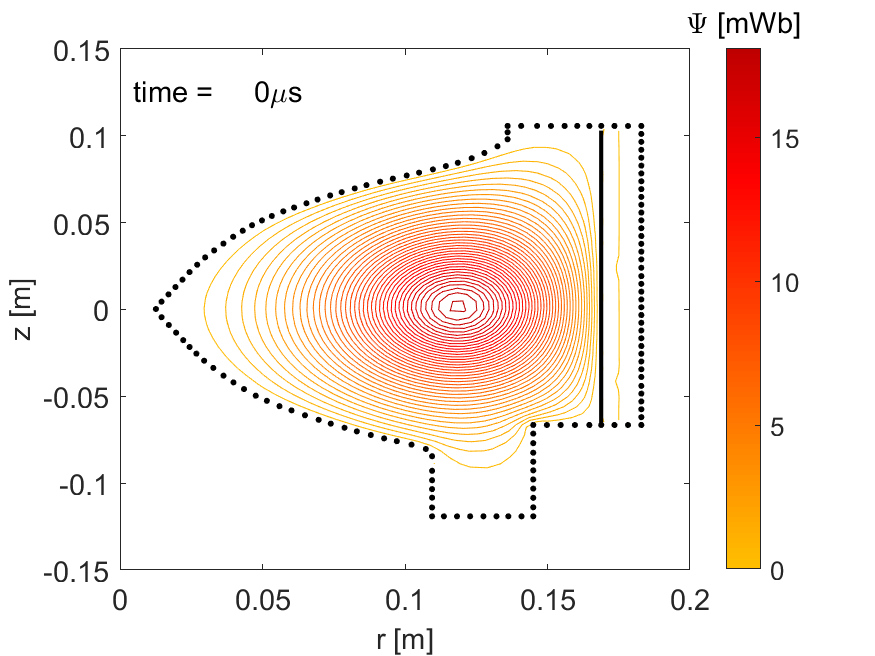}}\hfill{}\subfloat[]{\includegraphics[width=7cm,height=7cm]{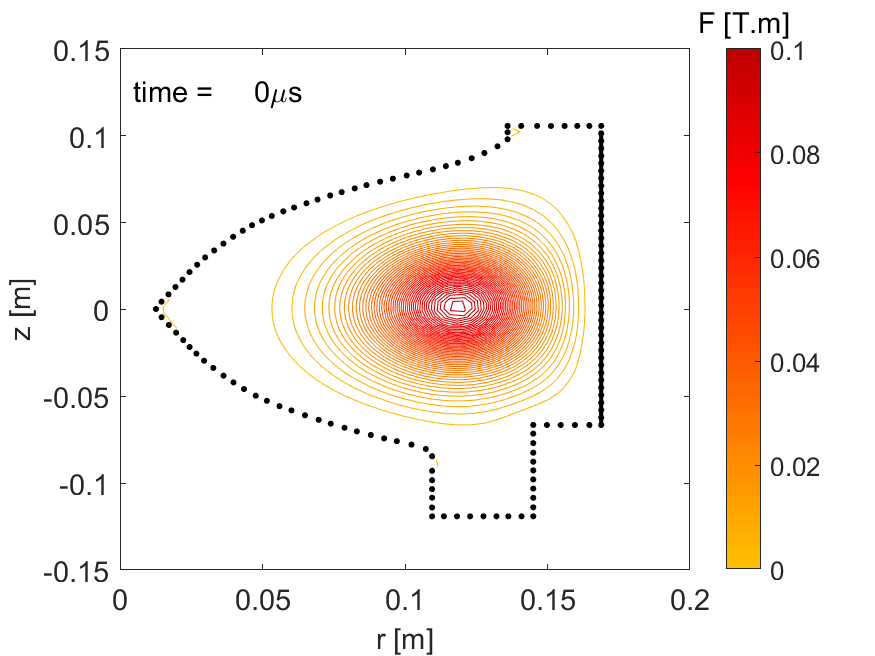}}
\par\end{centering}
\centering{}\subfloat[]{\includegraphics[width=10cm,height=6.5cm]{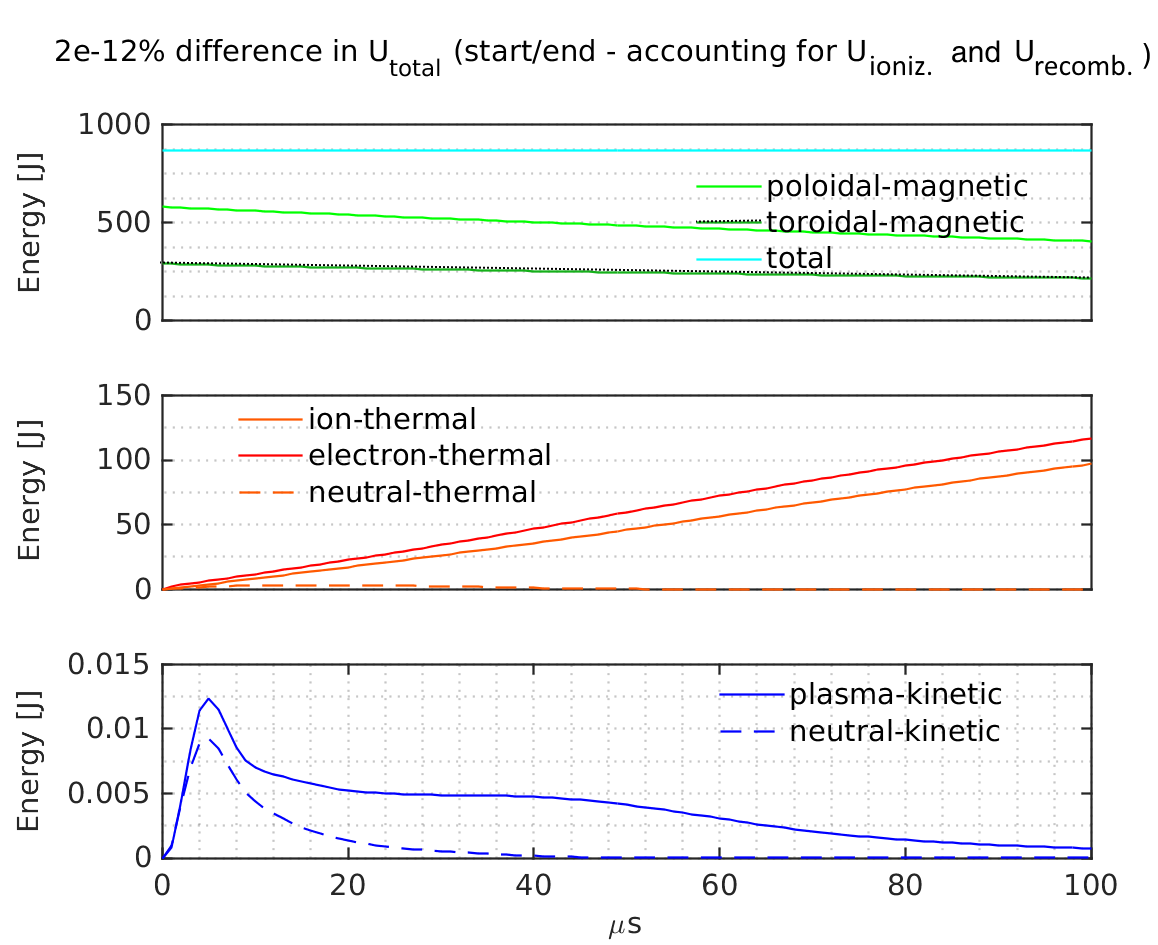}}\hfill{}\subfloat[]{\includegraphics[width=6cm,height=5cm]{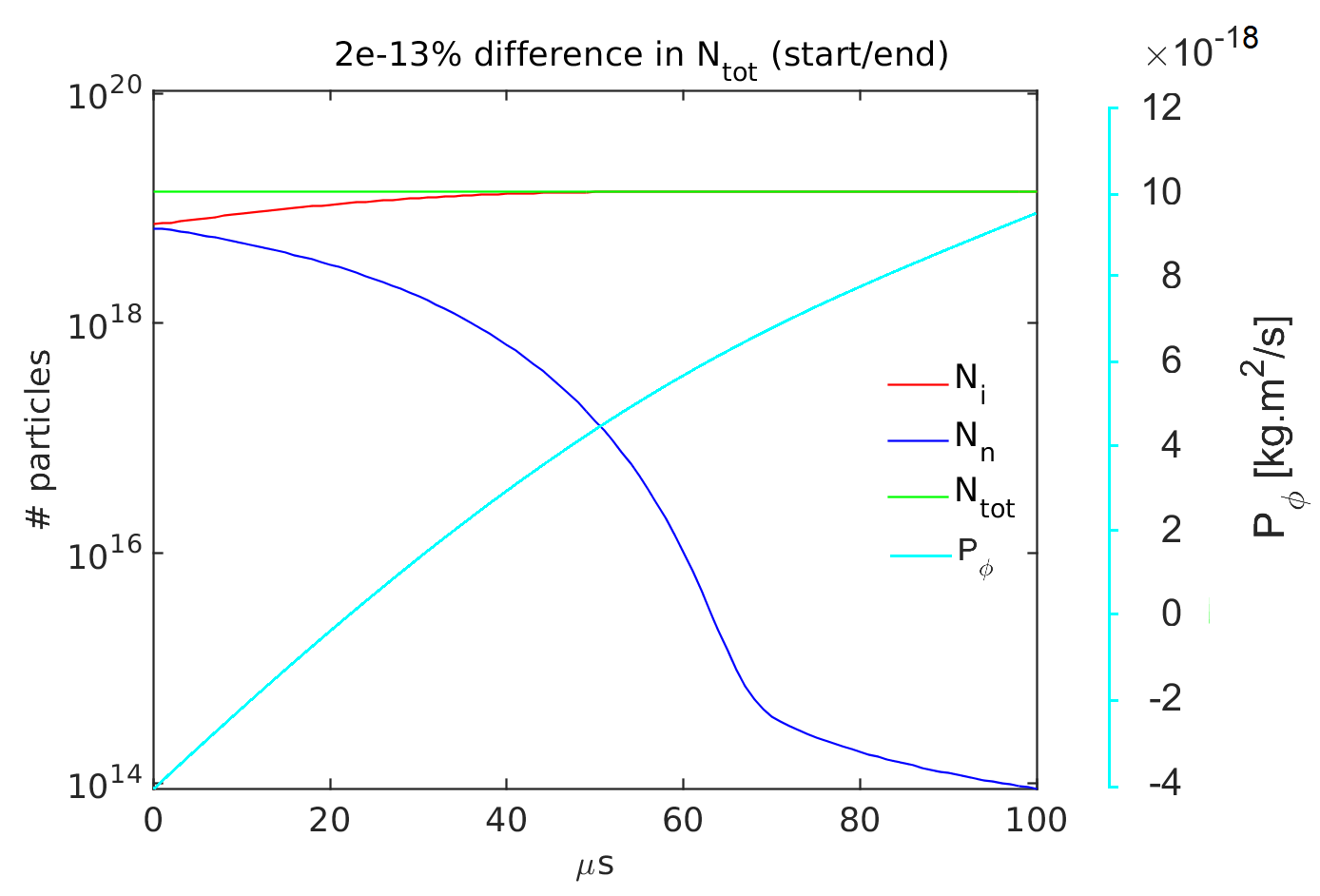}}\caption{\label{fig:Energy_N_cons}$\,\,\,\,$Contours of $\psi$ and $f$
for the initial equilibrium (figures (a) and (b)), and illustration
of energy conservation (c), and particle and angular momentum conservation
(d) for MHD simulation with neutral fluid}
\end{figure}
To demonstrate conservation properties, an MHD simulation in which
an initial Grad-Shafranov equilibrium was allowed to decay resistively
over 100 $\upmu$s was conducted. The simulation domain is a representation
of the CT confinement region of the SMRT plasma injector (see section
\ref{sec:Neutral_SMRT}). Contours of $\psi$ and $f$ for the initial
equilibrium are indicated in figures \ref{fig:Energy_N_cons}(a) and
(b).  Figure \ref{fig:Energy_N_cons}(c) indicates the partition of
energy, and how total system energy is conserved for a simulation
where a neutral fluid is evolved along with the plasma fluid. In this
case, the initial neutral fluid number density  was approximately
equal to the initial plasma fluid number density, with spatially uniform
initial distributions. 

The only explicitly applied boundary conditions for the simulation
were $v_{r}|_{\Gamma}=v_{z}|_{\Gamma}=v_{nr}|_{\Gamma}=v_{nz}|_{\Gamma}=0\mbox{ and }\psi|_{\Gamma}=0$,
so that the thermal and Poynting fluxes are zero through the boundary,
as described in \cite{SIMpaper,thesis}. Initial spatially uniform
temperatures were $T_{i}=T_{e}=T_{n}=0.02$eV. When the electron fluid
energy lost through ionization and recombination (the model assumes
that in the radiative recombination reaction, the electron thermal
energy is lost to the photon emitted, which leaves the optically thin
system without further interaction) processes are accounted for, it
is evident that total energy is conserved to numerical precision.
This simulation indicate that electron thermal energy lost to photons
as a result of recombination is insignificant over 100 $\upmu$s,
around 10 mJ or 0.001\% of the total system energy. However, to demonstrate
system energy conservation to approximately numerical precision, this
energy must be accounted for. Electron thermal energy expended due
to ionization processes ($(\gamma-1)\left(\Gamma_{i}^{ion}\phi_{ion}\right)$
{[}J/m$^{3}$/s{]}) is more significant, around 30 J over 100 $\upmu$s.
 The partitions of magnetic energy and the thermal and kinetic plasma
fluid energies follow the trends outlined in \cite{SIMpaper,thesis}
for case without neutral fluid evolution.

Figure \ref{fig:Energy_N_cons}(d) shows how total particle count
($N_{tot})$ is conserved to numerical precision for the same simulation.
Initial neutral particle count ($N_{n}$) was around equal to the
ion inventory ($N_{i}$). By the end of the simulation, neutral particles
account for around one in 100,000 of the total number of particles,
due to ionization. This is the main reason for the reduction over
time of neutral fluid kinetic and thermal energy.

Net angular momentum of the plasma and neutral fluids ($P_{\phi}$)
is conserved to numerical precision, as also indicated in figure \ref{fig:Energy_N_cons}(d).
No boundary conditions are explicitly applied to $v_{\phi}$ or to
$v_{n\phi}$, so the natural boundary conditions $\left(\nabla_{\perp}\,\omega\,\right)|_{\Gamma}=\left(\nabla_{\perp}\,\omega_{n}\,\right)|_{\Gamma}=0$
(here $\omega=v_{\phi}/r$ and $\omega_{n}=v_{n\phi}/r$ are the plasma
fluid and neutral fluid angular speeds respectively) are automatically
imposed, as described in \cite{SIMpaper,thesis}. 

In physical systems, net energy is \emph{not} conserved if there is
heat and electromagnetic flux through the system boundary. The simulations
of CT formation and magnetic compression presented in section \ref{sec:Simulation-results-with}
allow thermal flux out of the system, and CT formation and magnetic
compression is modelled by adding magnetic energy to the system (again,
the details of the model are presented in \cite{SIMpaper,thesis}),
so energy is not conserved. However, having a numerical model that
conserves the energy of an isolated system, as exemplefied above,
lends confidence to the results obtained, especially when the model
is applied to novel physical regimes. Angular momentum conservation
is not physical if there is friction between the fluids and the boundary
wall. Angular momentum is not conserved in the simulations presented
in section \ref{sec:Simulation-results-with}, because all velocity
components (for the plasma as well as the neutral fluids) are set
to zero at the boundary. 

\section{Simulation results with neutral fluid\label{sec:Simulation-results-with}}

\subsection{Neutral fluid interaction in SMRT geometry\label{sec:Neutral_SMRT} }

\begin{figure}[H]
\subfloat[]{\raggedright{}\includegraphics[scale=0.5]{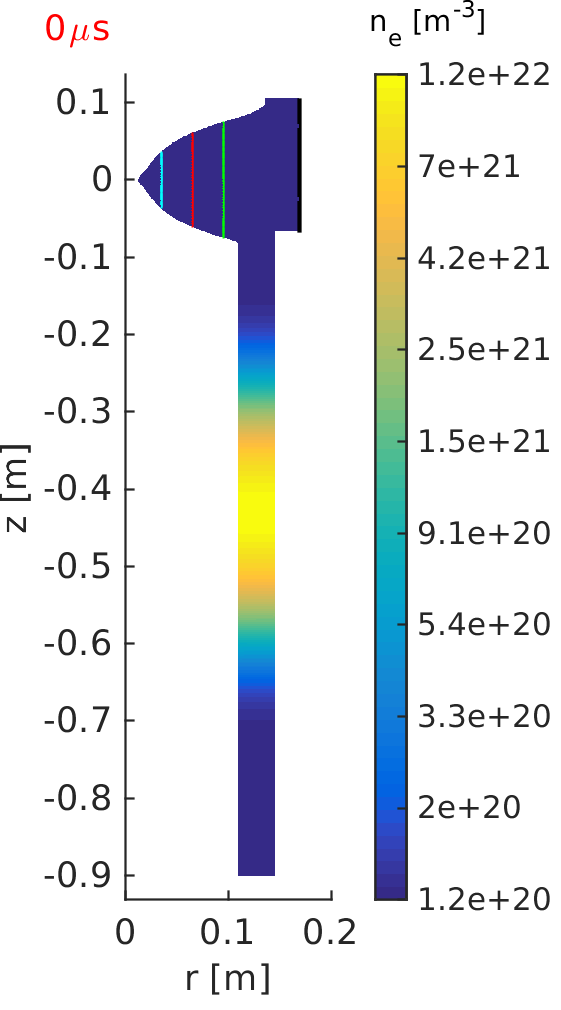}}\hfill{}\subfloat[]{\raggedright{}\includegraphics[scale=0.5]{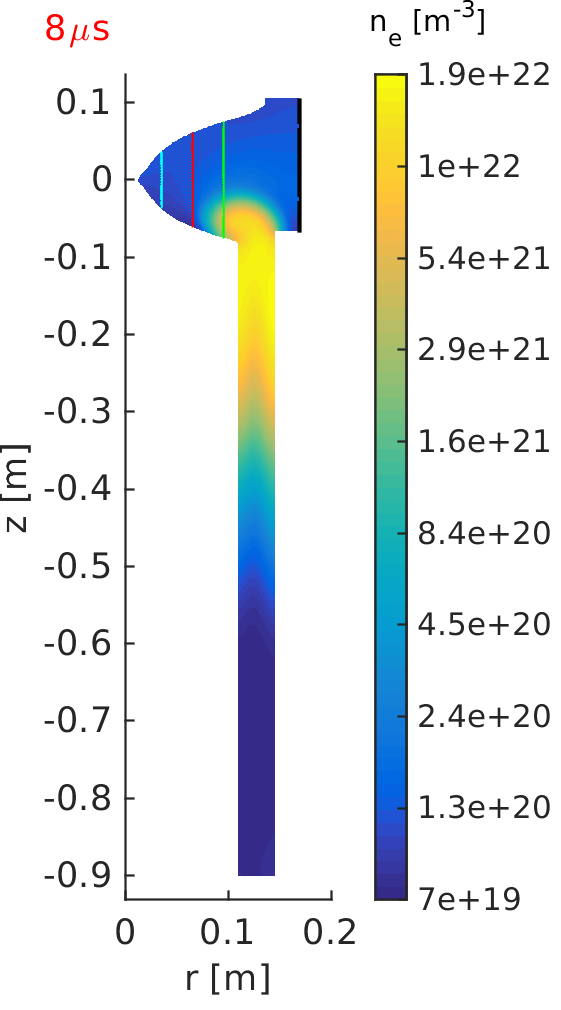}}\hfill{}\subfloat[]{\raggedright{}\includegraphics[scale=0.5]{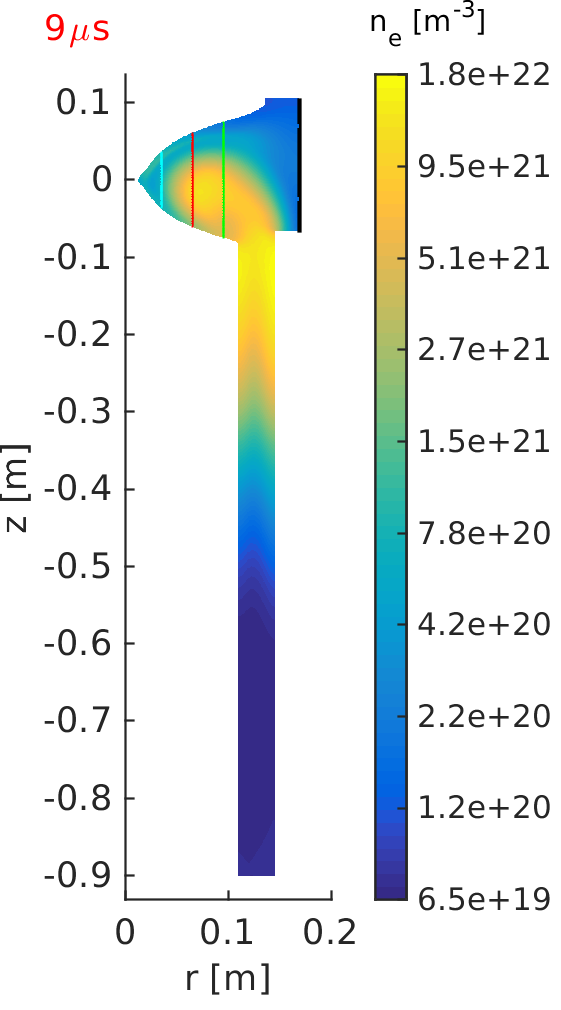}}

\subfloat[]{\raggedright{}\includegraphics[scale=0.5]{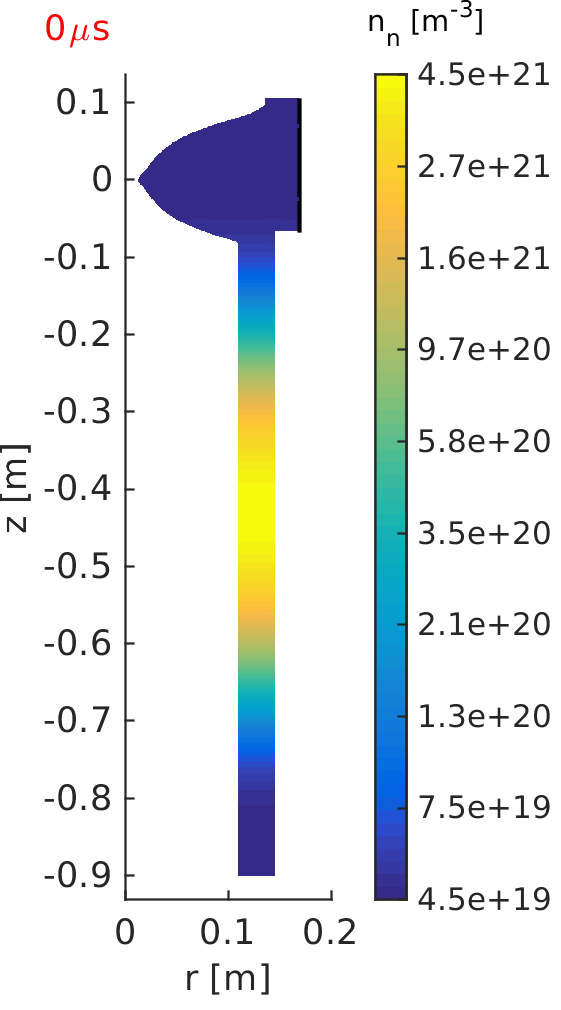}}\hfill{}\subfloat[]{\raggedright{}\includegraphics[scale=0.5]{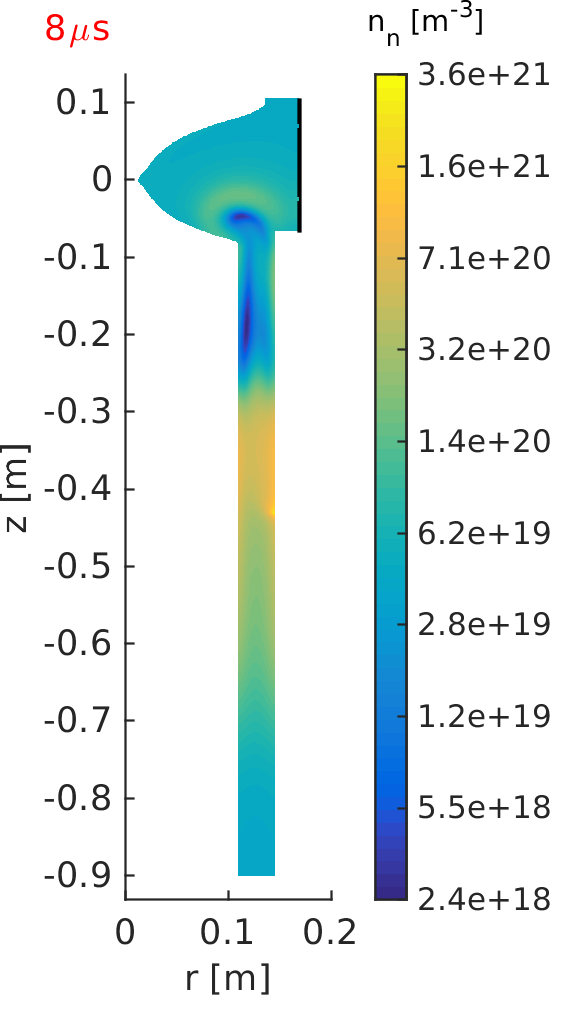}}\hfill{}\subfloat[]{\raggedright{}\includegraphics[scale=0.5]{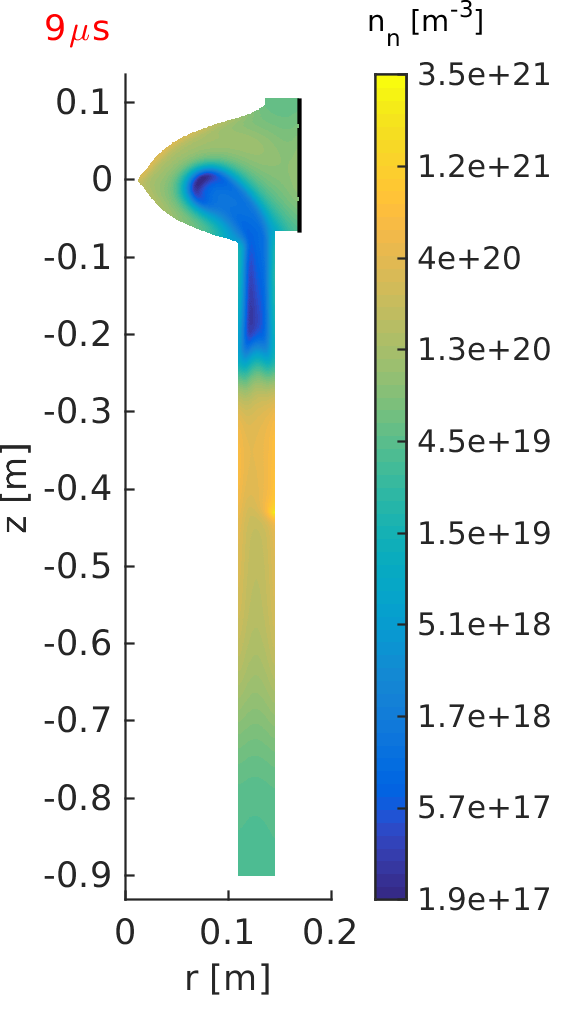}}

\caption{\label{fig: neut_bub_0}$\,\,\,\,$Electron (figures (a)-(c)) and
neutral fluid ((d)-(f)) density profiles at various times from a simulation
of CT formation in the SMRT plasma injector}
\end{figure}

In this section, results from simulations of CT formation and magnetic
compression are presented. The operation of magnetised Marshall guns,
and of the SMRT plasma injector in particular, is described in detail
in \cite{thesis,exppaper}, but a brief overview will be given here.
\begin{figure}[H]
\subfloat[]{\raggedright{}\includegraphics[scale=0.5]{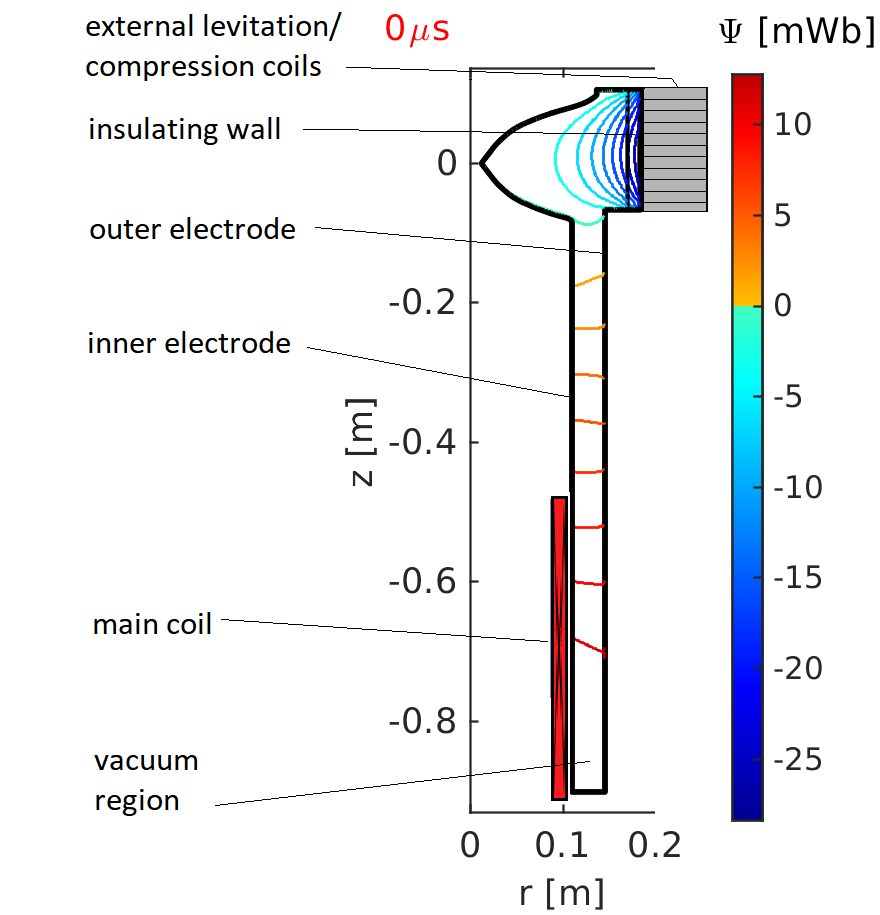}}\hfill{}\subfloat[]{\raggedright{}\includegraphics[scale=0.5]{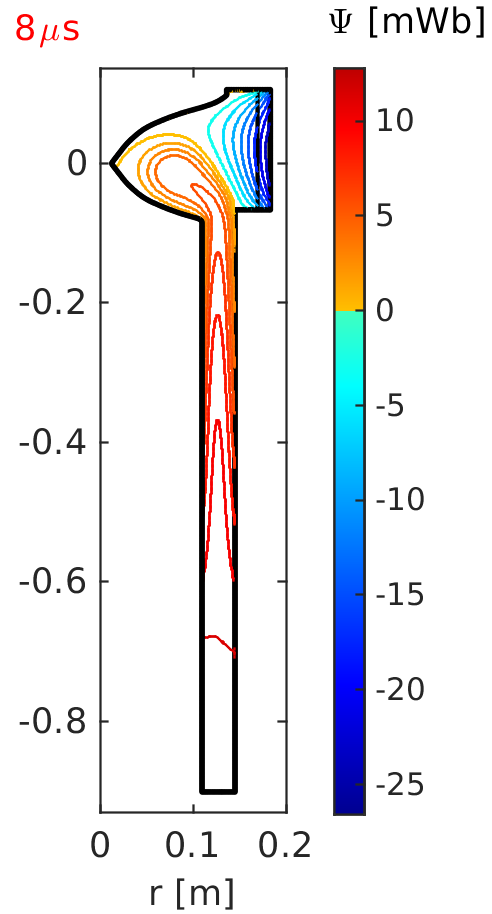}}\hfill{}\subfloat[]{\raggedright{}\includegraphics[scale=0.5]{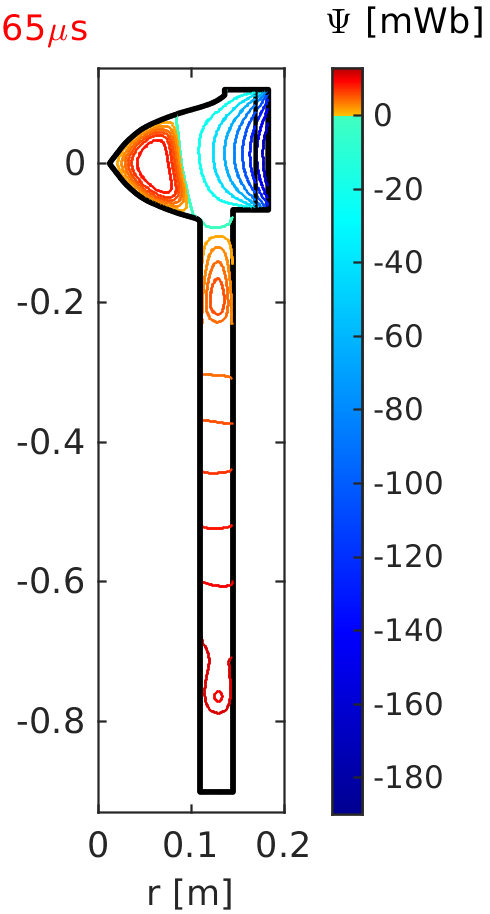}}

\caption{\label{fig:MHDform}$\psi$ contours from MHD simulation, with CT
formation (figures (a) and (b)) and magnetic compression (c)}
\end{figure}
 Poloidal flux contours from an MHD simulation of the magnetic compression
experiment are shown in figure \ref{fig:MHDform}. Simulation times
are notated in red at the top left of the figures. Referring to figure
\ref{fig:MHDform}(a), the outer radius of the hollow cylindrical
inner formation electrode (negative potential, up to 10kV at CT formation)
and inner radius of the anular outer formation electrode (grounded)
of the SMRT magnetized Marshall gun are represented by the vertical
lines at r=11cm and r=15cm respectively. The gun has a height of around
one meter. There are 100 turns of wire (AWG 4) located inside the
inner electrode (main coil); around 70 A steady state current (duration
3s) in the coil results in approximately 15mWb gun flux. The field
from the main coil soaks into the stainless steel walls that comprise
the inner and outer electrodes, as indicated in figure \ref{fig:MHDform}(a).
While the inner and upper walls of the CT confinement region are made
of aluminum, the outer wall of the CT confinement region is insulating,
as indicated in figure \ref{fig:MHDform}(a). This is a feature particular
to the SMRT device that allows magnetic flux due to toroidal current
in the stack of 11 levitation/compression coils to penetrate into
the CT confinement region. In the simulations, vacuum field only is
solved for in the insulating region, and the plasma dynamics are solved
for in the remaining solution domain (the solutions for $\psi$ and
$f$ in the two domains are coupled). The CT is formed into the levitation
field where it is levitated radially off the wall (figure \ref{fig:MHDform}(b)).
CT formation is initiated by puffing gas into the vacuum region (the
gas valves are located at $z=-$0.43 m), and energising the formation
electrodes. Plasma is formed initially near the gas valves, and is
advected rapidly by the $\mathbf{J}_{r}\times\mathbf{B}_{\phi}$ force
into the CT containment region. Here, $\mathbf{J}_{r}$ is the radial
formation current density across the plasma between the electrodes,
and $\mathbf{B}_{\phi}$ is the toroidal field due to the axial formation
current in the electrodes. The magnetic field associated with the
main coil is resistively pinned to the electrodes, and partially frozen
into the conducting plasma. The field is advected upwards with the
plasma, and reconnects near the entrance to the CT containment region,
forming closed CT flux surfaces by around 20$\upmu$s. The CT is rapidly
compressed radially inwards when the compression capacitor banks are
fired (compression current is up to 1MA, with a rise time of 20 $\upmu$s).
In this simulation, compression is initiated at $45\upmu\mbox{s}$,
and peak compression is at $65\upmu\mbox{s}$ (figure \ref{fig:MHDform}(c)).
Details of the models used to simulate CT formation, levitation and
compression can be found in \cite{SIMpaper,thesis}. 
\begin{table}[H]
\centering{}%
\begin{tabular}{|c|c|c|c|c|}
\hline 
\textbf{\small{}$\mathbf{neutralfluid}$} & {\small{}$\mathbf{N_{0}[\mbox{\textbf{m}}^{-3}]}$} & \textbf{\small{}$\mathbf{\boldsymbol{\sigma}}_{\mathbf{N}}^{\mathbf{2}}[\mbox{\textbf{m}}^{2}]$} & \textbf{\small{}$\mathbf{\boldsymbol{\zeta}_{n}[\mbox{\textbf{m\ensuremath{\mathbf{^{\mathbf{2}}}}/s}}]}$} & \textbf{\small{}$\mathbf{add_{N}}$}\tabularnewline
\hline 
{\small{}1} & {\small{}$4.5\times10^{20}$} & {\small{}0.01} & {\small{}90} & {\small{}1}\tabularnewline
\hline 
 &  &  &  & \tabularnewline
\hline 
{\small{}$\mathbf{vary_{\chi_{N}}}$} & \textbf{\small{}$\mathbf{vary_{\nu_{N}}}$} & \textbf{$\boldsymbol{\chi_{Nmax}}$$[\mbox{\textbf{m\ensuremath{\mathbf{^{\mathbf{2}}}}/s}}]$} & \textbf{$\boldsymbol{\nu_{Nmax}}$$[\mbox{\textbf{m\ensuremath{\mathbf{^{\mathbf{2}}}}/s}}]$} & {\small{}$\boldsymbol{\chi_{CX}}$}\tabularnewline
\hline 
{\small{}1} & 1 & $5\times10^{4}$ & $1\times10^{4}$ & {\small{}1}\tabularnewline
\hline 
\end{tabular}\caption{\label{tab:Sim parametersNeut} $\,\,\,\,$Neutral-relevant code input
parameters }
\end{table}
The code inputs in table \ref{tab:Sim parametersNeut} are related
to the neutral fluid dynamics for the simulation  presented here.
When interaction between the plasma and a neutral fluid is evolved,
code input parameter $neutralfluid$ is set equal to one. Analogous
to the case described in \cite{SIMpaper,thesis} for the initial plasma
distribution, the initial static neutral fluid distribution is determined
by a Gaussian profile centered around the location of the plasma injector
gas valves at $z=-0.43$m (refer to figure \ref{fig: neut_bub_0}(a)),
with variance $\sigma_{N}^{2}$ determining the degree of neutral
fluid spread around the gas valves, and neutral number density scaling
$N_{0}.$ $\zeta_{n}[\mbox{m}^{2}\mbox{/s}]$ is the coefficient of
neutral fluid density diffusion, which is required for numerical stability.
$add_{N}=1$ implies that neutral fluid is added to the simulation
domain at the location of the gas valves throughout the simulation.
Physically, the gas valves remain open for up to around a millisecond
after they are first opened. In general, simulations including neutral
dynamics have $vary_{\nu_{N}}$ and $vary_{\chi_{N}}$ set to one,
so that the analytical closures given by the Chapman-Enskog formulae
for the neutral fluid viscous and thermal diffusion coefficients are
used. However, if code input $\chi_{CX}$ is also set to one, as it
is for this simulation, the modified expression for $\chi_{N}$ is
used to determine thermal diffusion for the neutral fluid: 
\begin{equation}
\chi_{n}\approx\frac{75\sqrt{\pi}}{64}\frac{V_{thn}^{2}}{\nu_{cn}+\nu_{cx}}=\frac{75\sqrt{\pi}}{64}\frac{V_{thn}}{\frac{1}{\lambda_{mfp}}+\sigma_{cx}n_{n}}\label{eq:544.1}
\end{equation}
This expression \cite{Meier,MeierPhd} arises from the consideration
that, if the charge exchange collision frequency is higher than the
frequency for neutral-neutral scattering collisions, neutral thermal
conductivity should be reduced. Here, $V_{thn}$ is the thermal speed
of the neutral particles, $\nu_{cn}=V_{thn}/\lambda_{mfp}$ is the
neutral-neutral scattering collision frequency, and $\nu_{cx}=V_{thn}/\lambda_{cx}$
is the charge exchange frequency, where $\lambda_{cx}\approx1/(\sigma_{cx}n_{n})$
is the mean free path for charge exchange collisions. According to
equation \ref{eq:544.1}, $\chi_{n}$ will be limited by whichever
frequency dominates. If code input $\chi_{CX}$ is set to zero, $\chi_{n}\rightarrow(75\sqrt{\pi}/64)\,(V_{thn}^{2}/\nu_{cn}$),
the standard Chapman-Enskog expression \cite{farside_plasma}. It
is found that the expression in equation \ref{eq:544.1} results in
an increase of maximum $T_{n}$ of around 10\% compared with cases
where $\chi_{CX}$ is set to zero and $vary_{\chi_{N}}$ is set to
one. Constant coefficients are used if $vary_{\nu_{N}}$ and $vary_{\chi_{N}}$
are set to zero. $\chi_{Nmax}$ and $\nu_{Nmax}$ determine the upper
limits, required for moderately long timesteps, applied to the neutral
fluid thermal and viscous diffusion coefficients. 
Figures \ref{fig: neut_bub_0}(a) and (d) show the initial distributions
for plasma density, represented by $n_{e}$, and neutral fluid density
$n_{n}$, from a simulation of CT formation in the SMRT plasma injector,
on which the magnetic compression experiment \cite{thesis,exppaper}
was conducted. The initial density distributions are Gaussian profiles,
centered around the gas puff locations at $z=-0.43$m, with a higher
variance for the neutral fluid distribution, representing that the
neutral gas has diffused around the gas puff valve locations, while
the initial plasma distribution, is more localised to the gas puff
locations. The initial neutral particle inventory, determined by $\sigma_{N}^{2}$
and $N_{0}$, was over half the initial plasma particle inventory
for this simulation. Note that $n_{e}=Z_{eff}\,n_{i}$, where $Z_{eff}$,
the volume-averaged ion charge, is equal to 1.3 for this simulation.
As shown in figures \ref{fig: neut_bub_0}(b) and (c), plasma is starting
to enter the CT containment region at 8$\upmu$s and $9\upmu$s. A
front of neutral fluid precedes the plasma as it is advected upwards
(figures \ref{fig: neut_bub_0}(e) and (f)). Note that neutral particles
are being added at the gas puff valve locations by the outer boundary
at $z=-0.43$m. In the experiment, the gas valves are opened at $t\sim-400\upmu$s,
and remain open for $\sim1\mbox{ms}$, so that cold neutral gas is
being added to the vacuum vessel throughout the simulation, at a rate
that can be estimated and assigned to the simulated neutral particle
source terms. 
\begin{figure}[H]
\subfloat[]{\raggedright{}\includegraphics[width=7cm,height=5cm]{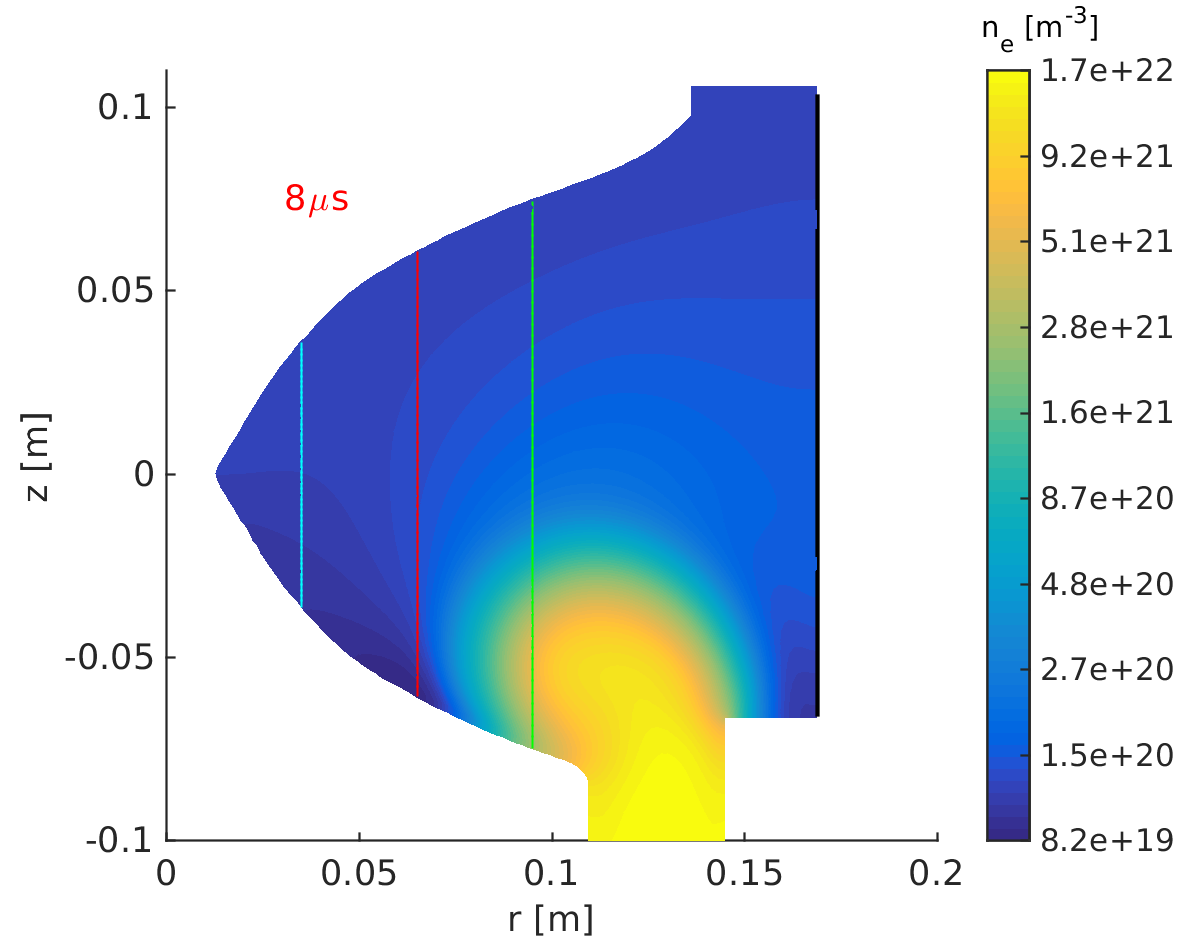}}\hfill{}\subfloat[]{\raggedright{}\includegraphics[width=7cm,height=5cm]{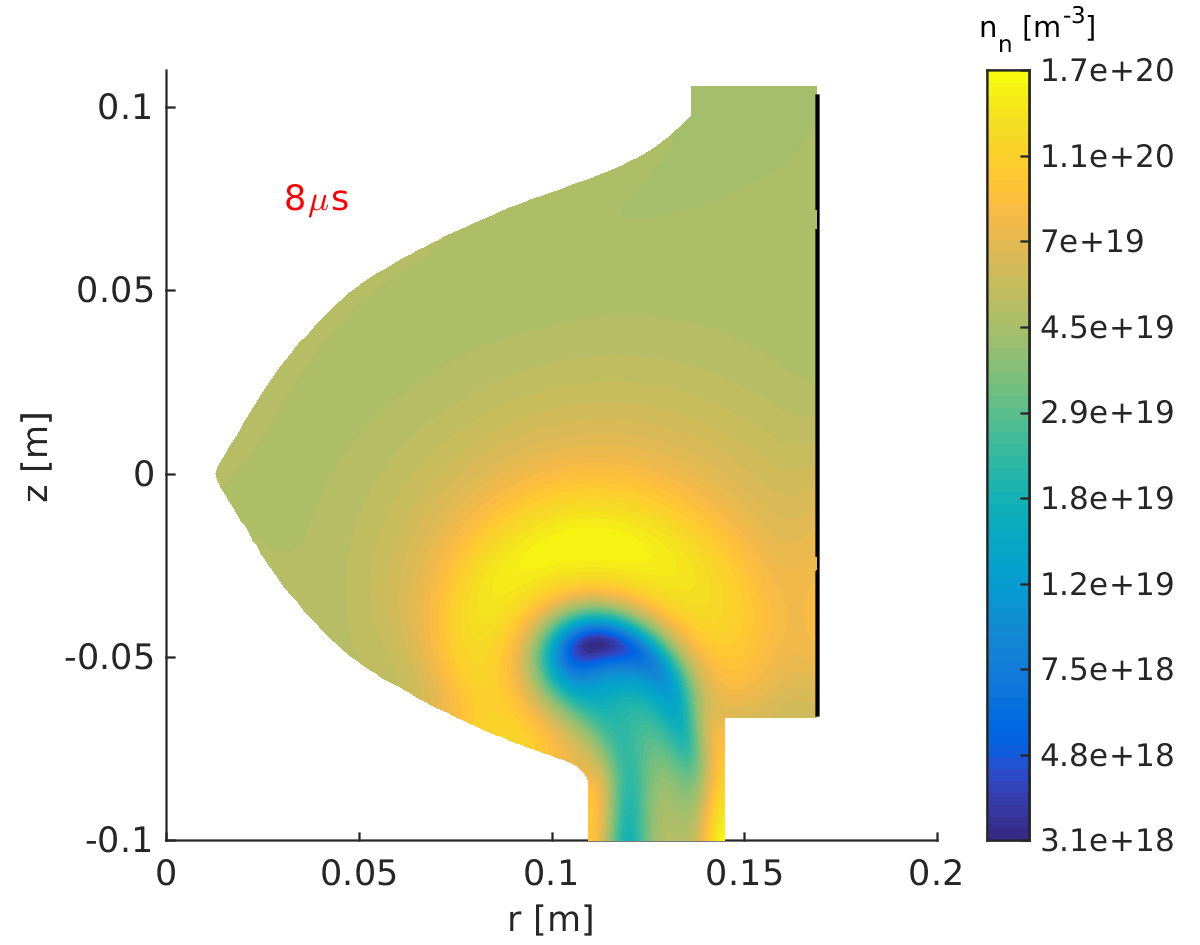}}

\subfloat[]{\raggedright{}\includegraphics[width=7cm,height=5cm]{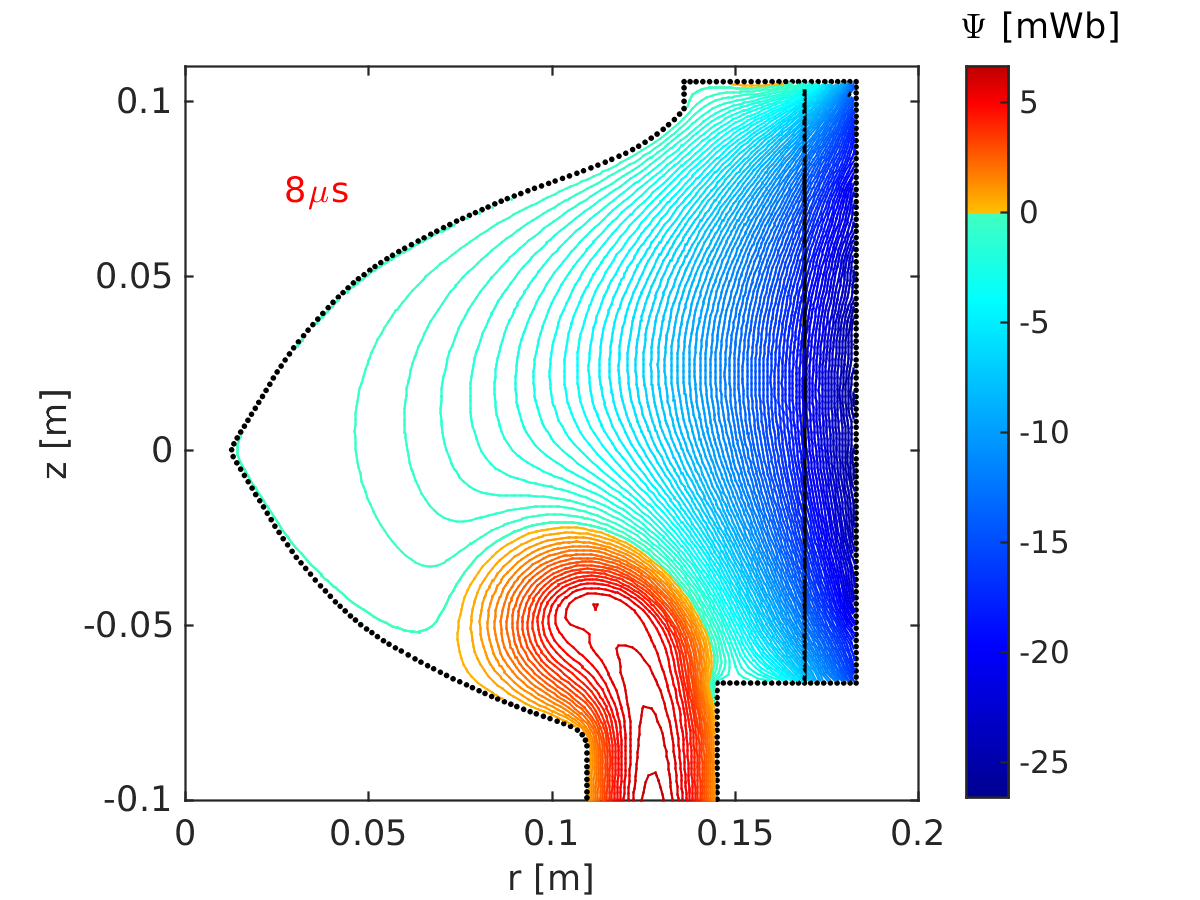}}\hfill{}\subfloat[]{\raggedright{}\includegraphics[width=7cm,height=5cm]{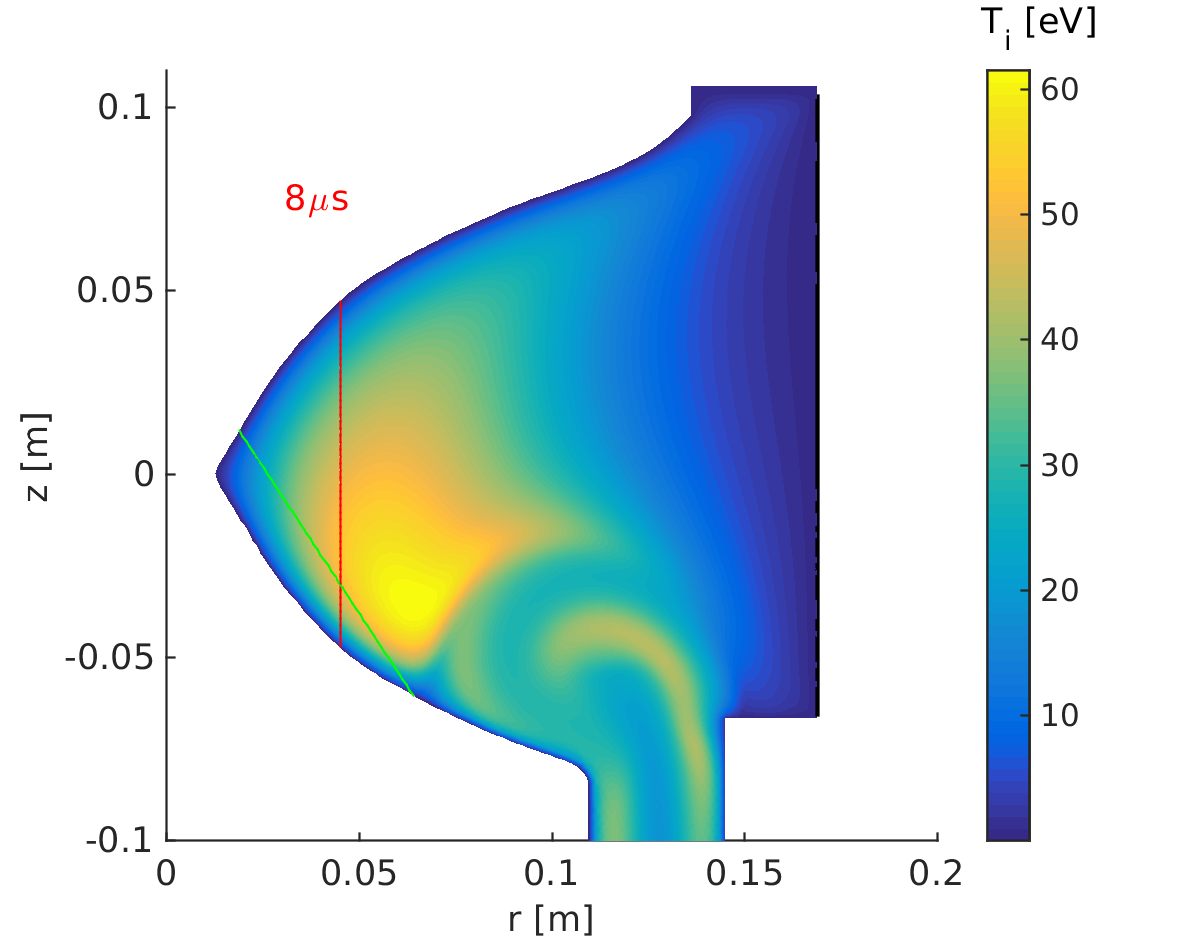}}

\subfloat[]{\raggedright{}\includegraphics[width=7cm,height=5cm]{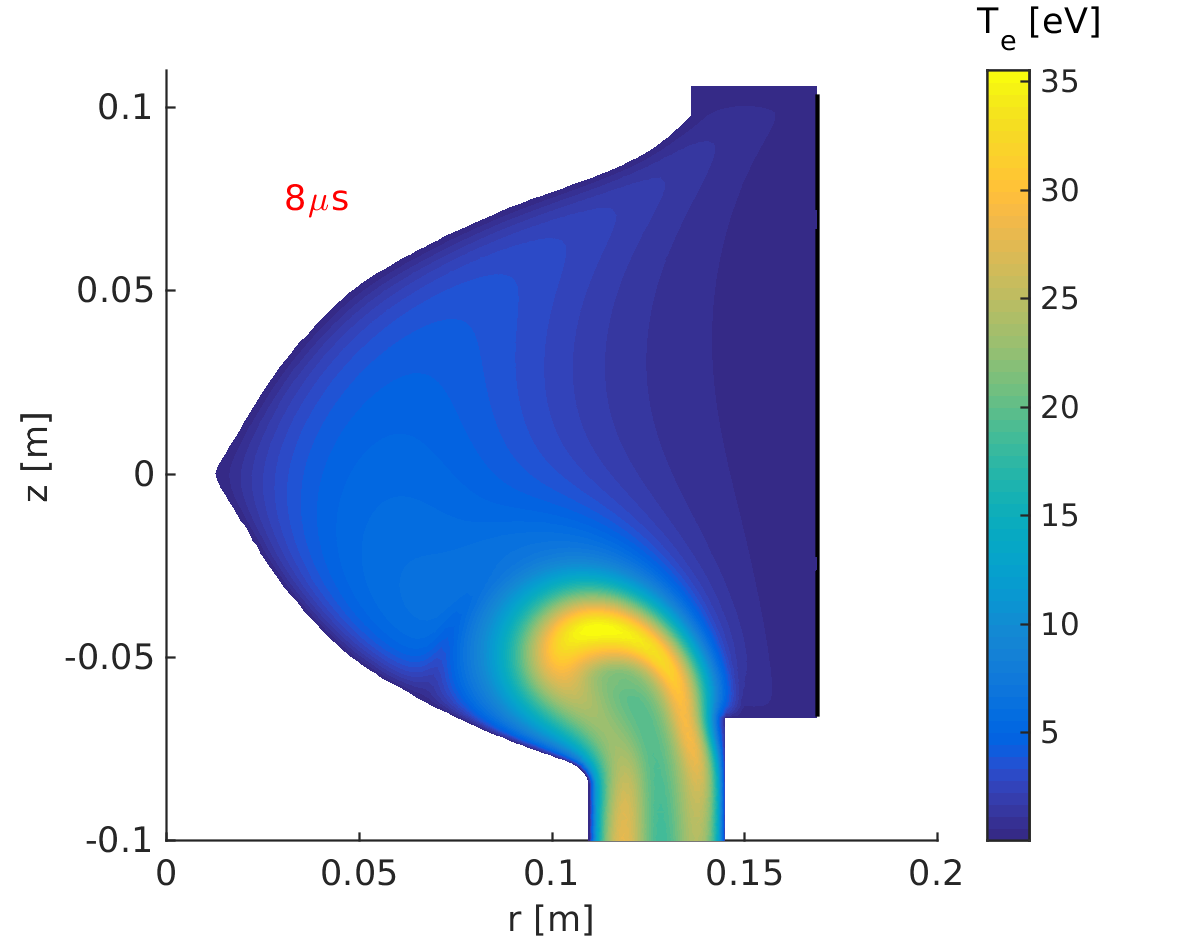}}\hfill{}\subfloat[]{\raggedright{}\includegraphics[width=7cm,height=5cm]{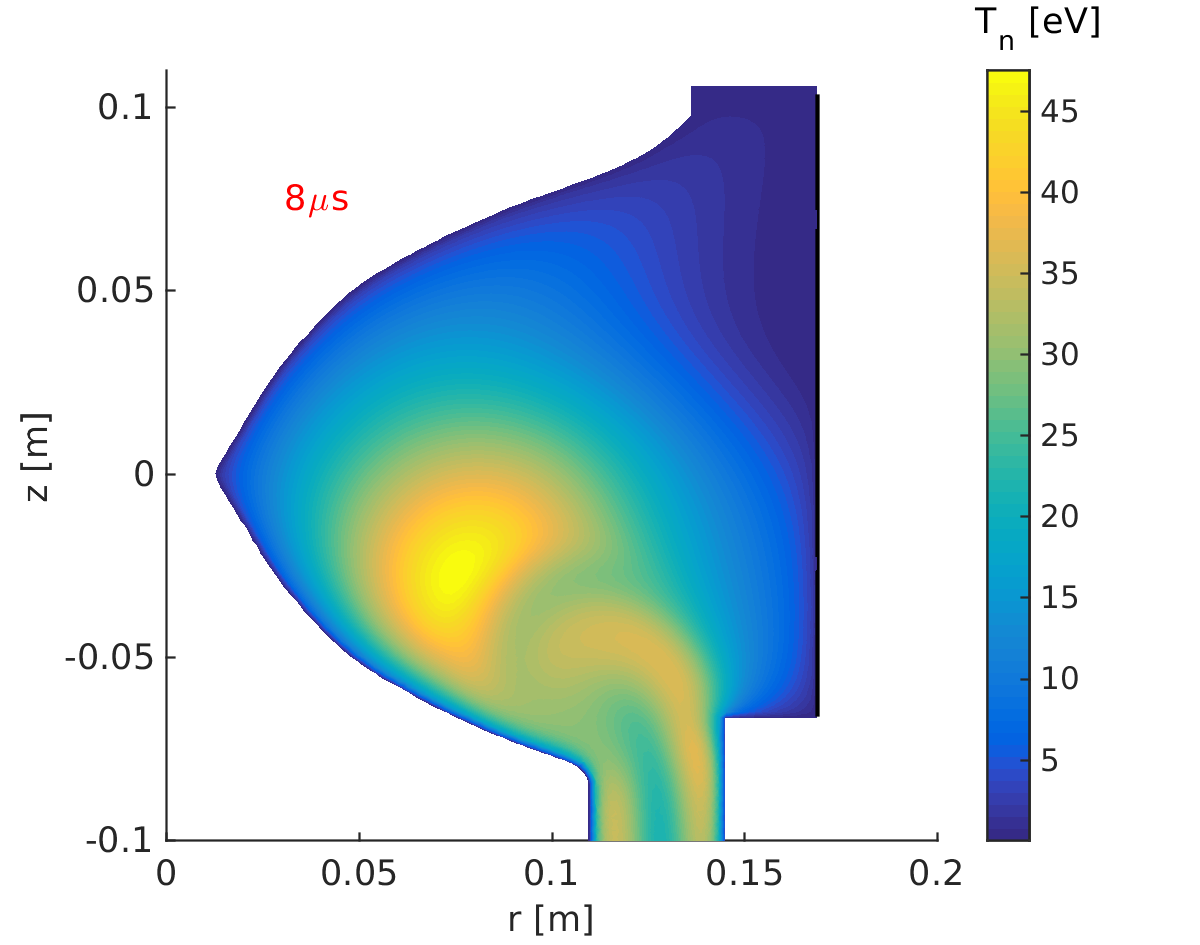}}

\caption{\label{fig: neut_bub_1}$\,\,\,\,$Electron and neutral fluid density
profiles (figures (a), (b)), poloidal flux contours (c), and ion (d),
electron (e) and neutral fluid (f) temperature profiles at 8$\upmu$s,
from a simulation of CT formation in the SMRT plasma injector}
\end{figure}
Figures \ref{fig: neut_bub_1}(a) and (b) show close-up views of $n_{e}$
and $n_{n}$ at 8$\upmu$s. Figures \ref{fig: neut_bub_1}(c) and
(d) show $\psi$ contours and the distribution of $T_{i}$ at the
same time. Ions are hot due to viscous heating. Ohmic heating in combination
with heat exchange with ions results in hot electrons (figure \ref{fig: neut_bub_1}(e)).
Note that neutral fluid density is low where $T_{e}$ is high due
to ionization (figure \ref{fig: neut_bub_1}(b)). Due to charge exchange
reactions, neutral fluid temperature tends to equilibrise with ion
temperature (figures \ref{fig: neut_bub_1} (d) and (f)), and can
become hotter than ions if the thermal diffusion for neutral fluid
is set to be lower than ion thermal diffusion. In general, when $\chi_{\parallel i}$
and $\chi_{\parallel e}$ are fixed at moderate experimentally relevant
values, such as $\chi_{\parallel e}\sim16000\,[\mbox{m}^{2}\mbox{/s}],\,\chi_{\parallel i}\sim5000\,[\mbox{m}^{2}\mbox{/s}],\,\chi_{\perp e}\sim240\,[\mbox{m}^{2}\mbox{/s}]$,
and $\chi_{\perp i}\sim120$ $[\mbox{m}^{2}\mbox{/s}]$ for this simulation,
and $\chi_{N}$ is determined by equation \ref{eq:544.1}, with $\chi_{Nmax}\gtrsim5\times10^{4}$,
as is the case for this simulation, it is found that $T_{n}<T_{i}$.
\begin{figure}[H]
\subfloat[]{\raggedright{}\includegraphics[width=7cm,height=5cm]{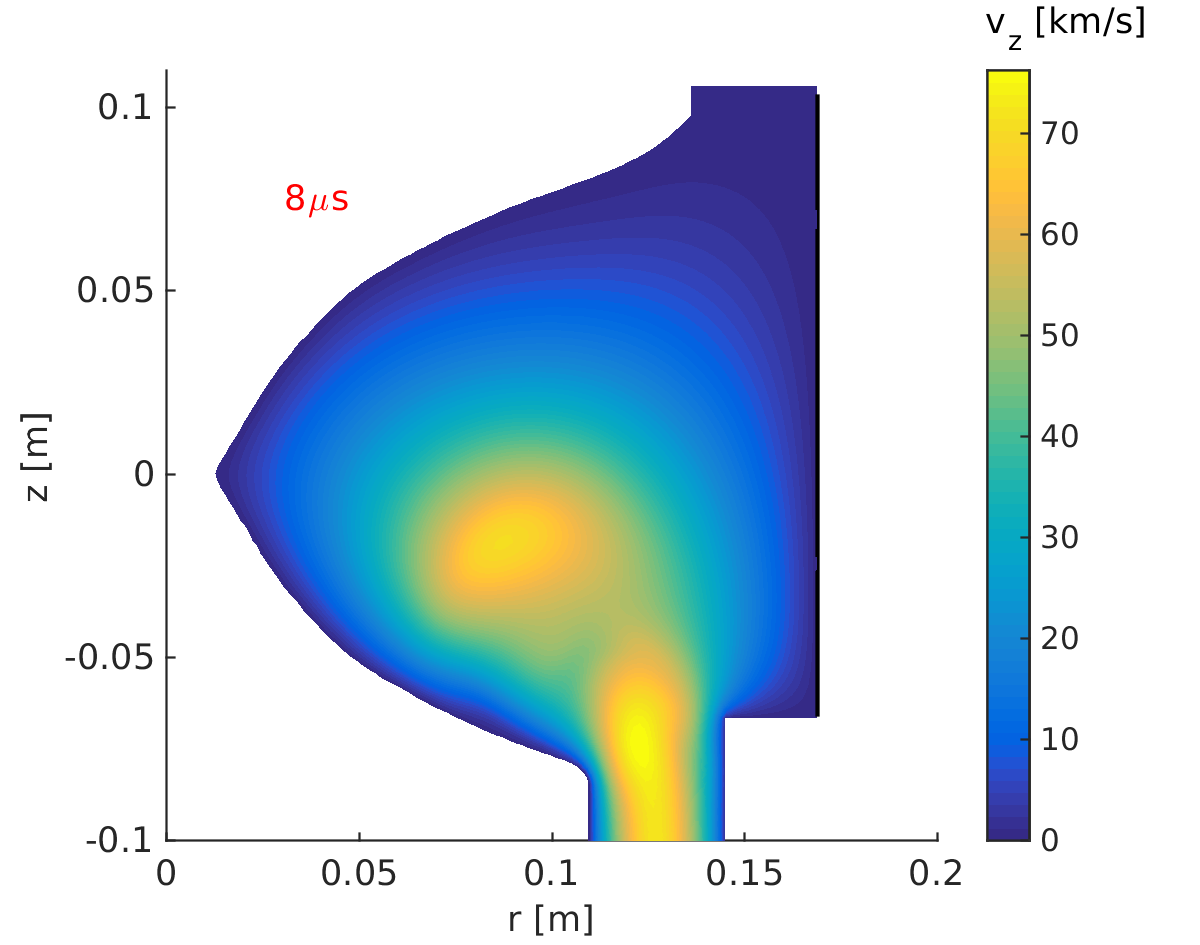}}\hfill{}\subfloat[]{\raggedright{}\includegraphics[width=7cm,height=5cm]{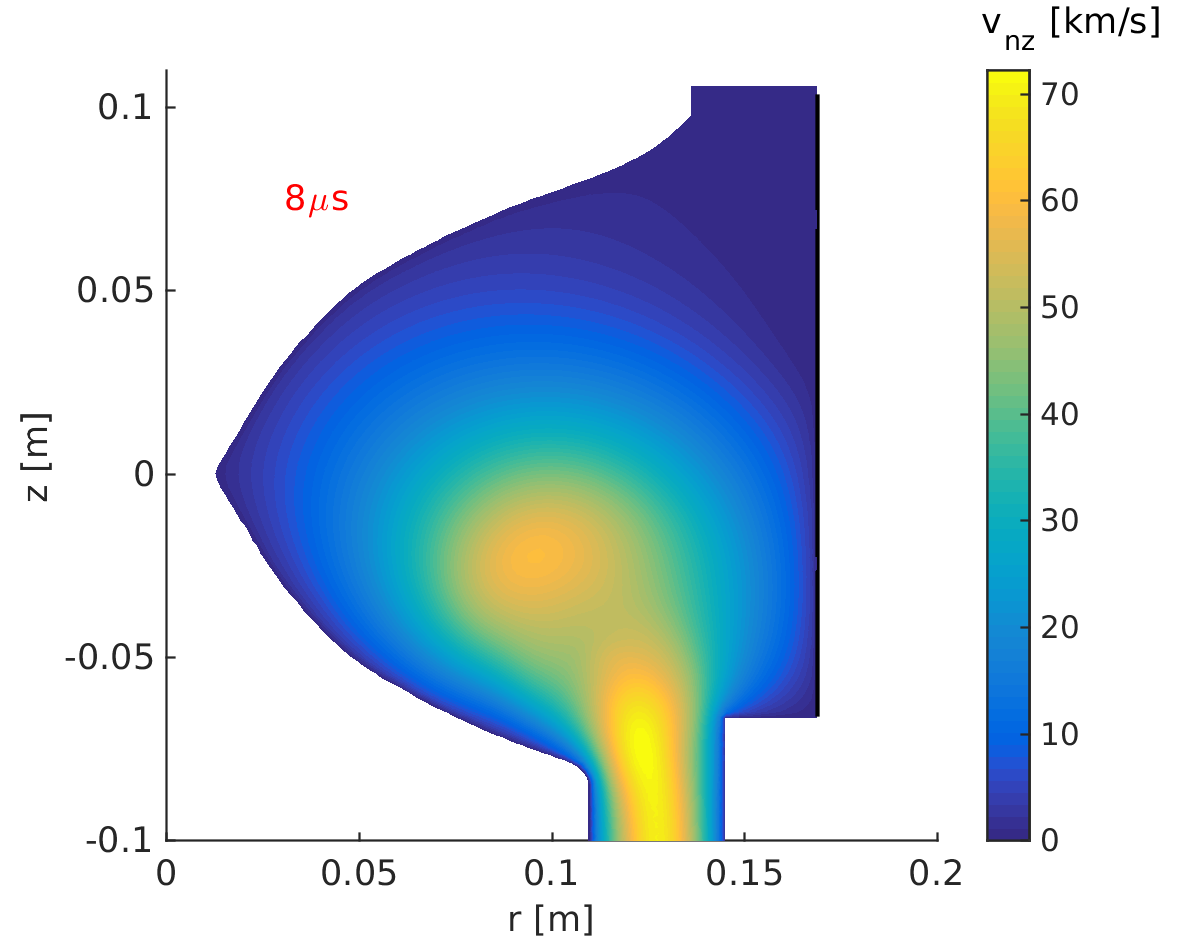}}

\subfloat[]{\raggedright{}\includegraphics[width=7cm,height=5cm]{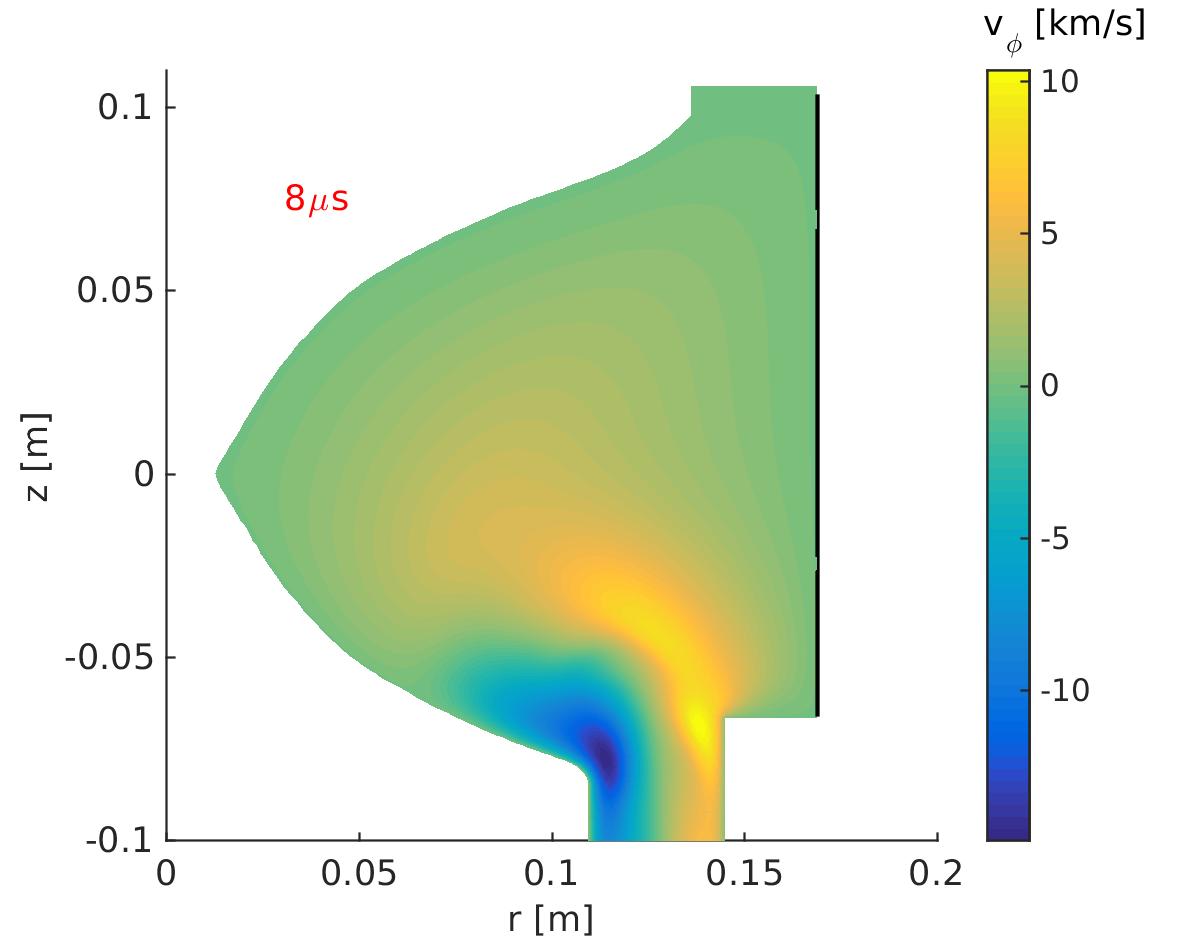}}\hfill{}\subfloat[]{\raggedright{}\includegraphics[width=7cm,height=5cm]{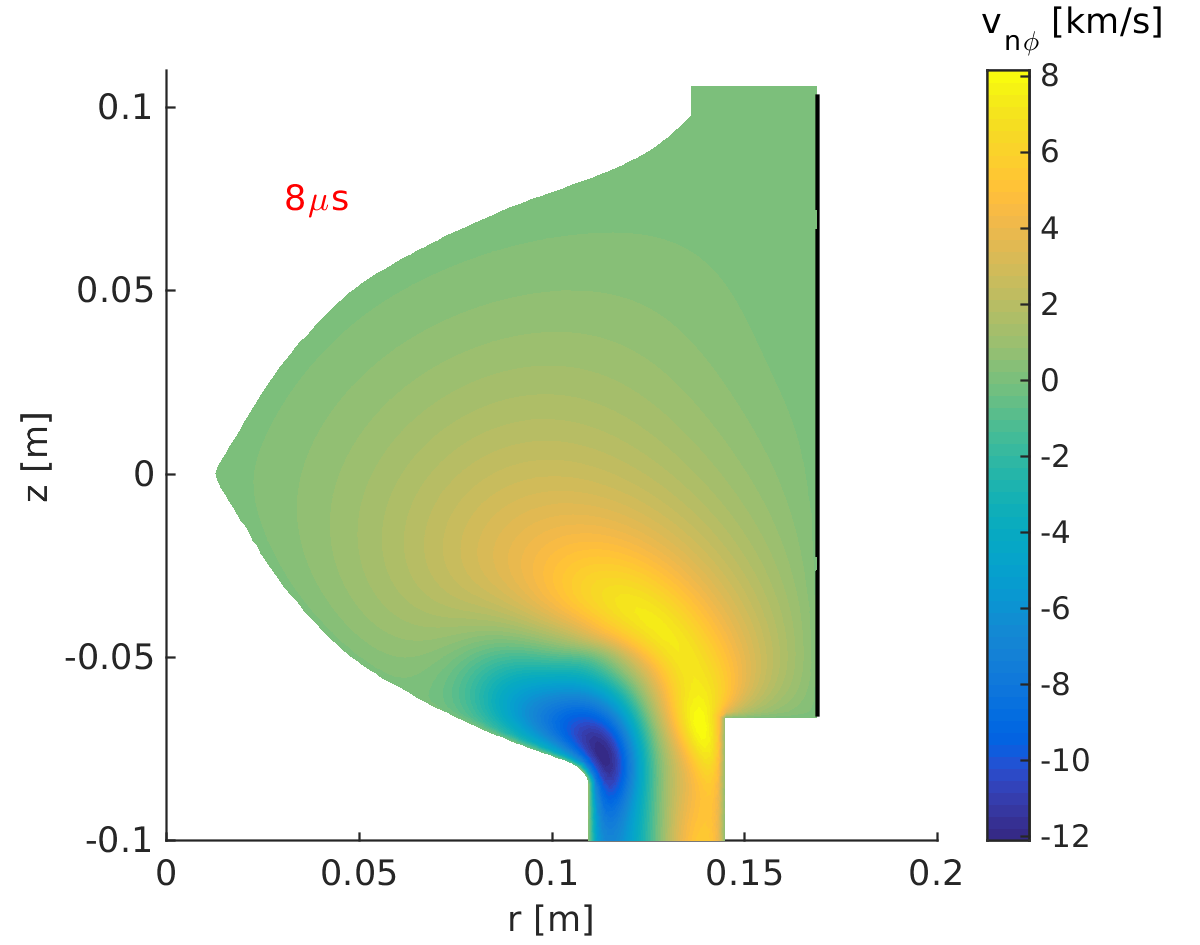}}

\caption{\label{fig: neut_bub_2}$\,\,\,\,$Profiles of plasma (figures (a),
(c)) and neutral fluid ((b), (d)) axial and azimuthal velocity components
at 8$\upmu$s, from a simulation of CT formation in the SMRT plasma
injector}
\end{figure}
Figures \ref{fig: neut_bub_2}(a) and (b) shows profiles of axial
velocity at $8\upmu$s for the plasma fluid and neutral fluid respectively,
while azimuthal velocity profiles are presented at the same output
time in figures \ref{fig: neut_bub_2}(c) and (d). Plasma acceleration
leads to neutral fluid acceleration, due to frictional forces associated
with charge-exchange reactions, and due to momentum exchange arising
from recombination processes. It can be seen how the neutral fluid
attains nearly the same velocity magnitudes as the plasma fluid.

\subsubsection{Effect of inclusion of the $Q_{e}^{rec}$ term\label{subsec:Effect-of-inclusion}}

In the derivations presented in \cite{Meier,MeierPhd}, $Q_{e}^{rec}$
(the volumetric rate of thermal energy transfer from electrons to
photons and neutral particles due to radiative recombination) is not
evaluated because its derivation with the moment-taking method leads
to an integral that cannot be evaluated. It is suggested that this
term can be dropped if the loss of electron thermal energy due to
recombination is not expected to play an important role in the energy
balance \cite{Meier,MeierPhd}. However, from equation \ref{eq:533.0},
it can be seen that $Q_{e}^{rec}=Q_{i}^{rec}(T_{e}/T_{i})$, so that
in cases where $T_{i}\sim T_{e},$ it may seem unreasonable to neglect
$Q_{e}^{rec}$ while retaining $Q_{i}^{rec}$. The $Q_{e}^{rec}$
term is included as an undetermined energy sink/source for the electron/neutral
fluids respectively in \cite{Meier,MeierPhd}, without scaling by
the factor $m_{e}/m_{n}$ in the neutral fluid energy equation (\ref{eq:533.1}),
and is ignored when the equations are implemented to code. As discussed
in section \ref{subsec:Econ_ionrecomb}, from looking at the kinematics
of the radiative recombination reaction, it is more physical to neglect
$Q_{e}^{rec}$ as an energy source for the neutral fluid (most of
the electron thermal energy is transferred to the photon), but include
it as an energy sink for the electron fluid.  It is interesting however
to note the effect of including the term as an energy source for the
neutral fluid without the scaling factor $m_{e}/m_{n}$ - in this
(unphysical) scenario it is assumed, as presented in \cite{Meier},
that all the electron thermal energy lost during radiative recombination
is transferred to the neutral particle.
\begin{figure}[H]
\subfloat[]{\raggedright{}\includegraphics[width=7cm,height=5cm]{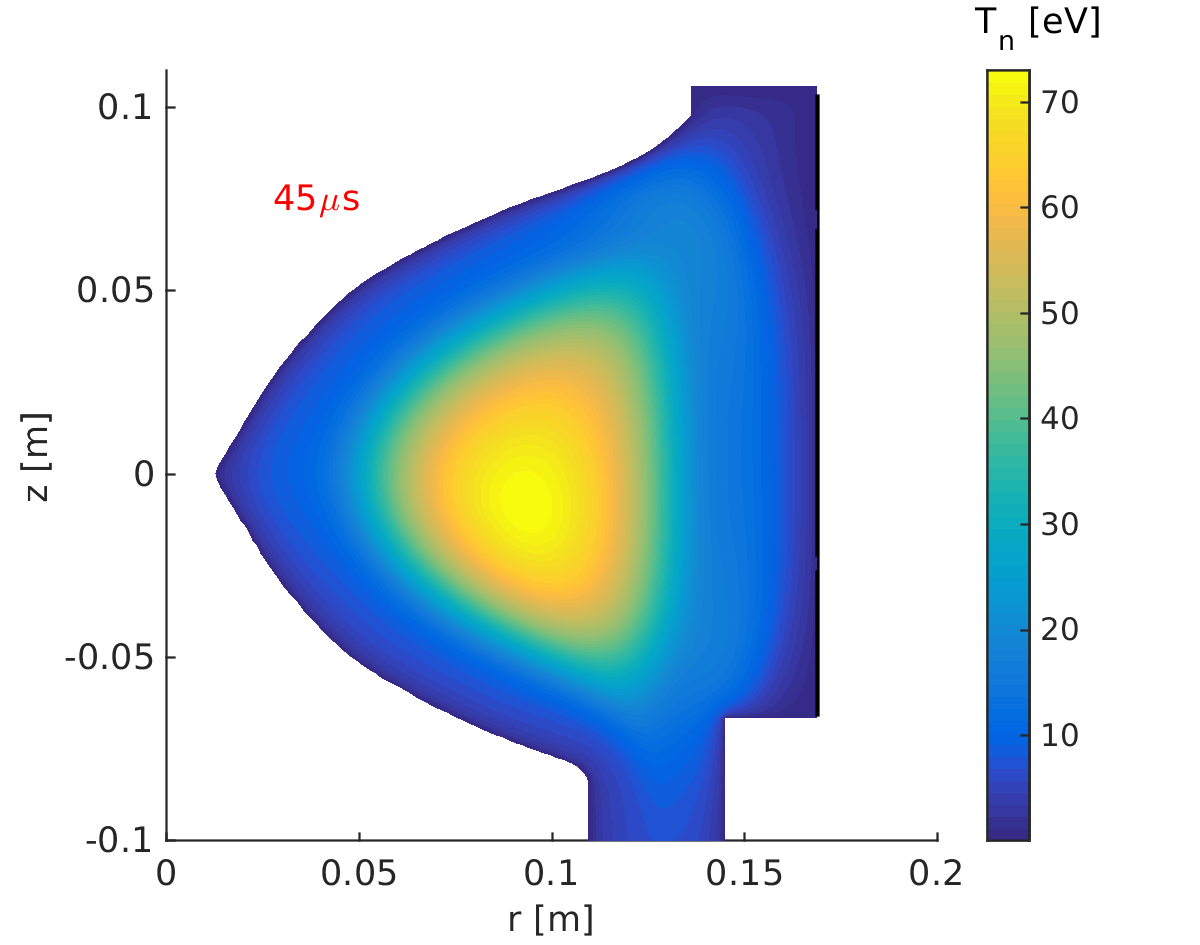}}\hfill{}\subfloat[]{\raggedright{}\includegraphics[width=7cm,height=5cm]{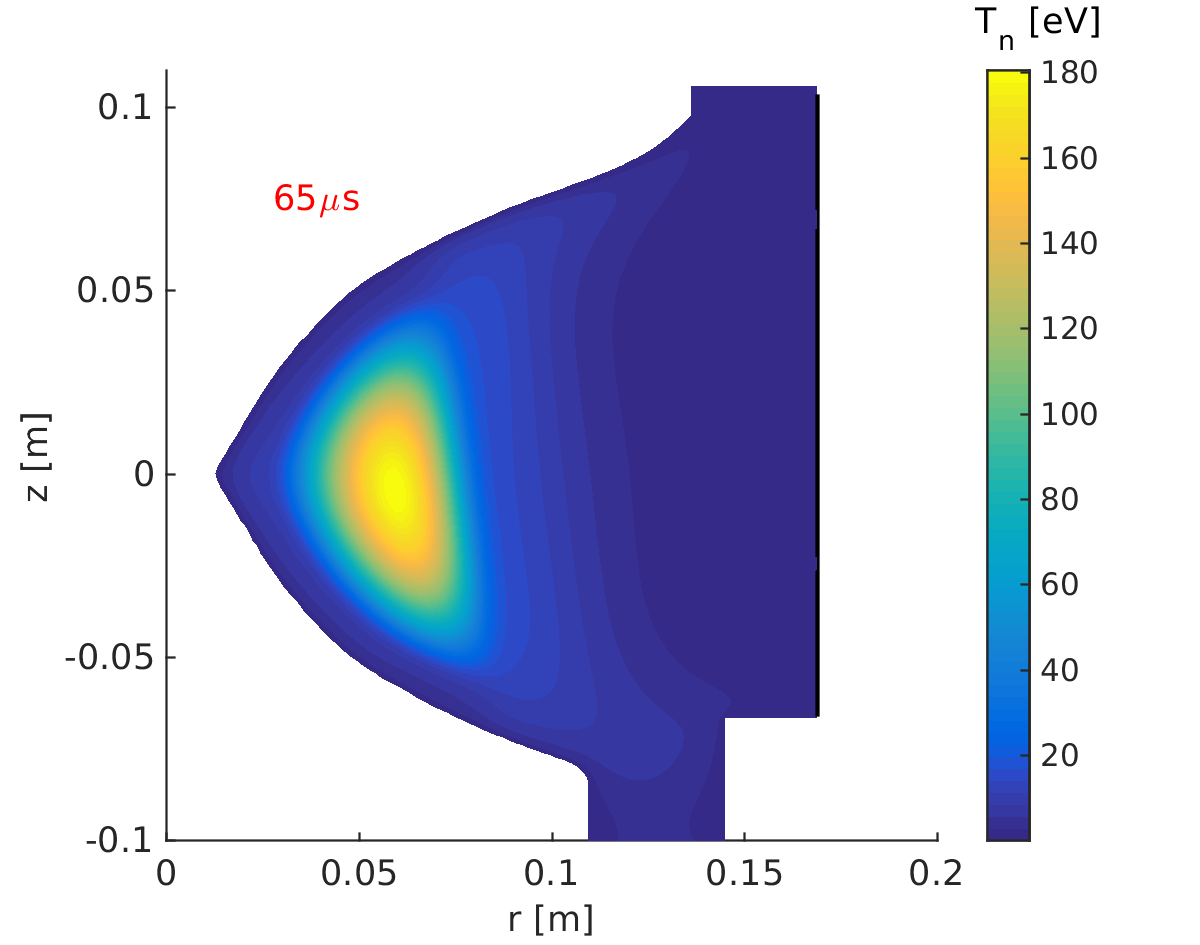}}

\subfloat[]{\raggedright{}\includegraphics[width=7cm,height=5cm]{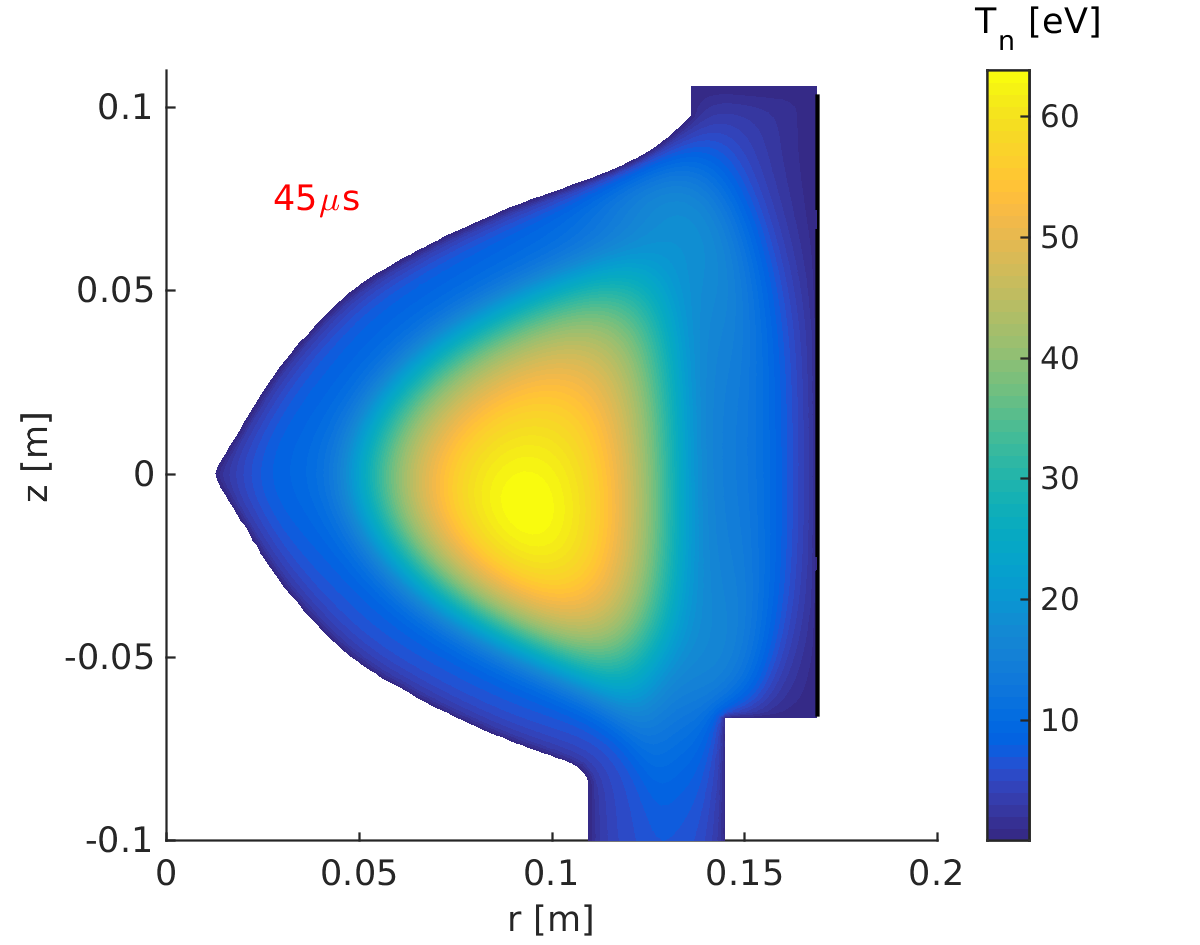}}\hfill{}\subfloat[]{\raggedright{}\includegraphics[width=7cm,height=5cm]{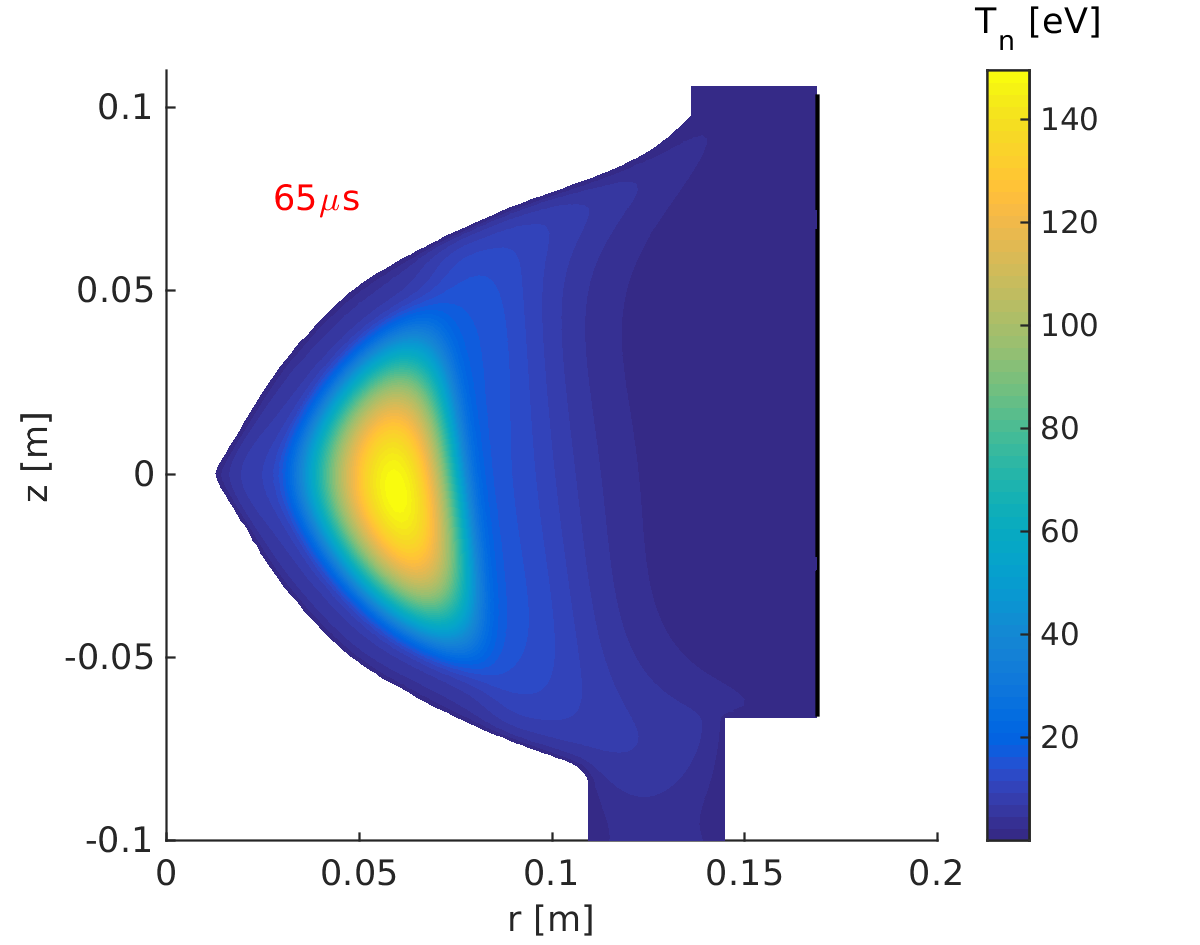}}

\caption{\label{fig: Qerec}$\,\,\,\,$Effect of inclusion of the $Q_{e}^{rec}$
term, which is included in simulations pertaining to figures (a) and
(b) and omitted in simulations pertaining to figures (c) and (d) }
\end{figure}
Figures \ref{fig: Qerec}(a) and (b) show profiles of $T_{n}$ from
a simulation in which the $Q_{e}^{rec}$ term is included in the energy
equations for the electron and neutral fluids, at $45\upmu$s just
prior to magnetic compression, and at peak compression at $65\upmu$s.
Figures \ref{fig: Qerec}(c) and (d) show $T_{n}$ profiles at the
same times from a simulation that is identical except that the $Q_{e}^{rec}$
term is not included in the electron and neutral fluid energy equations.
It can be seen how peak $T_{n}$ increases by around 20\% at $45\upmu$s,
and by around 30\% at $65\upmu$s, when the $Q_{e}^{rec}$ term is
included. Note that if $\chi_{Nmax}$ is increased from $5\times10^{4}${[}m$^{2}$/s{]}
to $1\times10^{5}${[}m$^{2}$/s{]}, that peak $T_{n}$ increases
by around 30\% at $45\upmu$s, and by around 80\% at $65\upmu$s when
the $Q_{e}^{rec}$ term is included. From equations \ref{eq:530.001},
\ref{eq:531.0}, and \ref{eq:533.0}, it can be seen that 
\[
Q_{e}^{rec}\propto Z_{eff}^{3}\,n_{i}^{2}\sqrt{T_{e}}
\]
Hence, $Q_{e}^{rec}$ is high in regions where plasma density and
electron temperature are high. The increase in $T_{n}$ when $Q_{e}^{rec}$
is included (without the physical scaling factor $m_{e}/m_{n}$) is
particularly noticeable in such regions, for example near the CT core
at peak magnetic compression, where the rate of ionization is high
and hence $n_{n}$, and thermal energy associated with neutral particles,
is low. 

Not shown here, peak electron temperature falls by around 1\% when
the $Q_{e}^{rec}$ term is included in the electron fluid energy equation.
$Q_{e}^{rec}$ appears as a (physical) thermal energy sink in the
electron fluid energy equation, but the reduction in $T_{e}$ when
$Q_{e}^{rec}$ is included is negligible, even in regions where $Q_{e}^{rec}$
is high, due to the relatively high levels of electron thermal energy
in such regions. In the regimes studied, it turns out that the $Q_{e}^{rec}$
term can be dropped from the electron energy equation without significantly
affecting electron temperature.

\subsection{Neutral fluid interaction in SPECTOR geometry\label{sec:Neutral_SPECTOR} }

SPECTOR is a magnetized Marshall gun that is similar in principle
to the SMRT injector, except that the insulating wall in SMRT is replaced
with a conducting wall (no CT levitation and compression), and that
up to 0.5 MA current is driven up the central shaft, increasing the
CT toroidal field \cite{spectPoster}. Toroidal field at the CT core
is up to $0.5\mbox{T}$. The low CT aspect ratio, and the $q$ profile,
define the CTs as spherical tokamaks. Initial gun flux is up to 30
mWb.  It is usual to observe a significant rise in electron density
at around $500\upmu$s on the SPECTOR plasma injector, and it was
thought that this may be a result of neutral gas, that remains concentrated
around the gas valve locations after CT formation, diffusing up the
gun. Ionization of the neutral particles would lead to CT fueling
and an increase in observed electron density. The model for interaction
between plasma and neutral fluids was applied to study the issue. 

\begin{figure}[H]
\subfloat[]{\raggedright{}\includegraphics[scale=0.5]{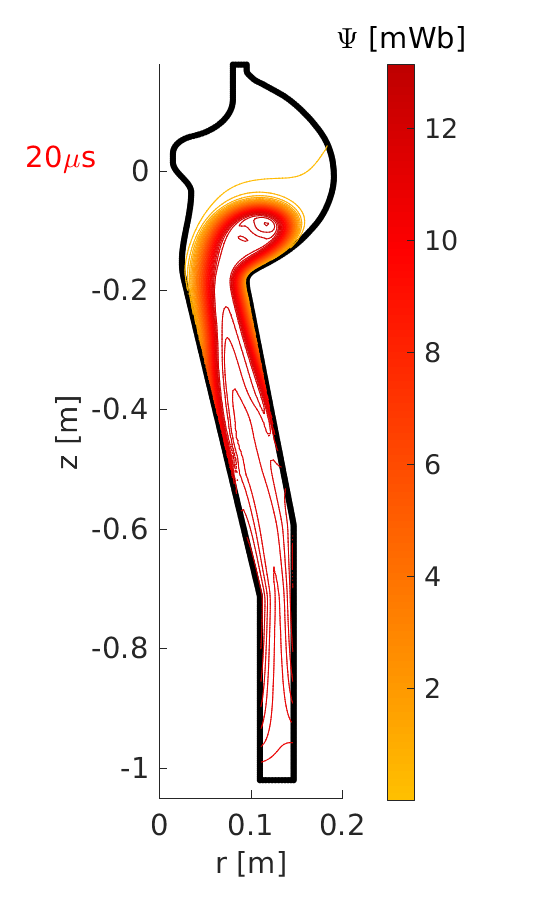}}\hfill{}\subfloat[]{\raggedright{}\includegraphics[scale=0.5]{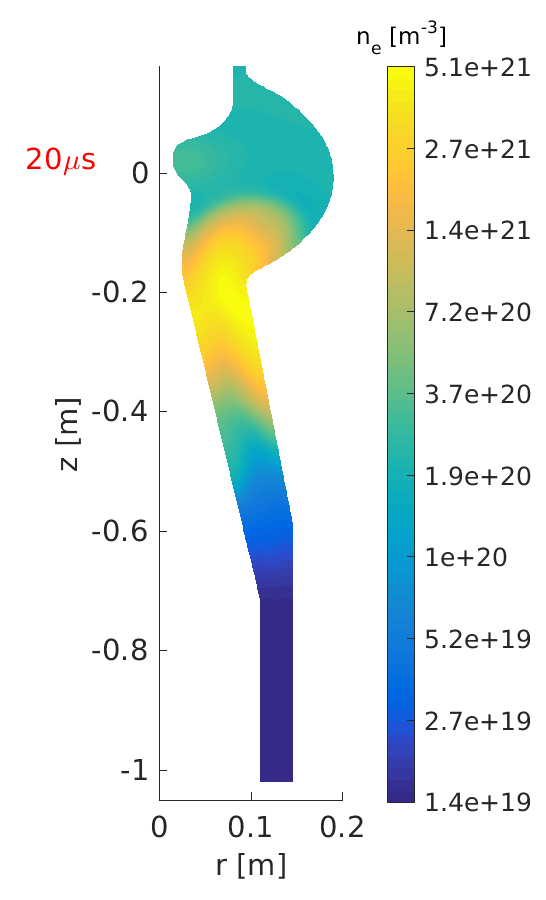}}\hfill{}\subfloat[]{\raggedright{}\includegraphics[scale=0.5]{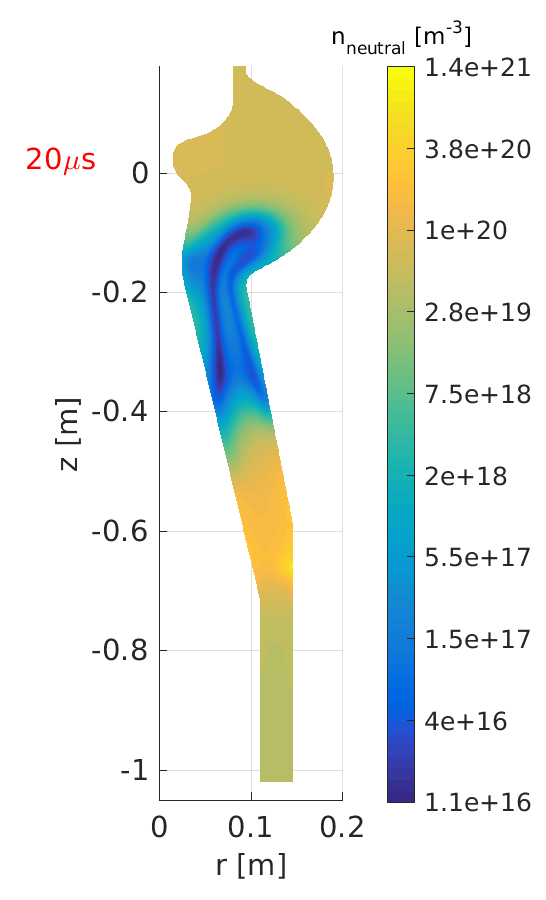}}

\subfloat[]{\raggedright{}\includegraphics[scale=0.5]{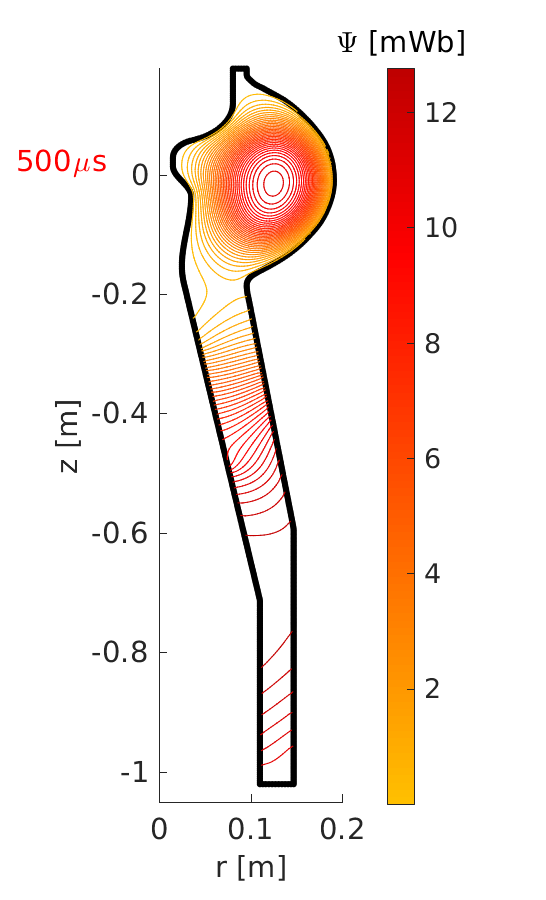}}\hfill{}\subfloat[]{\raggedright{}\includegraphics[scale=0.5]{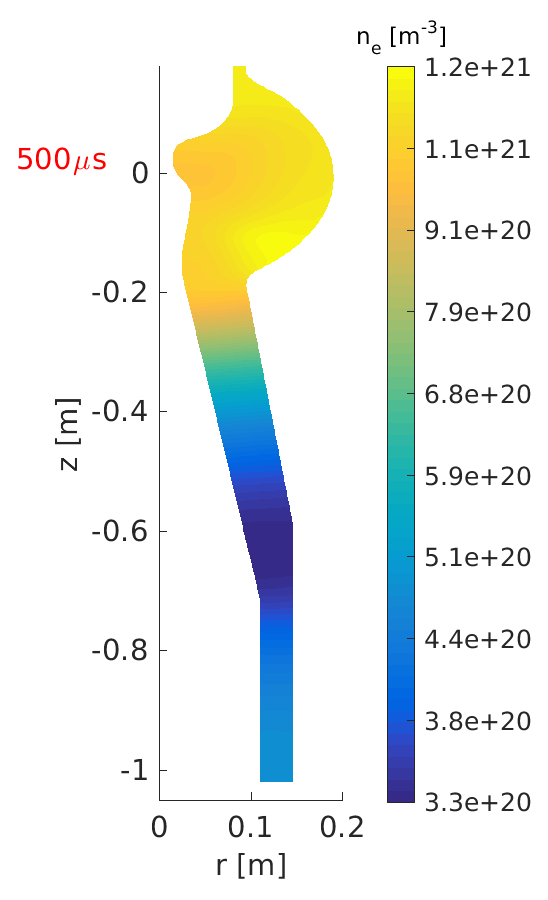}}\hfill{}\subfloat[]{\raggedright{}\includegraphics[scale=0.5]{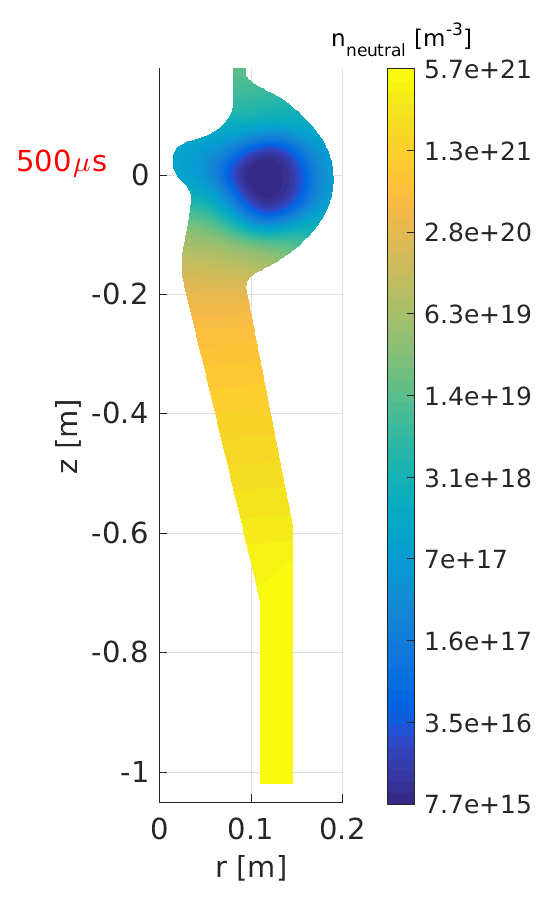}}

\caption{\label{fig:SPEC_psi_n_nN}$\,\,\,\,$Poloidal flux contours (figures
(a), (d)) and profiles of electron ((b), (e)) and neutral fluid ((c),
(f)) densities at various times from a simulation of CT formation
in the SPECTOR plasma injector}
\end{figure}
Figures \ref{fig:SPEC_psi_n_nN}(a), (b) and (c) show $\psi$ contours
and profiles of $n_{e}$ and $n_{n}$ at $20\upmu$s, as plasma enters
the CT containment region. Profiles of the same quantities are shown
in figures \ref{fig:SPEC_psi_n_nN}(d), (e) and (f) at $500\upmu$s,
around the time when the density rise is usually observed. It can
be seen how neutral fluid density is highest at the bottom of the
gun barrel (figure \ref{fig:SPEC_psi_n_nN}(f)) - any neutral gas
advected or diffusing upwards is ionized. A region of particularly
high electron density is apparent just above, and outboard of, the
entrance to the containment region (figure \ref{fig:SPEC_psi_n_nN}(e))
- this is due to the fueling effect arising from neutral gas diffusion.
\\
\begin{figure}[H]
\subfloat[]{\raggedright{}\includegraphics[width=7cm,height=6.3cm]{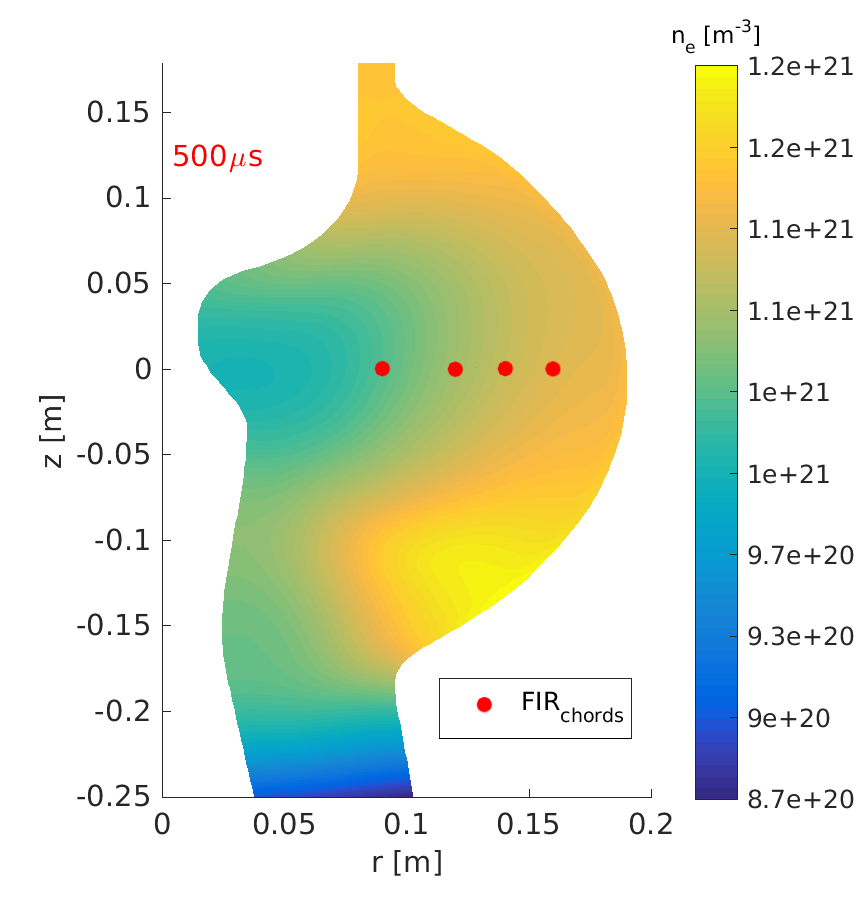}}\hfill{}\subfloat[]{\raggedright{}\includegraphics[width=7cm,height=6.3cm]{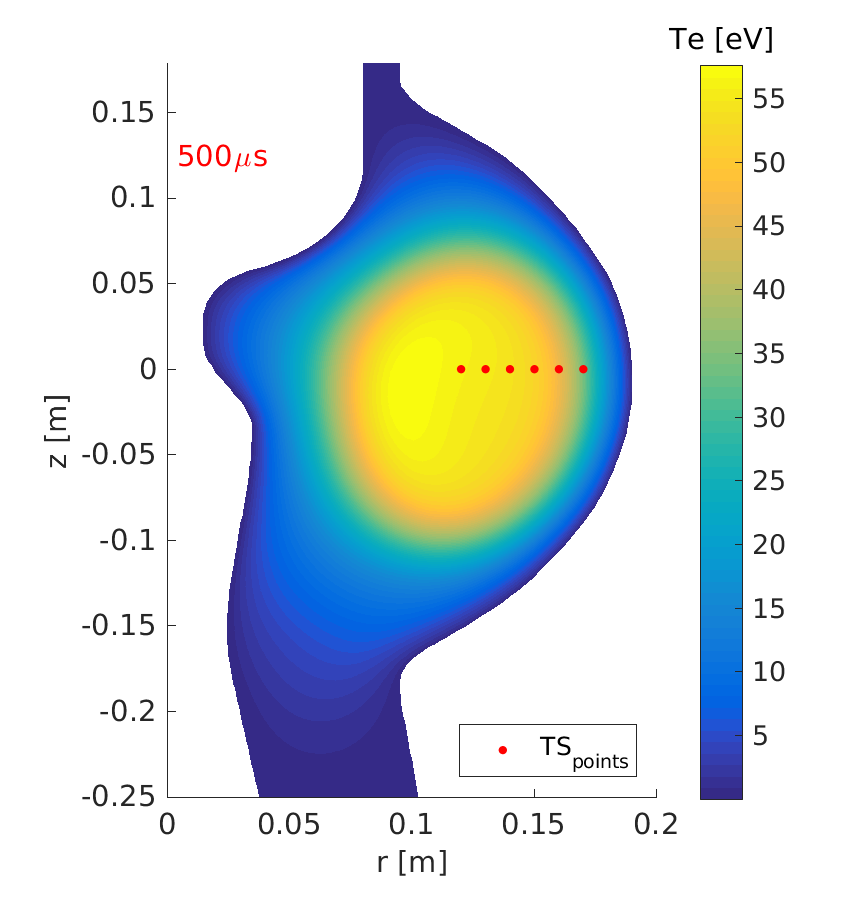}}

\caption{\label{fig: Spect_n_Te}$\,\,\,\,$Profiles of electron density (figure
(a)) and electron temperature (b) at $500\upmu$s from a simulation
of CT formation in the SPECTOR plasma injector}
\end{figure}
The region of particularly high electron density is more defined in
figure \ref{fig: Spect_n_Te}(a), in which cross-sections of the horizontal
chords representing the lines of sight of the FIR (far-infrared) interferometer
are also depicted. The electron temperature profile at 500$\upmu$s
is shown in figure \ref{fig: Spect_n_Te}(b). Referring to figure
\ref{fig:SPEC_psi_n_nN}(f), it can be seen how neutral fluid density
is low in regions of high $T_{e}$ as a result of ionization.

\begin{figure}[H]
\subfloat[]{\raggedright{}\includegraphics[width=7cm,height=5cm]{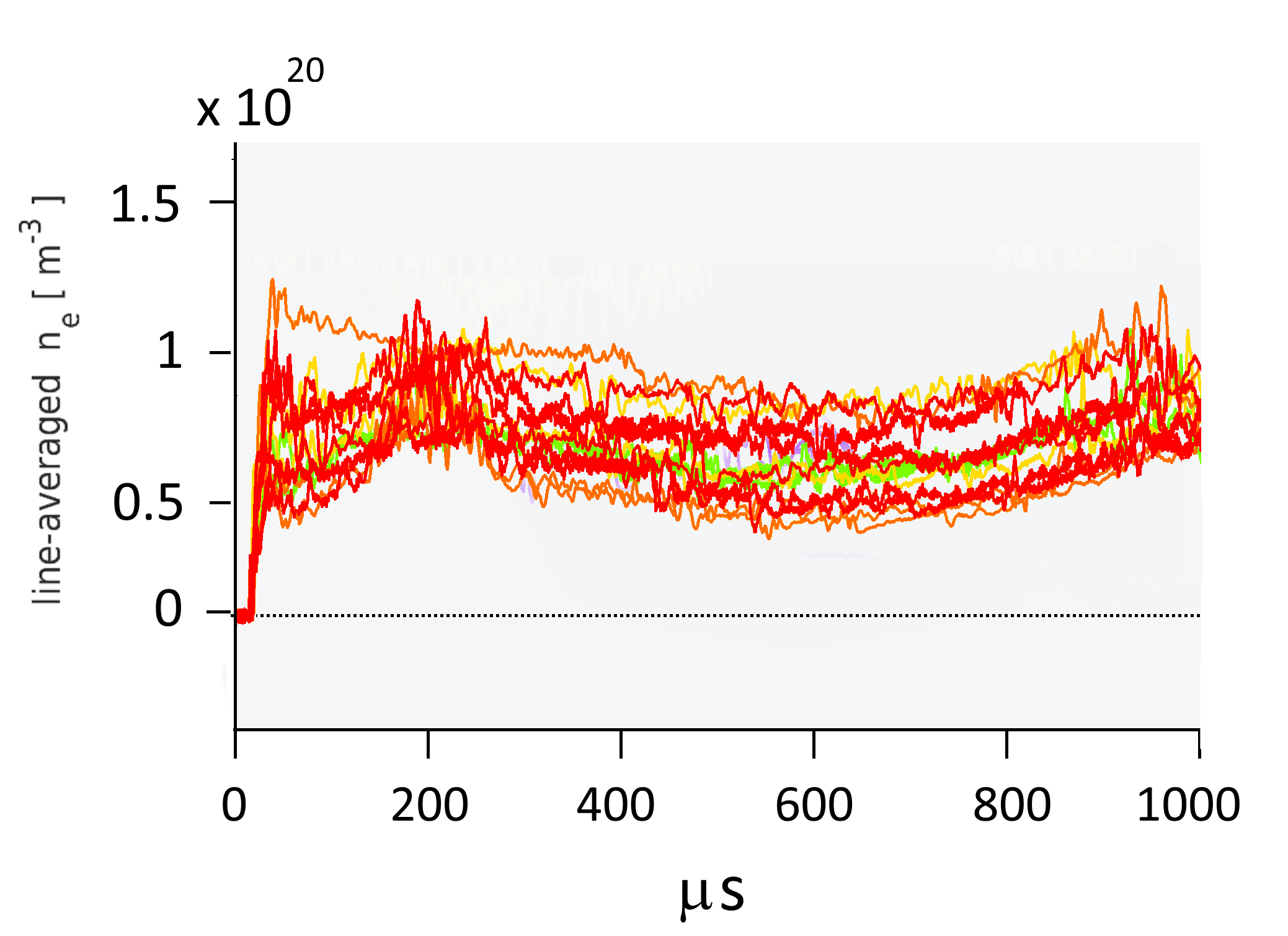}}\hfill{}\subfloat[]{\raggedright{}\includegraphics[width=7cm,height=5cm]{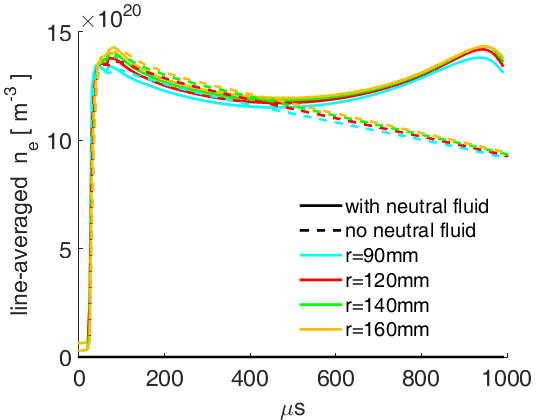}}

\caption{\label{fig:Spector_nN}$\,\,\,\,$Effect of neutral fluid dynamics
in SPECTOR geometry. Measured (a) and simulated (b) electron density
rise at around $500\upmu$s}
\end{figure}
Figure \ref{fig:Spector_nN}(a) shows line-averaged electron density
measured along the interferometer chord at $r=140$ mm from a selection
of several shots in SPECTOR. It can be seen how density starts to
rise at around 500 to $600\upmu$s. Figure \ref{fig:Spector_nN}(b)
shows the simulated diagnostic for line-averaged electron density
along the chords indicated in figure \ref{fig: Spect_n_Te}(a). The
density rise is qualitatively reproduced when neutral fluid is included
in the simulation. Similar simulations without the inclusion of neutral
fluid do not indicate this density rise (dashed lines in figure \ref{fig:Spector_nN}(b)).
The density rise was not observed in the magnetic compression experiment
because CT lifetimes were shorter than the time it takes for a sufficient
amount of neutral gas to diffuse upwards toward the containment region.
The simulations presented in figure \ref{fig:Spector_nN}(b) were
run with artificially high plasma density in order to allow for an
increased simulation timestep and moderately short simulation run-times
(note that timestep scales inversely with Alfven speed, and hence
scales with $\sqrt{n_{i}}$ as a consequence of the explicit time-advance
scheme that is implemented in the code). Note that the electron temperatures
indicated in figure \ref{fig: Spect_n_Te}(b) are underestimations
of the actual temperatures due to the overestimation of density in
the simulation - the main goal of these simulations was to demonstrate
that the inclusion of neutral fluid interaction can qualitatively
model the observed electron density increase.

The CT fueling and cooling effect of neutral gas diffusing up the
gun is thought to be related to the unusually significant increases
in CT lifetime and electron temperature observed when a biased electrode
was inserted 11mm into the CT edge on the SPECTOR plasma injector
\cite{Spector_biasing}. Electrode biasing involves the insertion
of an electrode, that is biased relative to the vessel wall near the
point of insertion, into the edge of a magnetized plasma. This leads
to a radially directed electric field between the probe and the wall.
The resultant $\mathbf{J}_{r}\times\mathbf{B}$ force imposed on the
plasma at the edge of the CT confinement region varies with distance
between the probe and the wall, because $E_{r}$, as well as the magnetic
field, vary in that region. The associated torque overcomes viscous
forces, spinning up the edge plasma, and results in shearing of the
particle velocities between the probe and the wall. The sheared velocity
profile is thought to suppress the growth of turbulent eddies that
advect hot plasma particles to the wall, thereby reducing this plasma
cooling mechanism. In general, high confinement modes induced by probe
biasing share features of those initiated by various methods of heating,
including a density pedestal near the wall (near the probe radius
for probe biasing), diminished levels of recycling as evidenced by
reduced $\mbox{H}_{\alpha}$ emission intensity, and increased particle
and energy confinement times. 

In contrast to most of the biasing experiments conducted on tokamak
plasmas, where electron density is found to increase as a consequence
of biasing, electron density was markedly reduced in the SPECTOR edge
biasing experiment. This density reduction is thought to be due to
the effect of the transport barrier impeding the CT fueling associated
with neutral gas diffusing up the gun. CT lifetimes and temperatures
were found to increase by factors of around 2.4 with edge biasing.
The scale of the improvements observed is significantly greater than
that associated with prior biasing experiments (mainly in tokamaks,
in which continuous uncontrolled plasma fueling by cold particles
did not occur), and it is thought that this result is associated with
a reduction of the cooling effect associated with CT fueling. 

\section{Conclusions\label{sec:conclusions}}

It has been shown how the terms that determine the source rates of
species momentum and energy due to ionization and recombination can
be derived from basic principles rather than the more formal and involved
process of taking successive moments of the collision operators pertaining
to the reactions. The latter method must be used to determine terms
relating to charge exchange reactions. Only the former method enables
determination of $Q_{e}^{rec}$, which prescribes the volumetric rate
of thermal energy transfer from electrons to photons and neutral particles
due to radiative recombination. This approach also allows for determination
of the terms that must be included in the MHD equations when external
particle sources are present, for example neutral gas that is continuously
injected through the machine gas valves in the SMRT injector. From
examination of the kinematics of the radiative recombination reaction,
it appears reasonable to neglect $Q_{e}^{rec}$ as an energy source
for the neutral fluid (most of the electron thermal energy is transferred
to the photon), but include it as an energy sink for the electron
fluid. It turns out that the inclusion of this term as an energy sink
in the electron energy equation does not lead to any significant reduction
in electron fluid temperature in the regime examined. The three and
two component conservation equations have been developed for plasma
and neutral fluids. The two fluid equations have been implemented
to the DELiTE code framework, and applied to the study of CT formation
in the SMRT and SPECTOR plasma injectors.

The energy conservation property of the DELiTE code framework is maintained
when the neutral fluid model is included, as long as the electron
thermal energy expended on ionization and recombination processes
is accounted for. With appropriate boundary conditions, net neutral
fluid angular momentum is conserved in simulations, along with net
plasma angular momentum, and conservation of net particle count is
maintained. The initial motivation for the study of plasma/neutral
interaction was that inclusion of neutral fluid interaction in the
model was expected to lead to net ion cooling; this concept has been
disproved. Due to bulk inertial effects, increased net initial particle
inventory results in reduced ion viscous heating, regardless of whether
part of the initial inventory includes neutral particles. Charge exchange
reactions lead to ion-neutral heat exchange, but the ion temperature
is relatively unchanged by charge exchange reactions in hot-ion regions
where electron temperature and ionization rates are also high and
neutral fluid density is consequently low. 

The model for plasma/neutral fluid interaction was used to clarify
the cause for the increase in electron density that is routinely observed
well after CT formation in the SPECTOR plasma injector. Residual neutral
gas remaining near the gas valves after formation diffuses up the
gun to the CT containment area where it is ionized and becomes an
uncontrolled continuous source of CT fueling and cooling. This insight
helps account for the exceptionally significant increase in electron
temperature, and noticeably  reduced electron density, observed during
the electrode edge biasing experiment conducted on SPECTOR \cite{Spector_biasing}.
The model implementation is a good testbed for further studies and
improvements.

\section{Acknowledgments}

Funding was provided in part by General Fusion Inc., Mitacs, University
of Saskatchewan, and NSERC. We would like to thank to Ivan Khalzov,
Meritt Reynolds, Eric Meier, and Richard Marchand for useful discussions.
We acknowledge the University of Saskatchewan ICT Research Computing
Facility for computing time. 

\appendix

\section{Appendix: Development of scattering collision terms\label{sec:Appendix:AScattering-collision}}

The Boltzmann equation for species $\alpha$ is 
\[
\frac{\partial f_{\alpha}}{\partial t}+\nabla\cdot(\mathbf{V}\,f_{\alpha})+\nabla_{v}\cdot\left(\left(\frac{q_{\alpha}}{m_{\alpha}}(\mathbf{E}(\mathbf{r},t)\mathbf{+V}\mathbf{\times B}(\mathbf{r},t))\right)\,f_{\alpha}\right)=\frac{\partial f_{\alpha}}{\partial t}|_{collisions}=C_{\alpha}(f)
\]
where $\alpha=i,\,e,$ or $n$ denote ions, electrons, or neutral
particles. If $\alpha=n$, then the acceleration term vanishes, because
$q_{n}=0$. The collision operator can be split into parts pertaining
to elastic scattering collisions and reacting collisions: $C_{\alpha}(f)=C_{\alpha}^{scatt.}(f)+C_{\alpha}^{react.}(f)$.
Boltzmann's collision operator for neutral gas is 

\begin{equation}
C_{\alpha}^{scatt.}(f)=\underset{\sigma}{\Sigma}\,C_{\alpha\sigma}^{scatt.}(f_{\alpha},\,f_{\sigma})\label{eq:465.1}
\end{equation}
Here, $C_{\alpha\sigma}^{scatt.}(f_{\alpha},\,f_{\sigma})$, the rate
of change of $f_{\alpha}$ due to collisions of species $\alpha$
with species $\sigma$, considers only binary collisions and is therefore
bilinear because $C_{\alpha\sigma}^{scatt.}$ is a linear function
of both its arguments \cite{HazeltineWaelbroeck,bellan fundamentals}.
In plasmas, where long-range Coulomb interactions lead to Debye shielding,
a many-body effect, collisions are not strictly binary. However, in
a weakly coupled plasma, the departure from bilinearity is logarithmic,
and can be neglected to a good approximation since the logarithm is
a relatively weakly varying function \cite{farside_plasma}. The collisional
process for the elastic collisions described by $C_{\alpha}^{scatt.}$
conserves particles, momentum and energy at each point \cite{HazeltineWaelbroeck}.
Particle conservation is expressed by
\begin{equation}
\int C_{\alpha\sigma}^{scatt.}d\mathbf{V}=\int C_{\alpha}^{scatt.}d\mathbf{V}=0\label{eq:466}
\end{equation}
$i.e.,$ scattering collisions have no $0^{th}$ moment effect. To
find the contribution of scattering collisions to the rates of change
of momentum of the ions, electrons and neutral particles, the first
moments of the operators for scattering collisions can be taken. The
total friction force (collisional momentum exchange) acting on species
$\alpha$ due to the net effect of the frictional interaction with
each of species $\sigma$ is defined as
\begin{equation}
\mathbf{R}_{\alpha}=\underset{\sigma}{\Sigma}\mathbf{R}_{\alpha\sigma}=\int m_{\alpha}\underset{\sigma}{\Sigma}\,C_{\alpha\sigma}^{scatt.}\mathbf{V}d\mathbf{V}=\int m_{\alpha}C_{\alpha}^{scatt.}\mathbf{V}d\mathbf{V}\label{eq:467-1}
\end{equation}
Here, $\mathbf{R}_{\alpha\sigma}$ is the frictional force exerted
by species $\sigma$ on species $\alpha$. We define 
\begin{equation}
\mathbf{c}_{\alpha}(\mathbf{r},t)=\mathbf{V-\mathbf{v}_{\alpha}}(\mathbf{r},t)\label{eq:354}
\end{equation}
as the species random velocity relative to $\mathbf{v}_{\alpha}$,
the species fluid velocity \cite{bittencourt}. Noting that the statistical
average of a random quantity is zero, and the average of an averaged
value is unchanged, and using equations \ref{eq:354}, \ref{eq:466},
and \ref{eq:467-1}, we can write: $\mathbf{R}_{\alpha}=\int m_{\alpha}C_{\alpha}^{scatt.}\mathbf{c}_{\alpha}d\mathbf{V}$.
In a system with just two species, ions and electrons, $\mathbf{R}_{\alpha}$
is given as \cite{bellan fundamentals,farside_plasma}
\begin{equation}
\mathbf{R}_{e}=-\mathbf{R}_{i}=\eta'ne\mathbf{J}-0.71n\nabla T_{e}\label{eq:355-3}
\end{equation}
Here, $\eta'$ {[}$\Omega$-m{]} is the plasma resistivity, $n$ {[}m$^{-3}${]}
is the number density, $e$ {[}C{]} is the electron charge, $\mathbf{J}$
{[}A/m$^{2}${]} is the current density, and $T_{e}$ {[}J{]} is the
electron temperature.  Ignoring the thermal force for simplicity,
this is: $\mathbf{R}_{e}=-\mathbf{R}_{i}=\eta'ne\mathbf{J}=\nu_{ei}\rho_{e}(\mathbf{v}_{i}-\mathbf{v}_{e})$,
where $\nu_{ei}$ {[}s$^{-1}${]} is the electron-ion collision frequency
and $\rho_{e}$ {[}kg/m$^{3}${]} is the electron fluid density. When
there are just two species, then, since $\mathbf{R}_{\alpha}=\underset{\sigma}{\Sigma}\mathbf{R}_{\alpha\sigma}$(equation
\ref{eq:467-1}) and $\mathbf{R}_{\sigma\sigma}=0$ (a fluid does
not exert friction on itself), the equivalent notation $\mathbf{R}_{e}\equiv\mathbf{R}_{ei}$
and $\mathbf{R}_{i}\equiv\mathbf{R}_{ie}$ can be introduced for the
frictional forces due to scattering collisions. 

The notation $\left(\frac{\partial X}{\partial t}\right)_{scatt.}$
is introduced to represent, the rate of change of any quantity $X$
that pertains to scattering collisions. The first moment of the first
term of the Boltzmann equation is $\frac{\partial\left(\mathbf{v}_{\alpha}m_{\alpha}n_{\alpha}\right)}{\partial t}$.
Since
\[
\left(\frac{\partial\left(\mathbf{v}_{\alpha}m_{\alpha}n_{\alpha}\right)}{\partial t}\right)_{scatt.}=\rho_{\alpha}\left(\frac{\partial\mathbf{v}_{\alpha}}{\partial t}\right)_{scatt.}+\mathbf{v}_{\alpha}\left(\frac{\partial\rho_{\alpha}}{\partial t}\right)_{scatt.}
\]
and scattering collisions are not a source of particles $\left(\Rightarrow\left(\frac{\partial\rho_{\alpha}}{\partial t}\right)_{scatt.}=0\right)$,
the terms $\rho_{\alpha}\left(\frac{\partial\mathbf{v}_{\alpha}}{\partial t}\right)_{scatt.}$
can be expressed as:
\begin{equation}
\rho_{\alpha}\left(\frac{\partial\mathbf{v}_{\alpha}}{\partial t}\right)_{scatt.}=\mathbf{R}_{\alpha}=\underset{\sigma}{\Sigma}\mathbf{R}_{\alpha\sigma}\label{eq:520.1}
\end{equation}
Note that $\mathbf{R}_{\alpha\sigma}=-\mathbf{R}_{\sigma\alpha}$;
the frictional force exerted by species $\alpha$ on species $\sigma$
is balanced by the frictional force exerted by species $\sigma$ on
species $\alpha$. Just as $\mathbf{R}_{ei}=\nu_{ei}\rho_{e}(\mathbf{v}_{i}-\mathbf{v}_{e})$
and $\mathbf{R}_{ie}=\nu_{ie}\rho_{i}(\mathbf{v}_{e}-\mathbf{v}_{i})$,
the forms for the charged-neutral friction forces are $\mathbf{R}_{in}=\nu_{in}\rho_{i}(\mathbf{v}_{n}-\mathbf{v}_{i})$
and $\mathbf{R}_{en}=\nu_{en}\rho_{e}(\mathbf{v}_{n}-\mathbf{v}_{e})$,
where $\nu_{\alpha\sigma}\sim\frac{m_{\alpha}}{m_{\sigma}}\nu_{\sigma\alpha}$
is the frequency for scattering of particles of species $\alpha$
from particles of species $\sigma$ \cite{bellan fundamentals}. 

To find the contribution of scattering collisions to the rates of
change of energy of the ions, electrons and neutral particles, the
second moments of the operators for scattering collisions can be taken.
Collisional energy conservation requires that $Q_{L\alpha\sigma}+Q_{L\sigma\alpha}=0$,
where 
\begin{equation}
Q_{L\alpha\sigma}=\int C_{\alpha\sigma}^{scatt.}(\frac{1}{2}m_{\alpha}V^{2})\,d\mathbf{V}\label{eq:521.2}
\end{equation}
 is the rate at which species $\sigma$ collisionally transfers energy
to species $\alpha$. The $L$ subscript is to indicate that the kinetic
energy of both species is measured in the same ($eg.$ Laboratory)
frame. $Q_{\alpha\sigma}=\int C_{\alpha\sigma}^{scatt.}(\frac{1}{2}m_{\alpha}c_{\alpha}^{2})\,d\mathbf{V}$
is the rate at which species $\sigma$ collisionally transfers energy
to species $\alpha$ in the rest frame of species $\alpha$ (the frame
moving with velocity $\mathbf{v}_{\alpha}$) \cite{HazeltineWaelbroeck,farside_plasma}.
The total rate of collisional energy transfer to species $\alpha$
in the rest frame of species $\alpha$ is 
\begin{equation}
Q_{\alpha}=\underset{\sigma}{\Sigma}Q_{\alpha\sigma}\label{eq:467-2}
\end{equation}
Using equation \ref{eq:354} in equation \ref{eq:521.2}, it can be
shown that 
\[
Q_{L\alpha\sigma}=\int C_{\alpha\sigma}^{scatt.}(\frac{1}{2}m_{\alpha}(\mathbf{c}_{\alpha}+\mathbf{v}_{\alpha})^{2})\,d\mathbf{V}=Q_{\alpha\sigma}+\mathbf{v}_{\alpha}\cdot\mathbf{R}_{\alpha\sigma}
\]
so that: 
\begin{align}
\underset{\sigma}{\Sigma}Q_{L\alpha\sigma}=\int C_{\alpha}^{scatt.}(\frac{1}{2}m_{\alpha}V^{2})\,d\mathbf{V} & =Q_{\alpha}+\mathbf{v}_{\alpha}\cdot\mathbf{R}_{\alpha} & \mbox{ (use eqns. \ref{eq:465.1}, \ref{eq:467-2} \& \ref{eq:467-1})}\label{eq:468-1}
\end{align}
Since $\mathbf{R}_{\alpha}=\underset{\sigma}{\Sigma}\mathbf{R}_{\alpha\sigma}$
(equation \ref{eq:467-1}), and $Q_{\alpha}=\underset{\sigma}{\Sigma}Q_{\alpha\sigma}$(equation
\ref{eq:467-2}), equation \ref{eq:468-1} becomes
\begin{align}
\int C_{\alpha}^{scatt.}(\frac{1}{2}m_{\alpha}V^{2})\,d\mathbf{V} & =Q_{\alpha}+\mathbf{v}_{\alpha}\cdot\mathbf{R}_{\alpha}=\underset{\sigma}{\Sigma}Q_{\alpha\sigma}+\mathbf{v}_{\alpha}\cdot\underset{\sigma}{\Sigma}\mathbf{R}_{\alpha\sigma}\label{eq:521.1}
\end{align}
The second moment of the first term of the Boltzmann equation is
\[
\frac{\partial}{\partial t}\left(\frac{1}{2}\rho_{\alpha}v_{\alpha}^{2}+\frac{p_{\alpha}}{\gamma-1}\right)=\frac{v^{2}}{2}\frac{\partial\rho_{\alpha}}{\partial t}+\rho_{\alpha}\mathbf{v}_{\alpha}\cdot\frac{\partial\mathbf{v}_{\alpha}}{\partial t}+\frac{1}{\gamma-1}\frac{\partial p_{\alpha}}{\partial t}
\]
where $\gamma=5/3$ is the adiabatic gas constant. Since $\left(\frac{\partial\rho_{\alpha}}{\partial t}\right)_{scatt.}=0$
($i.e.,$ scattering collisions are not a source of particles), this
leads to\\
\[
\frac{1}{\gamma-1}\left(\frac{\partial p_{\alpha}}{\partial t}\right)_{scatt.}=\left(\frac{\partial}{\partial t}\left(\frac{1}{2}\rho_{\alpha}v_{\alpha}^{2}+\frac{p_{\alpha}}{\gamma-1}\right)\right)_{scatt.}-\rho_{\alpha}\mathbf{v}_{\alpha}\cdot\left(\frac{\partial\mathbf{v}_{\alpha}}{\partial t}\right)_{scatt.}
\]
Using equations \ref{eq:521.1} and \ref{eq:520.1}, this implies
that: 
\begin{equation}
\frac{1}{\gamma-1}\left(\frac{\partial p_{\alpha}}{\partial t}\right)_{scatt.}=\underset{\sigma}{\Sigma}Q_{\alpha\sigma}+\cancel{\mathbf{v}_{\alpha}\cdot\underset{\sigma}{\Sigma}\mathbf{R}_{\alpha\sigma}}-\cancel{\mathbf{v}_{\alpha}\cdot\underset{\sigma}{\Sigma}\mathbf{R}_{\alpha\sigma}}\label{eq:521.11}
\end{equation}

\section{Appendix: Development of expressions for $\left(\frac{\partial p_{\alpha}}{\partial t}\right)_{ire}$\label{sec:Appendix:B} }

Using equations \ref{eq:531.40} and \ref{eq:531.51}, equation \ref{eq:531.39}
can be re-expressed as:

{\footnotesize{}
\begin{align*}
\frac{1}{\gamma-1}\left(\frac{\partial p_{\alpha}}{\partial t}\right)_{ire}=\frac{1}{\gamma-1}\underset{k}{\Sigma}\left(\xi_{\alpha k}S_{\alpha k}\underset{j}{\Sigma}T_{0jk}\right)+\frac{1}{2}\underset{k}{\Sigma}\left(S_{\alpha k}\underset{j}{\Sigma}\left(m_{jk}v_{0jk}^{2}\right)\right)-\frac{1}{2}m_{\alpha}v_{\alpha}^{2}\underset{k}{\Sigma}\left(S_{\alpha k}\right)-m_{\alpha}n_{\alpha}\mathbf{v}_{\alpha}\cdot\left(\frac{\partial\mathbf{v}_{\alpha}}{\partial t}\right)_{ire}
\end{align*}
}{\small{}
\begin{align}
\Rightarrow\left(\frac{\partial p_{\alpha}}{\partial t}\right)_{ire} & =\underset{k}{\Sigma}\left(\xi_{\alpha k}S_{\alpha k}\underset{j}{\Sigma}T_{0jk}\right)+(\gamma-1)\biggl[\frac{1}{2}\underset{k}{\Sigma}\left(S_{\alpha k}\underset{j}{\Sigma}\left(m_{jk}v_{0jk}^{2}\right)\right)\nonumber \\
 & \,\,\,\,\,\,-\frac{1}{2}m_{\alpha}v_{\alpha}^{2}\underset{k}{\Sigma}\left(S_{\alpha k}\right)-\mathbf{v}_{\alpha}\cdot\left(\underset{k}{\Sigma}\left(S_{\alpha k}\underset{j}{\Sigma}\left(m_{jk}\mathbf{v}_{0jk}\right)\right)-m_{\alpha}\mathbf{v}_{\alpha}\left(\underset{k}{\Sigma}S_{\alpha k}\right)\right)\biggr]\nonumber \\
 & =\underset{k}{\Sigma}\left(\xi_{\alpha k}S_{\alpha k}\underset{j}{\Sigma}T_{0jk}\right)+(\gamma-1)\left(\frac{1}{2}m_{\alpha}v_{\alpha}^{2}\underset{k}{\Sigma}\left(S_{\alpha k}\right)+\underset{k}{\Sigma}\left(S_{\alpha k}\underset{j}{\Sigma}\left(m_{jk}\left(\frac{1}{2}v_{0jk}^{2}-\mathbf{v}_{\alpha}\cdot\mathbf{v}_{0jk}\right)\right)\right)\right)\nonumber \\
\Rightarrow\left(\frac{\partial p_{\alpha}}{\partial t}\right)_{ire} & =\underset{k}{\Sigma}\left(\xi_{\alpha k}S_{\alpha k}\underset{j}{\Sigma}T_{0jk}\right)+(\gamma-1)\left(\underset{k}{\Sigma}\left(S_{\alpha k}\left(\frac{1}{2}m_{\alpha}v_{\alpha}^{2}+\underset{j}{\Sigma}\left(m_{jk}\left(\frac{1}{2}v_{0jk}^{2}-\mathbf{v}_{\alpha}\cdot\mathbf{v}_{0jk}\right)\right)\right)\right)\right)\nonumber \\
\nonumber \\
\text{{\color{red}}}\label{eq:532}
\end{align}
}For the ions and electrons (all sources), and for the neutral source
terms corresponding to ionization and external sources, where $\underset{k}{\Sigma}\left(S_{\alpha k}\underset{j}{\Sigma}\left(m_{jk}\mathbf{v}_{0jk}\right)\right)\rightarrow m_{\alpha}\underset{k}{\Sigma}\left(S_{\alpha k}\mathbf{v}_{0k}\right)$,
$\underset{k}{\Sigma}\left(S_{\alpha k}\underset{j}{\Sigma}\left(m_{jk}v_{0jk}^{2}\right)\right)\rightarrow m_{\alpha}\underset{k}{\Sigma}\left(S_{\alpha k}v_{0k}^{2}\right)$,
and $\underset{j}{\Sigma}T_{0jk}\rightarrow T_{0k}$, this general
expression can be simplified to
\begin{align*}
\left(\frac{\partial p_{\alpha}}{\partial t}\right)_{ire} & =\underset{k}{\Sigma}\left(\xi_{\alpha k}S_{\alpha k}T_{0k}\right)+(\gamma-1)\left(\frac{1}{2}m_{\alpha}\underset{k}{\Sigma}\left(S_{\alpha k}\left(v_{\alpha}^{2}-2\mathbf{v}_{\alpha}\cdot\mathbf{v}_{0k}+v_{0k}^{2}\right)\right)\right)\\
\Rightarrow\left(\frac{\partial p_{\alpha}}{\partial t}\right)_{ire} & =\underset{k}{\Sigma}\left(\xi_{\alpha k}S_{\alpha k}T_{0k}\right)+(\gamma-1)\left(\frac{1}{2}m_{\alpha}\underset{k}{\Sigma}\left(S_{\alpha k}\left(\mathbf{v}_{\alpha}-\mathbf{v}_{0k}\right)^{2}\right)\right)
\end{align*}
However, the more complicated form of equation \ref{eq:532} must
be retained for the recombination neutral source term.

\section{Appendix: 3-fluid MHD equations\label{sec:Appendix: 3-fluid-MHD-equations}}

Here we define $\left(\frac{\partial X}{\partial t}\right)_{CE},\,\left(\frac{\partial X}{\partial t}\right)_{scatt.},\,\left(\frac{\partial X}{\partial t}\right)_{react.},\,\left(\frac{\partial X}{\partial t}\right)_{ext.},\,\left(\frac{\partial X}{\partial t}\right)_{ire},\,\mbox{ and }\left(\frac{\partial X}{\partial t}\right)_{cx}$
as, respectively, the time-rates of change of any quantity $X$ that
arise due to the combination of \emph{collisions} and \emph{external}
sources, to \emph{scattering} collisions, to \emph{reacting} collisions\emph{,
}to\emph{ external} sources, to the combination of \emph{ionization},
\emph{recombination} and \emph{external} sources, and to \emph{charge}
\emph{exchange} collisions. These quantities are related as 
\[
\left(\frac{\partial X}{\partial t}\right)_{CE}=\left(\frac{\partial X}{\partial t}\right)_{scatt.}+\left(\frac{\partial X}{\partial t}\right)_{react.}+\left(\frac{\partial X}{\partial t}\right)_{ext.}=\left(\frac{\partial X}{\partial t}\right)_{scatt.}+\left(\frac{\partial X}{\partial t}\right)_{ire}+\left(\frac{\partial X}{\partial t}\right)_{cx}
\]
 For $X=n_{\alpha}$, scattering and charge exchange collisions can
be neglected so that, referring to equation \ref{eq:531.2}: 
\begin{align}
\left(\frac{\partial n_{i}}{\partial t}\right)_{CE} & =\Gamma_{i}^{ion}-\Gamma_{n}^{rec}\nonumber \\
\left(\frac{\partial n_{e}}{\partial t}\right)_{CE} & =\Gamma_{i}^{ion}-\Gamma_{n}^{rec}\label{eq:540.0}\\
\left(\frac{\partial n_{n}}{\partial t}\right)_{CE} & =\Gamma_{n}^{rec}-\Gamma_{i}^{ion}+\Gamma_{n}^{ext}\nonumber 
\end{align}
Combining equations \ref{eq:521-1}, \ref{eq:531.6}, and \ref{eq:537},
and using the identity $\mathbf{v}_{in}=\mathbf{v}_{i}-\mathbf{v}_{n}$,
the complete set of terms that arise in the species momentum equations
due to scattering and reactive collisions, and an external neutral
particle source, can be assembled:

\begin{align}
\left(\frac{\partial\mathbf{v}_{i}}{\partial t}\right)_{CE} & =\frac{1}{\rho_{i}}\left(\mathbf{R}_{ie}+\mathbf{R}_{in}-\Gamma_{i}^{ion}m_{i}\mathbf{v}_{in}-\Gamma^{cx}m_{i}\mathbf{v}_{in}-\mathbf{R}_{ni}^{cx}+\mathbf{R}_{in}^{cx}\right)\nonumber \\
\left(\frac{\partial\mathbf{v}_{e}}{\partial t}\right)_{CE} & =\frac{1}{\rho_{e}}\left(\mathbf{R}_{ei}+\mathbf{R}_{en}+\Gamma_{i}^{ion}m_{e}(\mathbf{v}_{n}-\mathbf{v}_{e})\right)\label{eq:541}\\
\left(\frac{\partial\mathbf{v}_{n}}{\partial t}\right)_{CE} & =\frac{1}{\rho_{n}}\biggl(\mathbf{R}_{ni}+\mathbf{R}_{ne}+\Gamma_{n}^{rec}(m_{i}\mathbf{v}_{i}+m_{e}\mathbf{v}_{e}-m_{n}\mathbf{v}_{n})\nonumber \\
 & \,\,\,\,\,\,\,+\Gamma_{n}^{ext}m_{n}(\mathbf{v}_{n0}-\mathbf{v}_{n})+\Gamma^{cx}m_{i}\mathbf{v}_{in}+\mathbf{R}_{ni}^{cx}-\mathbf{R}_{in}^{cx}\biggr)\nonumber 
\end{align}
Similarly, combining equations \ref{eq:522-1}, \ref{eq:533.1}, and
\ref{eq:540.1}, the equivalent set of terms in the species energy
equations are:{\small{}
\begin{align}
\left(\frac{\partial p_{i}}{\partial t}\right)_{CE} & =\left(\gamma-1\right)\biggl[Q_{ie}+Q_{in}+\frac{m_{i}}{m_{n}}\,Q_{n}^{ion}-Q_{i}^{rec}+\frac{1}{2}m_{i}(\Gamma_{i}^{ion}+\Gamma^{cx})v_{in}^{2}\nonumber \\
 & -\mathbf{v}_{in}\cdot\mathbf{R}_{in}^{cx}+Q_{in}^{cx}-Q_{ni}^{cx}\biggr]\nonumber \\
\left(\frac{\partial p_{e}}{\partial t}\right)_{CE} & =\left(\gamma-1\right)\left(Q_{ei}+Q_{en}+\frac{m_{e}}{m_{n}}\,Q_{n}^{ion}-Q_{e}^{rec}+\Gamma_{i}^{ion}\left(\frac{1}{2}m_{e}\left(\mathbf{v}_{e}-\mathbf{v}_{n}\right)^{2}-\phi_{ion}\right)\right)\label{eq:542}\\
\left(\frac{\partial p_{n}}{\partial t}\right)_{CE} & =\left(\gamma-1\right)\biggl[Q_{ni}+Q_{ne}+\frac{m_{i}}{m_{n}}\,Q_{i}^{rec}+\frac{m_{e}}{m_{n}}\,Q_{e}^{rec}-Q_{n}^{ion}+\Gamma_{n}^{rec}\biggl(\frac{1}{2}m_{n}v_{n}^{2}+\frac{1}{2}m_{i}v_{i}^{2}+\frac{1}{2}m_{e}v_{e}^{2}-m_{i}\mathbf{v}_{n}\cdot\mathbf{v}_{i}\nonumber \\
 & -m_{e}\mathbf{v}_{n}\cdot\mathbf{v}_{e}\biggr)+\Gamma_{n}^{ext}\frac{1}{2}m_{n}\left(\mathbf{v}_{n}-\mathbf{v}_{n0}\right)^{2}+\Gamma^{cx}\frac{1}{2}m_{i}v_{in}^{2}+\mathbf{v}_{in}\cdot\mathbf{R}_{ni}^{cx}-Q_{in}^{cx}+Q_{ni}^{cx}\biggr]+\Gamma_{n}^{ext}T_{n0}\nonumber 
\end{align}
}{\small\par}

\subsection{Mass conservation}

Mass conservation (equation \ref{eq:531.2}) for species $\alpha$
can be written as:
\begin{align*}
\frac{\partial n_{\alpha}}{\partial t} & =-\nabla\cdot(n_{\alpha}\mathbf{v}_{\alpha})+\left(\frac{\partial n_{\alpha}}{\partial t}\right)_{CE}
\end{align*}

\subsection{Momentum conservation}

The expression for species momentum conservation (equation \ref{eq:531.401})
can be re-expressed as:
\[
\frac{\partial\mathbf{v}_{\alpha}}{\partial t}=-(\mathbf{v}_{\alpha}\cdot\nabla)\mathbf{v}_{\alpha}+\frac{1}{\rho_{\alpha}}\left(-\nabla p_{\alpha}-\nabla\cdot\underline{\mathbf{\boldsymbol{\pi}}}_{\alpha}+q_{\alpha}n_{\alpha}\left(\mathbf{E}+\mathbf{v}_{\alpha}\times\mathbf{B}\right)+\mathbf{R}_{\alpha}\right)\mbox{}
\]
The term $\mathbf{R}_{\alpha}/\rho_{\alpha}$, which arose by taking
the first moment of the collision operator for scattering collisions
only, is replaced with the appropriate term from equation \ref{eq:541}
to yield the momentum equations in the three-fluid system:
\begin{align*}
\frac{\partial\mathbf{v}_{\alpha}}{\partial t} & =-(\mathbf{v}_{\alpha}\cdot\nabla)\mathbf{v}_{\alpha}+\frac{1}{\rho_{\alpha}}\left[-\nabla p_{\alpha}-\nabla\cdot\overline{\boldsymbol{\pi}}_{\alpha}+q_{\alpha}n_{\alpha}\left(\mathbf{E}+\mathbf{v}_{\alpha}\times\mathbf{B}\right)\right]+\left(\frac{\partial\mathbf{v}_{\alpha}}{\partial t}\right)_{CE}
\end{align*}

\subsection{Energy conservation}

The expression for species energy conservation is \cite{farside_plasma}:
\[
\frac{\partial p_{\alpha}}{\partial t}=-\mathbf{v}_{\alpha}\cdot\nabla p_{\alpha}-\gamma p_{\alpha}\nabla\cdot\mathbf{v}_{\alpha}+(\gamma-1)\left(-\underline{\mathbf{\boldsymbol{\pi}}}_{\alpha}:\nabla\mathbf{v}_{\alpha}-\nabla\cdot\mathbf{q}_{\alpha}+Q_{\alpha}\right)
\]
The term $\left(\gamma-1\right)Q_{\alpha}$, which arose by taking
the second moment of the collision operator for scattering collisions
only, is replaced with the relevant term from equation \ref{eq:542},
to obtain the energy equations in the three-fluid system: 
\begin{align*}
\frac{\partial p_{\alpha}}{\partial t} & =-\mathbf{v}_{\alpha}\cdot\nabla p_{\alpha}-\gamma p_{\alpha}\nabla\cdot\mathbf{v}_{\alpha}+(\gamma-1)\left[-\overline{\boldsymbol{\pi}}_{\alpha}:\nabla\mathbf{v}_{\alpha}-\nabla\cdot\mathbf{q}_{\alpha}\right]+\left(\frac{\partial p_{\alpha}}{\partial t}\right)_{CE}
\end{align*}

\end{document}